\documentclass[aps,showpacs,prb,floatfix,twocolumn]{revtex4}
\usepackage{amsmath,amssymb,graphicx,bm,epsfig,color}

\newcommand{\eps}{\epsilon} \renewcommand{\a}{\alpha}
\renewcommand{\b}{\beta} 
 \newcommand{\vk}{{\mathbf{k}}}
\newcommand{\vK}{{\mathbf{K}}} \newcommand{\vq}{{\mathbf{q}}}
\newcommand{\vQ}{{\mathbf{Q}}} \newcommand{\vPhi}{{\mathbf{\Phi}}}
\newcommand{\vS}{{\mathbf{S}}} \newcommand{\cG}{{\cal G}}
\newcommand{\cF}{{\cal F}} \newcommand{\cD}{{\cal D}}
\newcommand{\Tr}{\mathrm{Tr}} \newcommand{\npsi}{\underline{\psi}}

 \newcommand{\om}{\omega}
\renewcommand{\Im}{\mathrm{Im}} \newcommand{\up}{\uparrow}
\newcommand{\down}{\downarrow}

\begin{document}

\title{Strongly Correlated Superconductivity: a plaquette
Dynamical mean field theory study}

\author{Kristjan Haule and Gabriel Kotliar}

\affiliation{Department of Physics, Rutgers University,
Piscataway, NJ 08854, USA}

\begin{abstract}
We use cluster Dynamical Mean Field Theory to
study the simplest models of correlated electrons, the Hubbard
model and the t-J model. We use a plaquette embedded in a medium
as a reference frame to compute and interpret the physical
properties of these models. We study various observables such as
electronic lifetimes, one electron spectra, optical
conductivities, superconducting stiffness, and the spin response
in both the normal and the superconducting state in terms of
correlation functions of the embedded cluster.  We find that the
shortest electron lifetime occurs near optimal doping where the
superconducting critical temperature is maximal. A second critical
doping connected to the change of topology of the Fermi surface is
also identified. The mean field theory provides a simple physical
picture of three doping regimes, the underdoped, the overdoped
and the optimally doped regime in terms of the physics of the
quantum plaquette impurity model. We compare the plaquette
Dynamical Mean Field Theory results with earlier resonating
valence bond mean field theories,
noting the improved
description of the momentum space anisotropy of the normal state
properties and the doping dependence of the coefficient of the
linear temperature dependence of the superfluid density in the
superconducting state.

\end{abstract}  \date{\today} \maketitle

\section{Introduction}

The origin and the nature of superconductivity in strongly correlated
materials is one of the greatest challenges in modern condensed matter
theory.  It received renewed attention with the discovery of the high
temperature superconductivity in copper oxide based materials. While
these materials have been studied intensively over the past decades
there is still no concensus as to what are the essential physical
ingredients responsible for the high temperature superconductivity
phenomena and how it should be modeled
\cite{NaturePhys,review1,review2,Fisher,review2.1,review3,review4,review5,review6,review7,review8,review9,review10,review11}.

P.W. Anderson proposed that the high temperature superconductivity
phenomena was intimately connected to the proximity to a parent Mott
insulating state \cite{Anderson1,Anderson11}.

Developing precise connections between the proximity to a Mott
insulator and high temperature superconductivity has proved to be a
difficult problem.  Suggestive conclusions have been reached using
slave boson methods \cite{Anderson_Zou,Gabi}, variational wave
functions \cite{Gros,Paramekanti}, and gauge theory techniques
\cite{review1}. However, lack of theoretical tools has made difficult
to prove that simple models are sufficient to explain the phenomenas
surrounding cuprates. For example it is still strongly debated whether
the existence of superconductivity with a high critical temperature
and a pseudogap is a genuine property of the models studies, or, an
artifact of the approximations employed to solve the model.

Over the past decade, significant progress in the field of correlated
electrons has been achieved through the development of Dynamical Mean
Field Theory \cite{DMFT-review,phys-today}.  In its single site
version, this method describes lattice models, in terms of a single
site impurity problem embedded in a medium.  The method has been very
successful in describing, and even predicting numerous properties of a
large number of materials
\cite{new-review,Held:2003,Held:2001:IJMP,Held:2001,Pu-nature,Pu-phonons,Am,Ce-optics,Cm-Pu}.
Cluster extensions of this method, Cluster Dynamical Mean Field Theory
(CDMFT) (for reviews see \onlinecite{DCA-review,new-review}), have
been proposed and are currently a subject of intensive investigations.

In this article  we apply  the cluster dynamical mean field
approach  to construct a mean field theory of the  simplest
models of strongly correlated materials, the one band Hubbard and
t-J models, using a $2\times 2$ cluster, namely the plaquette as
the basic mean field reference frame.

There are several motivations for constructing a mean field theory
based on a plaquette embedded in a dynamical bath of conduction
electrons : a) A plaquette embedded in a a self consistent medium can
describe the physics of singlet formation, which is very important in
the t-J and Hubbard model. There are two roads of singlet formation,
the Kondo effect, in which a spin can form a singlet with a bath of
conduction electrons, or, the superexchange mechanism which locks two
spins on a bond in a singlet state.  b) A plaquette in a medium is a
minimal unit to describe d-wave superconductivity and
antiferromagnetism on the same footing, given that their order
parameters (as well as that of other forms of order competing with
superconductivity) naturally fit on a plaquette.

From a methodological perspective, mean field theory allows to study
physical properties of different phases as a function of control
parameters, whether they are stable or metastable.  For example, we
will study the evolution of the superconducting state, together with
the underlying normal state, which appears as a metastable phase below
$T_C$. From a theoretical perspective metastable states are only
defined within a mean field theory, but they are of clear physical
relevance. Furthermore, comparison response or correlation functions
in both the normal and the superconducting state gives important clues
as to the mechanism of superconductivity.

 A clear understanding of the evolution of well defined mean
field phases of simplified  model is an important step towards
constructing the phase diagram of realistic Hamiltonians. Even if
a phase is not realized as the thermodynamically stable phase in
a mean field treatment of a simplified Hamiltonian, it could be
stabilized by adding additional longer range terms in the
Hamiltonian without significantly altering the short distance
properties described by the mean field theory.   Furthermore a
good understanding of the different mean field states can be
useful in elucidating  the results of numerical studies in larger
finite  clusters, since  complicated patterns  in a finite size
system, may be a reflection  of phase separation among different
competing mean field  phases.

The study of minimal models such as the t-J model or the Hubbard
model, describing a system near a Mott transition is an important
first step towards understanding real materials. From a study of
minimal models, one can learn what aspects follow from just the
proximity to a doping driven Mott transition. This is a necessary
step before the importance of other physical effects such as the
disorder or the electron phonon interactions certainly present in
the real materials, can be ascertained. A basic question yet to
be elucidated, is to which extent a minimal model of the doping
driven Mott transition such as the t-J model, describes at the
qualitatively level, the physical properties of the cuprates.  If
indeed the qualitative low energy physics of the cuprates results
from the proximity to a Mott insulating state, as described by a
minimal model of this phenomena, then the results can be refined
by including more realistic band structure, for example nearest
and next nearest neighbor hoppings, longer range interactions,
disorder and coupling to the lattice, as well as incorporating a
multiband situation which is needed to describe the physics in a
wider energy range.  It is possible to carry out these studies  in
the more realistic framework the combination of electronic
structure methods with DMFT, a subject which is left for future
studies. One should also ascertain the size of the corrections to
the mean field theory, by either expanding around mean field
theory\cite{Stan-exp} or by increasing the cluster size
\cite{Jarrel_Tc}.

Several studies have already shown that the Hubbard model treated
within cluster DMFT on a 2 by 2 plaquette successfully describes many
properties of the high temperature superconductors. For example the
competition of antiferromagnetism and superconductivity
\cite{Lichtenstein:2000,Jarrel-DCA,Capone_Nell,Senechal1,Senechal2},
the existence of a pseudogap at low doping
\cite{Jarrel-DCA2,Jarrel-pseudogap,Stanescu-Philips,tdstan,tudor,Kyung,Senechal3},
and the formation of Fermi arcs \cite{Parcollet,Civelli,tdstan,tudor}.

These phenomena involve short range non local correlations. In CDMFT
the approach to the Mott insulator is characterized by the growth of
the non local components of the self energy, which is responsible for
the phenomena of momentum space differentiation and the formation of
lines of zeros in the Green's function at zero temperature. Surprising
manifestations of strong correlations such as the transfer of optical
spectral weight upon condensation \cite{optics-Haule}, the existence
of an avoided quantum critical point \cite{qcp} underlying the
superconducting dome, and the presence of two distinct gaps
\cite{Marcello-submitted,Hanke} in the superconducting state of the
underdoped cuprates.  The approach describes well an anomalous
incoherent normal state \cite{Tremblay,optics-Haule} which is lifted
by the onset of superconductivity \cite{qcp,civelli}.

Other studies of the Hubbard model using large clusters at values of
$U \leq 8 t$ have focused on the convergence of the critical
temperature \cite{Jarrel_Tc}.  In a series of publications it has been
shown that the $d$ wave superconducting state is well described by
spin fluctuation theory \cite{treblay_SC,Jarrel_Tc,Maier_last}. To which extent
the physics of well defined quasiparticles interacting with spin
fluctuations responsible for the pairing, can be carried over to
strong coupling regime, is an important open problem, which can be
only be addressed by gaining a better understanding of the large U
limit of the Hubbard model, which is the focus of this article.

Hence we focus on understanding the physical content of the
plaquette mean field theory in the regime where the interaction
strength is large enough to drive a Mott transition at half
filling with a substantial Mott Hubbard gap. We gain insights by
comparing the superconducting state with the underlying normal
state. For example, we study the evolution of the Fermi arcs with
temperature, and trace the mechanism of superconductivity to the
optimization of the superexchange energy. We connect the maximum
critical temperature with anomalies at optimal doping, resulting
from a maximum in the inelastic scattering rate. The techniques
introduced in this paper, provides a simple interpretation of the
cuprate phase diagram in terms of the occupations of a small
number of cluster eigenstates or pseudoparticles which describe a
mean-field coarse-grained version of the important excitations of
the lattice system, and we use them to describe different
experimental probes, tunneling optics, neutron scattering, in
both the normal and the d-wave superconducting phase.
The superconducting state is characterized by two energy scales: one
increases with decreasing doping, and one decreases with decreasing
doping.  The first can be identified with the photoemission gap in the
antinodal region while the second can be identified with the slope of
the dirac cone along the Fermi surface. We investigate the effect of
the latter scale on the penetration depth.

The organization of the paper is the following: In section
\ref{formalism1} we summarize the formalism and introduce the models,
the cluster schemes and the impurity solvers, i.e.  the continuous
time Quantum Monte Carlo (CTQMC) \cite{Werner_1,ctqmc} and a
generalization of the non crossing approximation (NCA)
\cite{Bickers-review,tJEDMFT,tJEDMFT1,HaulefU}.
Section \ref{oneparticle} describes the evolution of the cluster
Green's functions and the self energies as a function of doping.  We
identify the existence of an anomalous scattering rate describing the
nodal region of the lattice model, which peaks at a characteristic
doping ${\delta_{2}}^{c}$ in the normal phase. The scattering rate is
drastically reduced in the superconducting state. We identify a second
characteristic doping ${\delta_{1}}^{c} $ at which another self energy
diverges, and connect this phenomena to the formation of lines of
zeroes in the Greens function.

One can view CDMFT in the superconducting phase as a
generalization of the Migdal Eliashberg theory to strongly
correlated electron systems, and
we present the frequency dependence of the superconducting order
parameter  in section \ref{supra}.  A great strength of the mean
field theory, is that it allows us to study the "normal" state
underlying the superconducting state and how it evolves with
temperature. This is done for the tunneling density of states in
section \ref{supra} for the optical conductivity in \ref{optics}
and for the magnetic properties in section \ref{magnetism}.  This
comparison between the mean field normal state and the mean field
superconducting state, establishes the superexchange as the main
pairing mechanism, as surmised in the RVB theory.

The pseudoparticles representing plaquette eigenstates, are not
only technical tools to set up strong coupling impurity solvers
but provide a physical picture of the excitations of the system
and we use them to interpret the CDMFT results in section
\ref{pseudoparticles}. We conclude with the connection between
our method and an earlier simpler mean field theory approach
based on the plaquette, the slave boson mean field theory and
closely related methods.
For related work advancing the RVB concepts using single site DMFT on
multiorbital models see Refs.~\onlinecite{CaponeM1,CaponeM2}.

\section{Formalism} \label{formalism1}

In this section we summarize the methodology used for our
investigation. Two minimal models of the proximity to a Mott
transition were considered: the Hubbard model and the t-J model.
There are several different versions of Dynamical Mean Field
Theory. For example in addition to standard DMFT, an extended version
of DMFT (EDMFT)
\cite{Quimiao,tJEDMFT,tJEDMFT1,Chitra,georges_old,georges_old2,georges_old3,new-review}
which replaces all the non local terms in the interaction (namely the
kinetic energy and the superexchange ) by a fermionic and a bosonic
bath has been proposed.  There are also numerous variants of cluster
Dynamical Mean Field Theory which differ by the dynamical medium
surrounding the plaquette (hybridization function of the impurity
model). Finally, the solution of the impurity model that results from
the CDMFT mapping, can be carried out with different impurity
solvers. In this work we use two complementary solvers, the non crossing
approximation (NCA) and a continuous time Quantum Monte Carlo (CTQMC)
method.

The goal of this article, is to highlight physical properties which
follow generally from the proximity to a Mott insulating state which
are captured by a local approach, namely cluster DMFT. For this
reason, we have focused on physics which emerges from both the Hubbard
and t-J model, and which is captured by all the different cluster
schemes (Cellular DMFT \cite{CDMFT}, Dynamical Cluster Approximation \cite{DCA} and their
extended versions). While we mention some quantitative differences
between these schemes, the stress is on qualitative main conclusions
that can be obtained with all quantum cluster schemes. In order to
keep the presentation clear and the article relatively concise we
provide only methodological details which are not available in the
literature.  To avoid unnecessary duplication, results for a given
physical quantity are presented with only one cluster scheme and
impurity solver, chosen to demonstrate more clearly a physical point.

\subsection{Models}

One of the more studied models in the field of strongly correlated
electrons is the Hubbard model defined by the Hamiltonian

\begin{equation}
  H = -\sum_{ij\sigma} t_{ij}c_{i\sigma}^\dagger c_{j\sigma}+
  \sum_{i} U n_{i \uparrow } n_{i \downarrow }
\end{equation}

It consist of a hopping term and an on-site repulsion. To be above
the Mott transition we take an on-site repulsion $U = 12t$.

A second model of great interest is  the t-J Hamiltonian
 \begin{equation}
  H = -\sum_{ij\sigma} t_{ij}c_{i\sigma}^\dagger c_{j\sigma}+
  \frac{1}{2}\sum_{ij} J_{ij}    \mathbf{S}_i\mathbf{S}_j .
\end{equation}

It contains two terms; first describes the kinetic energy which
delocalizes the holes introduced by doping, and the second
represents spin-spin interaction. In this work we take $J/t=0.3$.

In the t-J model,  a constraint forbidding all double occupancy
must be enforced, and will be treated exactly in this work.  In
the spirit of understanding general features of the proximity to
the Mott state we include only the nearest neighbor hopping $t=1$
($t'=0$).

\subsection{ Extended and standard DMFT }

In DMFT, the non local terms in the Hamiltonian coupling are replaced
by a a coupling to a bath of conduction electrons. In the Hubbard
model the only non local term is the kinetic energy, and this leads to
the standard DMFT mapping which is described in many reviews
\cite{DMFT-review}. In the t-J model also the superexchange
interaction connects different sites, and applying the DMFT philosophy
to that term also, leads to the Extended DMFT equations.

Here we outline the derivation of  the Extended version of the
cluster DMFT \cite{Quimiao,tJEDMFT,tJEDMFT1,Chitra}. We first
employ Hubbard-Stratonovich transformation to decouple the
non-local interaction term of the t-J model leading to the
following action \begin{widetext} \begin{eqnarray}
  S=\int_0^\beta d\tau\left\{ \sum_{\vk\sigma}c_{\vk\sigma}^\dagger(\tau)
  (\frac{\partial}{\partial\tau}-\mu+\eps_\vk)c_{\vk\sigma}(\tau)
  +\sum_i U n_{i\uparrow}(\tau)n_{i\downarrow}(\tau)
  \right.
  \left.
  +\sum_\vq\left[{\vPhi^\dagger}_\vq (\tau)\frac{2}{J_\vq}\vPhi_\vq(\tau)
  +i\; \vS_\vq\left(\vPhi_\vq^\dagger(\tau)+\vPhi_{-\vq}(\tau)\right)\right]
  \right\}.
\end{eqnarray} \end{widetext}
Here $\vPhi$ is the Hubbard-Stratonovich vector bosonic field which
decouples the spin-spin interaction,

 The many body theory described by the action above
 can be summarized in a functional:
\begin{eqnarray}
  \Gamma[\cG,\cD]&=&-\Tr\log(G_{0}^{-1}-\Sigma)-\Tr[\cG\Sigma]\nonumber\\
  &+&\frac{1}{2}\Tr\log(\cD_{0}^{-1}-\Pi)
  +\frac{1}{2}\Tr[\cD\Pi]+\Phi[\cG,\cD].
  \label{functional}
\end{eqnarray}
Here functional $\Phi$ of the exact Baym-Kadanoff functional contains
all two particle irreducible diagrams of an electron-boson system with
propagators $\cG,\cD $.  Extremizing the functional Eq.~(\ref{functional})
leads to the exact Dyson equations for this system. Cluster
approximations are obtained by restricting the functional to a
subset of trial Greens functions.  In the Cellular-DMFT
(C-DMFT)\cite{CDMFT,new-review},
the $\Phi$ functional is approximated as
follows: The full lattice is covered by non-overlapping clusters.
The functional within each cluster is treated exactly, i.e., if
two lattice points $i$ and $j$ are inside the same cluster
$\Phi^{C-DMFT}[{\cal G}_{ij},{\cal D}_{ij}] = \Phi^{exact}[{\cal
G}_{ij},{\cal D}_{ij}]$. If however, $i$ and $j$ are in different
clusters, $\Phi$ functional is set to zero. In this way, short
range correlations within the cluster are treated exactly, while
long range correlations are ignored.

Cluster approximations are obtained by replacing the exact
functional $\Phi$ in Eq.~(\ref{functional}) by its cluster
counterpart. The saddle point equations then become $\Sigma_{cluster}
= \delta\Phi(G_{cluster})/\delta G_{cluster}$,
$\Pi_{cluster}=-2\delta\Phi(D_{cluster})/\delta\Pi_{cluster}$.

The fluctuating bosons $\vPhi_\vq$  in the Extended DMFT
formalism allow to keep some out-of cluster short range
correlations and describe better the spin fluctuations by
allowing the cluster spin to relax more efficiently through its
direct exchange interaction with the bath.
We will see that this leads to higher superconducting critical
temperatures. Apart this quantitative difference, we did not find
any qualitative difference between extended version (which employs
bosons to describe spin fluctuations between the clusters) and
the results of the non-extended version of DMFT.

\subsection{Cluster Schemes and impurity models}

There are several cluster schemes in use, in the study of
correlated electron materials. The Dynamical Cluster
Approximation (DCA)\cite{DCA}  can be  thought of as a  coarse graining in
momentum space, obtained by relaxing the conservation of momentum. Rather than
treating the infinite number of lattice $\vk$ points and
corresponding Green's functions $\cG_\vk$, the  $\Phi$ functional
is approximated to depend only on the Green's function of a few
cluster momenta which we will denote by capital letters $\vK$ and
$\vQ$. The cluster Green's functions of the approximate functional
$\Phi[\cG_\vK,\cD_\vQ]$ are obtained by course graining the exact
Green's functions, i.e.,
$\cG_\vk\rightarrow\cG_\vK=\sum_{\vk\in\vK}{\cal G}_\vk$ and
$\cD_\vq\rightarrow\cD_\vQ=\sum_{\vq\in\vQ}\cD_\vQ$ where the sum
$\sum_{\vk\in\vK}$ is over those $\vk$ momenta in Brillouin zone
which correspond to certain cluster momenta $\vK$
(see Ref.~\onlinecite{DCA-review} and Ref.~\onlinecite{DCA}).

The results of this  paper were obtained with  both DCA and C-DMFT.
Again, all the qualitative features to be discussed in the next
sections can be seen with both methods.  Since DCA is a cluster
method with a simple interpretation in momentum space while C-DMFT
has a simple interpretation in real space,  the fact that the
qualitative physics emerges from both approaches, suggests that
the physical properties that we discuss in this article, are
genuine properties of cluster dynamical mean field theory on a
plaquette irrespectively of the specific cluster  scheme used.

We summarize the abbreviations used in the remaining of the text:
\begin{itemize}
\item CDMFT: cluster DMFT.
\item C-DMFT: Cellular DMFT \cite{CDMFT}.
\item DCA: Dynamical cluster approximation \cite{DCA}.
\item EC-DMFT: Extended version of Cellular DMFT
\item EDCA: Extended version of Dynamical cluster approximation \cite{Maier_EDCA,DCA}
\end{itemize}

A great advantage of all cluster DMFT formulations is that the
complicated functional Dyson equations for the self energies
and cluster response functions
can be witten in terms of an impurity model
\begin{widetext} \begin{eqnarray} Z=\int D[\psi^\dagger\psi]
\exp\left({-S_{cluster} -
   \int_0^\beta d\tau\int_0^\beta d\tau' \sum_{\vK}
   \underline{\psi}_\vK^\dagger(\tau)\underline{\Delta}_{\vK}(\tau,\tau')\underline{\psi}_{\vK}(\tau') +
   \frac{1}{2}\int_0^\beta d\tau\int_0^\beta d\tau' \sum_{\vQ}
   {\bf S}_\vQ(\tau){\chi_0^{-1}}_{\vQ}(\tau,\tau'){\bf S}_{\vQ}(\tau')}\right)
 \label{Zimp}
\end{eqnarray} \end{widetext}
which is numerically tractable and  where the effective Weiss
fields $\mathbf{\Delta}$ and ${\chi_0^{-1}}$ have to obey the
following self-consistency conditions
\begin{equation}
\mathbf{G}=\sum_{\vk}(i\omega-H_{\vk}-\mathbf{\Sigma}(i\omega))^{-1}
=(i\omega-E_{imp}-\mathbf{\Sigma}(i\omega)-\mathbf{\Delta}(i\omega))^{-1}
\label{SCC1}
\end{equation}
\begin{equation}
\mathbf{\chi}=\sum_{\vk}(\mathbf{M}(i\omega)+ J_{\vq})^{-1}
=(\mathbf{M}(i\omega)+\mathbf{\chi}_0^{-1}(i\omega))^{-1}
\label{SCC2}
\end{equation}
which merely express the fact that the cluster quantities, computed
from the impurity model
$1/(i\omega-E_{imp}-\mathbf{\Sigma}(i\omega)-\mathbf{\Delta}(i\omega))$,
have to coincide with the lattice local quantities when summing over
the reduced Brillouin zone.
Namely, in the C-DMFT, the lattice was
divided into non-overlapping clusters, hence the summations over $\vk$
run over the reduced Brillouin zone.
Here $\mathbf{M}$ plays the role of the spin
self-energy which is computed from the local susceptibility and Weiss
field by $\mathbf{M} = \mathbf{\chi}^{-1}-\mathbf{\chi}_0^{-1}$, as
evident from the Eq.~(\ref{SCC2}).

A special feature of the $2\times 2$ plaquette is worth stressing: the
cluster momentum $\vK$ is a good quantum number and therefore local
quantities like Green's function $\mathbf{G}$ or hybridization
$\mathbf{\Delta}$ take a diagonal form: \begin{eqnarray} \mathbf{G} =
\left(
\begin{array}{cccc} \underline{G}_{0,0}& 0 & 0 & 0\cr 0 &
\underline{G}_{\pi,0}& 0 \cr 0 & 0 & \underline{G}_{0,\pi}& 0\cr
0 & 0 & 0 & \underline{G}_{\pi,\pi}\cr \end{array} \right)
\label{Gmatrix} \end{eqnarray}
For large clusters, cellular DMFT would lead to off-diagonal terms in
the impurity action written in the basis of cluster momenta.  The
hybridization function in Eq.~(\ref{Zimp}) would take the form
$\underline{\psi}_\vK^\dagger
\underline{\Delta}_{\vK\vK'}\underline{\psi}_{\vK'}$. However, in the
$2\times 2$ case, both in C-DMFT and DCA, the hybridization function is
diagonal in cluster momentum.

\begin{figure}
\includegraphics[width=0.8\linewidth]{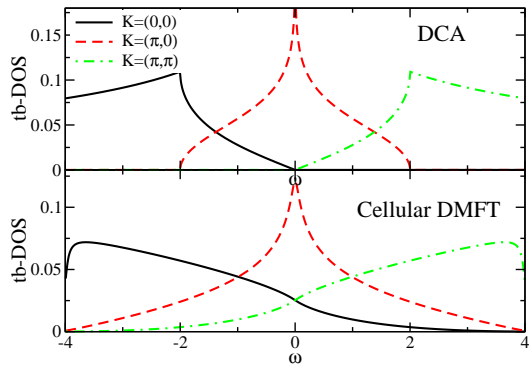}
\caption{
(Color online)
Tight-binding DOS for the three orbitals within DCA and
C-DMFT. Notice that the  tight-binding Hamiltonian within
C-DMFT Eq.~(\ref{H0}) contains off-diagonal elements
therefore DOS does not contain full information about the
non-interacting part of the Green's function $G_0$ ($G_0^{-1} =
G^{-1}+\Sigma$) }
\label{tbDOS}
\end{figure}

The DMFT mapping of the lattice model onto a plaquette in a medium,
allows us to make a connection between this problem and the
multiorbital Hubbard models which have been studied in connection with
the orbitally selective Mott transition \cite{OSMT1,OSMT2,OSMT3}.
This is defined by a set of bands, each one characterized by a local
density of a states, labeled by its cluster
wave-vector. Notice however that the interaction among the orbitals,
i.e. the Hubbard U term written in terms of
$\underline{\psi}_\vK^\dagger $ and $\underline{\psi}_\vK$ is more
complicated than what has been treated in the literature and deserves
further investigations.  The local density of states corresponding to
the different bands can be obtained by setting $U=0$ and evaluating
the non interacting Greens function $G_0$ corresponding to each
cluster wave vector. This is plotted in figure \ref{tbDOS}.

The formalism is easily extended to the  superconducting state
introducing Nambu notation
\begin{eqnarray}
  \npsi_\vK&=&\binom{c_{\vK\up}}{c_{-\vK\down}^\dagger}.
\end{eqnarray}
Assuming singlet  pairing, all the previous discussion carries
through, with the  cluster Greens functions and hybridization
functions taking the $2\times 2$ matrix  form:
\begin{equation}
\underline{G}_\vK(\tau)=-\langle T_\tau
  \npsi_\vK(\tau)\npsi_\vK^\dagger(0)\rangle=  \left(\begin{array}{cc}
    {\cal G}_{\vK\up}(\tau) & {\cal F}_{\vK}(\tau)\\
    {\cal F}_{\vK}^\dagger(\tau) & -{\cal G}_{-\vK\down}(-\tau)
  \end{array}\right).
\end{equation}
Here $\cal F_{\vK}$ is the anomalous component of the Green's
function. Hybridization $\underline{\Delta}_\vK$ becomes a matrix
as well \begin{equation}
  \underline{\Delta}_\vK(i\omega)=
  \left(\begin{array}{cc}
    \Delta_{\vK\up}(i\omega) & \Delta^{an}_\vK(i\omega)\\
    \Delta^{an\dagger}_\vK(i\omega) & -\Delta_{-\vK\down}(-i\omega)
  \end{array}\right)
\end{equation} and the impurity problem is off diagonal in Nambu
space.

In cluster momentum basis (see Eq.~(\ref{Gmatrix})), which we employed
in this work on the 2 by 2 plaquette, DCA and C-DMFT share the same
form of the impurity model, the only difference between the two
schemes lies in the form of the self consistency conditions.  This is
dictated by the form of the non-interacting part of the Hamiltonian
$H$ and the region of momentum summation.  In the DCA scheme the
non-interacting Hamiltonian $H_\vk$ is just the tight-binding energy
$\epsilon_\vk = -2t(\cos{k_x}+\cos{k_y})-4t'\cos{k_x}\cos{k_y}$. In
the self-consistency conditions Eqs.~(\ref{SCC1}) and (\ref{SCC2}),
the summation has to be performed only in the region of the patch
corresponding to each cluster momentum $\vK$ \cite{DCA-review}, i.e.,
\begin{equation} \underline{G}_\vK=\sum_{\vk\in\vK}
\left(\left(\begin{array}{cc} i\omega+\mu-\epsilon_\vk& 0\\ 0 &
i\omega-\mu+\epsilon_\vk
  \end{array}\right)
  -\underline{\Sigma}_\vK(i\omega)
  \right)^{-1}.
\end{equation}
The patches which correspond to different cluster momentum $\vK$ are
thus completely decoupled in the self-consistency condition. Their
coupling is only through the coulomb interaction.

In the real space C-DMFT, we can still define "orbitals" which
correspond to cluster momenta $\vK$ (see the form of local quantities
in Eq.(\ref{Gmatrix})), however, these orbitals are coupled through
both the Coulomb repulsion $U$ and the non-interacting Hamiltonian,
which takes the following form
\begin{widetext}
  \begin{equation} H_{\vk}=\left(
\begin{array}{cc|cc|cc|cc}
   \epsilon^0_\vk-\mu & 0 & i v^1_\vk & 0 & i v^2_\vk & 0 & v^0_\vk  & 0\cr
  0 & -\epsilon^0_\vk+\mu  & 0 & i v^1_\vk  & 0 & i v^2_\vk & 0 & -v^0_\vk \cr
\hline
  -i v^1_\vk&0&\epsilon^1_\vk-\mu& 0 & -v^0_\vk & 0& i v^4_\vk & 0\cr
  0& -i v^1_\vk& 0&  -\epsilon^1_\vk+\mu&  0& v^0_\vk & 0 & i v^4_\vk   \cr
\hline
 -i v^2_\vk & 0 & -v^0_\vk & 0 & \epsilon^2_\vk-\mu & 0 & i v^3_\vk & 0 \cr
 0& -i v^2_\vk & 0& v^0_\vk& 0 & -\epsilon^2_\vk+\mu&0& i v^3_\vk\cr
\hline v^0_\vk &0& -i v^4_\vk & 0& -i v^3_\vk & 0&
\epsilon^3_\vk-\mu   & 0\cr
 0 & -v^0_\vk& 0& -i v^4_\vk& 0 & -i v^3_\vk & 0&  -\epsilon^3_\vk+\mu\cr
\end{array} \right) \label{H0} \end{equation} \end{widetext}
where we defined \begin{eqnarray}
&&  \epsilon^0_\vk = -t(2+\cos{k_x}+\cos{k_y})-t'(1+\cos{k_x}\cos{k_y})\nonumber\\
&&  \epsilon^1_\vk = t(\cos{k_x}-\cos{k_y})+t'(1+\cos{k_x}\cos{k_y})\nonumber\\
&&  \epsilon^2_\vk = -t(\cos{k_x}-\cos{k_y})+t'(1+\cos{k_x}\cos{k_y})\nonumber\\
&&  \epsilon^3_\vk =  t(2+\cos{k_x}+\cos{k_y})-t'(1+\cos{k_x}\cos{k_y})\nonumber\\
&& v^0_\vk = t' \sin{k_x}\sin{k_y}\nonumber\\
&& v^1_\vk = \sin{k_x}(t+t'\cos{k_y})\nonumber\\
&& v^2_\vk = \sin{k_y}(t+t'\cos{k_x})\nonumber\\
&& v^3_\vk = \sin{k_x}(t-t'\cos{k_y})\nonumber\\
&& v^4_\vk = \sin{k_y}(t-t'\cos{k_y}).
\end{eqnarray}
The unit of distance choosen here is $a=1/2$ such that the summation
over the reduced Brillouin zone in Eqs.~(\ref{SCC1}) and (\ref{SCC2})
simply runs over $k_x\in[-\pi,\pi]$ and $k_y\in[-\pi,\pi]$. One can
readily show that this summation leads to diagonal form of local
quantities.

\begin{figure}
\includegraphics[width=0.8\linewidth]{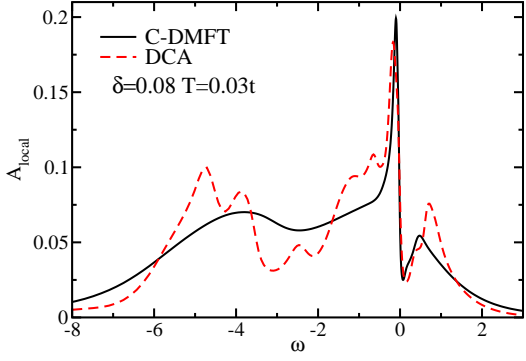}
\caption{(Color online)
  Comparison between the local spectral function computed in
  C-DMFT and  in  DCA  with  NCA  used as  impurity solver.}
\label{compA} \end{figure}

In figure \ref{compA} we compare the local spectral function of
the  t-J model in  the two cluster schemes. Notice the
similarities of the results in particular at low energies.
The spectral functions in both methods have a very similar
pseudogap. Hence in spite of quantitative differences, which will not
be investigated systematically in this paper, the qualitative physics,
which is the main focus of this article is present in both cluster
methods. Note, however that decoupling of orbitals in DCA method leads
to splitting of the Hubbard band into peaks which correspond to
excitations of the $2\times 2$ cluster. These finite size effects are
strongly reduced in C-DMFT method.

Here we comment on some quantitative differences between the
methods. The superconducting critical temperature is highest in EDCA
method and reaches the value $\sim 0.036t$ while it drops to $\sim
0.026t$ in EC-DMFT.  When the bosonic baths is switched-off, the real
space C-DMFT maximum critical temperature in both, the t-J model at
$J=0.3$ and the Hubbard model at $U=12t$ is around $\sim 0.01t$
. Notice that this value is close to the estimations in
Ref.~\onlinecite{Jarrel_Tc} for the critical temperature of the
Hubbard model in the thermodynamical limit for U = 4t. Namely, the
Hubbard model at $U=4t$ within large cluster DCA has $T_C\sim 0.023t$
\cite{Jarrel_Tc}. If we extrapolate this value to large $U=12t$
treating $T_C\propto J$ \cite{J_prop_T}, $T_C$ would drop to $\sim 0.008t$ which is
close to the C-DMFT result.

The existence of a finite transition temperature, the trends of the
superconducting transition temperature with doping and with the
strength of the superexchange interaction is a robuts propery of
plaquette DMFT and is common to all cluster schemes.
It would be interesting to understand the convergence properties
with cluster size within the different cluster schemes for the
t-J model, as was done for the Hubbard model at intermediate $U$
in Ref.~\onlinecite{Jarrel_Tc}
and in the classical limit in Ref.~\onlinecite{giulio}.

\subsection{Impurity Solvers}

At the heart of the cluster DMFT is the solution of the impurity
problem Eq.~(\ref{Zimp}).  In this work, we used two different
impurity solvers both based on the expansion of the impurity action
with respect to hybridization strength.  The first is the non-crossing
approximation (NCA) which sums up all diagrams with no crossing and is
conveniently formulated in slave particle approach \cite{HaulefU}.
The second is the recently implemented continuous time Monte Carlo Method
\cite{ctqmc,Werner_1} (CTQMC) which numerically samples the same type
of diagrams but sums up all diagrams using Monte Carlo importance
sampling.



\begin{figure}
\includegraphics[width=0.8\linewidth]{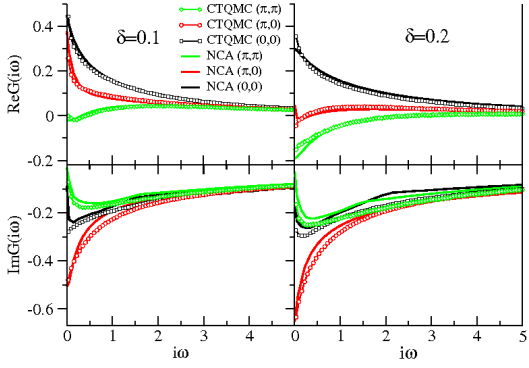}
\caption{(Color online)
  Comparison of NCA and CTQMC Green's functions on
imaginary
  axis for  several doping levels. We used real space C-DMFT.}
\label{compare} \end{figure}

The two impurity solvers are in good agreement with each other on
the imaginary axis, but the first method allows us to obtain real
frequency correlation functions which are unavailable in the QMC
approach.  Both approaches are well suited to study the regime of
intermediate temperatures and dopings, close to the tip of the
superconducting dome, separating overdoped and underdoped regions,
which is not easily accessible with other techniques.

Both impurity solvers used here require the introduction of the
cluster eigenstates obtained by the exact diagonalization of the
cluster, i.e., $H_{cluster}|m\rangle={E_m^{cluster}}|m\rangle$. To each
cluster eigenstate, a pseudoparticle $a_m$ can be assigned, i.e.,
\begin{equation} |m\rangle \equiv a_m^\dagger|0\rangle
\label{pseudo_def} \end{equation} to recast the cluster part of
the action to a quadratic form. The constraint \begin{eqnarray} Q
\equiv \sum_m a_m^\dagger a_m = \sum |m\rangle\langle m| = 1
\label{constraint} \end{eqnarray} which expresses the
completeness of the atomic eigenbase, has to be imposed.

The original problem can be exactly expressed in terms of
pseudoparticles $a_m$ with the only non-quadratic term of the
converted action being the hybridization between the cluster and
the medium \begin{widetext} \begin{eqnarray}
S_{eff}={\int_0^\beta\sum_m
a_m^\dagger(\tau)(\frac{\partial}{\partial\tau}+E^{cluster}_m-\lambda)a_m(\tau)
+
   \int_0^\beta d\tau \int_0^\beta d\tau'\sum_{mnm'n'}
   a_m^\dagger(\tau) a_n(\tau) D_{m n n'm'}(\tau-\tau')a_{n'}^\dagger(\tau')a_{m'}(\tau').
} \label{Zimp_pseudo} \end{eqnarray} denoted here by
\begin{eqnarray} D_{m_1 m_2 m_3 m_4}(i\omega) =\sum_\vK
(\mathbf{F}^{\vK\dagger})_{m_1
m_2}\underline{\Delta}_{\vK}(i\omega) (\mathbf{F}^{\vK})_{m_3
m_4}- \frac{1}{2} ({\bf S}_\vK)_{m_1
m_2}{\chi_0^{-1}}_{\vK}(i\omega) ({\bf S}_{\vK})_{m_3 m_4}
\end{eqnarray} \end{widetext} where \begin{eqnarray}
  (\mathbf{F}^\vK)_{mn}&=& \langle m| \npsi_\vK|n\rangle= \binom{\langle
  m|c_{\vK\up}|n\rangle}{\langle m|c_{-\vK\down}^{\dagger}|n\rangle}\\
  (\mathbf{S}_\vQ)_{mn}&=& \langle m| \mathbf{S}_\vQ|n\rangle.
\end{eqnarray} Note that the effective hybridization $D$ combines
both, the fermionic ($\underline{\Delta}_{\vK}$) and
bosonic bath (${\chi_0^{-1}}_{\vK}$) into the total effective
Weiss field felt by the cluster eigenstates (pseudoparticles). We
used lagrange multiplier $\lambda$ to enforce the constraint
(\ref{constraint}).

The continuous time quantum Monte Carlo method samples over the diagrams
generated by expanding the action $\int D[a^\dagger
a]\exp(-S_{cluster}-\Delta S)$ with respect to effective
hybridization $\Delta S$. Here $\Delta S$ stands for the second
term in Eq.~\ref{Zimp_pseudo}. The probability to visit each
diagram is proportional to its contribution to the partition
function which is computed by explicit evaluation of the cluster
trace $\int D[a^\dagger a]e^{-S_{cluster}} (-\Delta S)^k/k! $
keeping only single pseudoparticle in the system at each moment
in imaginary time. In this way, the constraint $Q=1$ is
explicitely taken into account. For more details, see
Ref.~\onlinecite{ctqmc}.

In the diagrammatic method, the constraint $Q=1$ is imposed by
letting the Lagrange multiplier $\lambda$ approach infinity.  The
physical observable can then be computed using the Abrikosov's
trick \cite{Abrikosov}
$\langle
A\rangle_{Q=1}=\lim_{\lambda\rightarrow\infty}\frac{\langle
QA\rangle}{\langle Q\rangle}$.

The coupling of the cluster to the medium, which simulates the rest of
the lattice, causes the cluster eigenstates to decay in time. Therefore
their spectral functions carry nontrivial frequency dependence and
important information about various physical processes such as the
RKKY interactions, the Kondo effect and d-wave superconductivity. The
corresponding pseudoparticle Green's function can be written in the
form
\begin{equation}
\overline{G}_{mn}(\omega) =
{(\omega+\lambda-E_{cluster}-\overline{\Sigma})^{-1}}_{mn}
\end{equation}
where $(E_{cluster})_{mn}=(E_{cluster})_m
\delta_{nm}$ is the energy of the cluster eigenstate and
$\lambda$ is Lagrange multiplier which will be sent to infinity
at the end of the calculation.

Although hybridization is a small quantity compared to other
scales in the problem, the perturbation is singular in the sense
that at zero temperature an infinite number of diagrams
substantially contribute to the solution of the problem. In
Ref.~\onlinecite{ctqmc} we showed a histogram (a distribution of
the perturbation order) which is peaked around $\langle
E_{kin}\rangle/T$ where $E_{kin}$ is average kinetic energy and
$T$ is temperature. An infinite resummation of diagrams is thus
necessary and the non-crossing diagrams are simplest to compute.

Just like in the  single site  Anderson and Kondo impurity problem
\cite{Bickers-review,tJEDMFT1}, the non-crossing approximation
works  well down to some breakdown temperature, which is slightly
below the superconducting transition temperature.  Although NCA is not
exact, this approximation has the virtue of yielding directly
real frequency information.
In Fig.~\ref{compare} we present  a typical comparison of the two
impurity solvers on imaginary axis for the cluster Green's functions
of the t-J model in normal state close to $T_c$.  This comparison
illustrates the degree of agreement within the two solvers on the
imaginary axis. Notice that all the qualitative features of the
evolution of the Greens functions with doping are seen
in both methods. Therefore we will use in this work, the strategy
of combining information from different solvers, in order to draw
conclusions as to the physical picture contained in the solution
of the cluster DMFT equations of the t-J and Hubbard model thus avoiding the
difficult problem of analytic continuation of imaginary time QMC
data.

In the non-crossing approximation, the pseudoparticle self-energies
are computed from
\begin{widetext} \begin{eqnarray}
  \overline{\Sigma}_{m^\prime m}(i\omega)=
 T \sum_{i\epsilon,nn^\prime}
 \overline{G}_{n^\prime n}(i\epsilon)
 \left\{
 D_{nmm'n'}(i\epsilon-i\omega)-D_{m'n'nm}(i\omega-i\epsilon)
  \right\}
\end{eqnarray} while the physical quantities like Green's
function and susceptibility are obtained by functional derivative
of the NCA Luttinger functional with respect to the hybridization
term and are given by \begin{eqnarray}
\underline{G}_\vK(i\omega)&=&- T\sum_{i\epsilon,mnm'n'}
(\mathbf{F^K})_{m'n'}\overline{G}_{n'n}(i\epsilon)\overline{G}_{mm'}(i\epsilon-i\omega)(\mathbf{F^{K\dagger}})_{nm}\\
\chi_\vQ^{\a\b}(i\omega)&=& T\sum_{i\epsilon,mnm'n'}
(\mathbf{S_\vQ^\a})_{m'n'}\overline{G}_{n'n}(i\epsilon)\overline{G}_{mm'}(i\epsilon-i\omega)(\mathbf{S_{-\vQ}^\b})_{nm}.
\end{eqnarray}
The above equations can be projected to the
physical subspace $Q=1$ only on the real axis. In the limit
$\lambda\rightarrow\infty$ they take the form \begin{eqnarray}
  \overline{\Sigma}_{m'm}(\omega)= \sum_{\vK,nn'}\int d\xi f(\xi)
  \overline{G}_{n'n}(\xi+\omega)\left\{
  \widehat{D}_{nmm'n'}(\xi)+\widehat{D}_{m'n'nm}(-\xi)
  \right\}
\end{eqnarray} \begin{eqnarray}
\underline{G}_\vK(\omega)&=&\sum_{mnm'n'}
(\mathbf{F^K})_{m'n'}(\mathbf{F^{K\dagger}})_{nm} \int d\xi
e^{-\beta\xi} \left[
  \overline{G}_{n'n}(\xi+\omega)\widehat{G}_{mm'}(\xi)
  -\widehat{G}_{n'n}(\xi)\overline{G}^*_{mm'}(\xi-\omega)
\right]\\
\chi^{\a\b}_\vQ(\omega)&=&\sum_{mnm'n'}
(\mathbf{S^\a_\vQ})_{m'n'}(\mathbf{S^\b_{\vQ}})_{nm} \int d\xi
e^{-\beta\xi} \left[
  \overline{G}_{n'n}(\xi+\omega)\widehat{G}_{mm'}(\xi)
  +\widehat{G}_{n'n}(\xi)\overline{G}^*_{mm'}(\xi-\omega)
\right] \end{eqnarray} \label{convolution} \end{widetext} Here we
used the following notation \begin{eqnarray}
{\widehat{D}}(\omega) = -\frac{1}{2\pi i}[{D}(\omega+i\delta)-{D}(\omega-i\delta)]\\
\widehat{G} = -\frac{1}{2\pi
i}[\overline{G}(\omega+i\delta)-\overline{G}(\omega-i\delta)].
\end{eqnarray}
The pseudoparticle quantities (Green's functions
$\widehat{G}$ and self-energies $\widehat{\Sigma}$) exponentially
vanish below a certain threshold energy (they have X-ray
singularity) which can be interpreted as the effective energy of
the many-body state associated with the pseudoparticle. These
thresholds can be removed by defining new quantities without
threshold \cite{Kroha}, i.e.,
\begin{eqnarray}
\widetilde{G}(\epsilon) = \widehat{G}(\epsilon)/f(-\epsilon)\\
\widetilde{\Sigma}(\epsilon) =
\widehat{\Sigma}(\epsilon)/f(-\epsilon) \end{eqnarray} Using these
quantities
we can rewrite the NCA equations as:

 \begin{widetext}
\begin{eqnarray}
  \widetilde{\Sigma}_{m'm}(\omega)= \sum_{\vK,nn'}\int d\xi \frac{f(\xi-\omega)f(-\xi)}{f(-\omega)}
  \widetilde{G}_{n'n}(\xi)\left\{
  \widehat{D}_{nmm'n'}(\xi-\omega)+\widehat{D}_{m'n'nm}(\omega-\xi)
  \right\}
\label{NCA_Sig} \end{eqnarray} \begin{eqnarray}
\Im\underline{G}_\vK(\omega)&=&-\pi\sum_{mnm'n'}
(\mathbf{F^K})_{m'n'}(\mathbf{F^{K\dagger}})_{nm} \int d\xi
\frac{f(\xi-\omega)f(-\xi)}{f(-\omega)}
  \widetilde{G}_{n'n}(\xi)\widetilde{G}_{mm'}(\xi-\omega)
\label{NCA_G}  \\
\Im \chi^{\a\b}_\vQ(\omega)&=&-\pi\sum_{mnm'n'}
(\mathbf{S^\a_\vQ})_{m'n'}(\mathbf{S^\b_{\vQ}})_{nm} \int d\xi
\frac{f(\xi-\omega)f(-\xi)}{b(-\omega)}
  \widetilde{G}_{n'n}(\xi)\widetilde{G}_{mm'}(\xi-\omega)
\label{NCA_Chi} \end{eqnarray} \end{widetext} At zero
temperature, the combination of the Fermi functions
$\frac{f(-\xi)f(\xi-\omega)}{f(-\omega)}=\frac{f(\xi)f(\omega-\xi)}{f(\omega)}$
is equal to unity in the interval $[min(0,\omega),max(0,\omega)]$
and zero outside.

These equations relate physical observables, like $G_\vK$ and
$\chi_\vQ$ to the pseudoparticle spectral functions. The later
represent coarse grained versions of the important many body
excitations of the system including fermionic quasiparticles and
bosonic collective modes. They have quantum numbers describing
their spin, number of particles, (which divided by the cluster
size, give the density), and a coarse grained momentum.

Relating several  experimental observables such as photoemission
spectra, tunneling spectra, and optical spectra, to the same set
of pseudoparticle spectral functions, gives   additional insights
into the important excitations of the system.

\section{Cluster one Particle Greens function,  Cluster Self Energy and Scattering Rate}
\label{oneparticle}

\begin{widetext}

\begin{figure}[h]
\includegraphics[width=1.\linewidth]{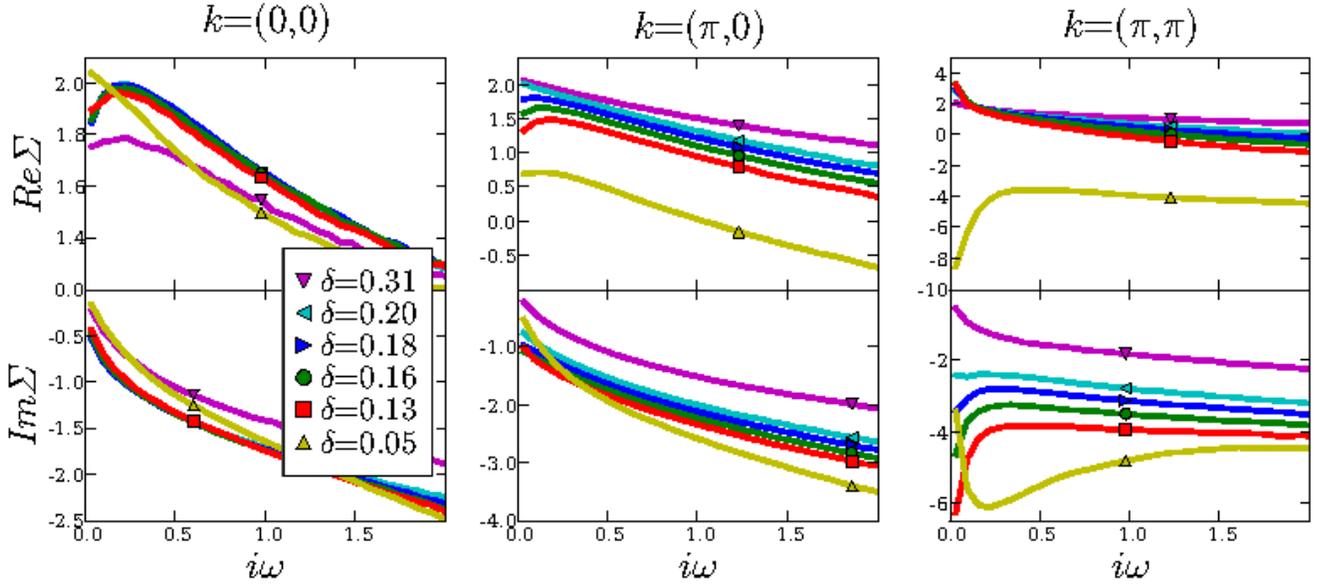}
\caption{ (Color online)
  C-DMFT cluster self-energies of the t-J model using CTQMC as
  the impurity solver. Temperature $T=0.01t$ and system is in the normal state.
Notice that $\Sigma_{00}$ is Fermi-liquid like (imaginary part vanishes at
zero matsubara frequency below the coherence temperature) in the
whole range of doping, $\Sigma_{\pi 0}$ is Fermi liquid in the
overdoped and underdoped regime while the scattering rate remains
of the order of unity in the optimally doped regime. Finally,
$\Sigma_{\pi\pi}$ is far the largest self-energy. Its real part
is so large that the orbital is gapped in all doping range
considered. The scattering rate is enormous and a pole appears on
the real axis around $\delta=0.1$. The pole is above $E_F$ at
very small doping, crosses $E_F$ at $\delta=0.1$ and goes below
$E_F$ for optimally doped and overdoped regime. This causes a
sign change of the real part of $\Sigma_{\pi\pi}$. The $(\pi\pi)$
orbital is thus in the Mott insulating state in most of the doping
range considered.
}
\label{Self_tj_100}
\end{figure}

\begin{figure}[h]
\includegraphics[width=1.\linewidth]{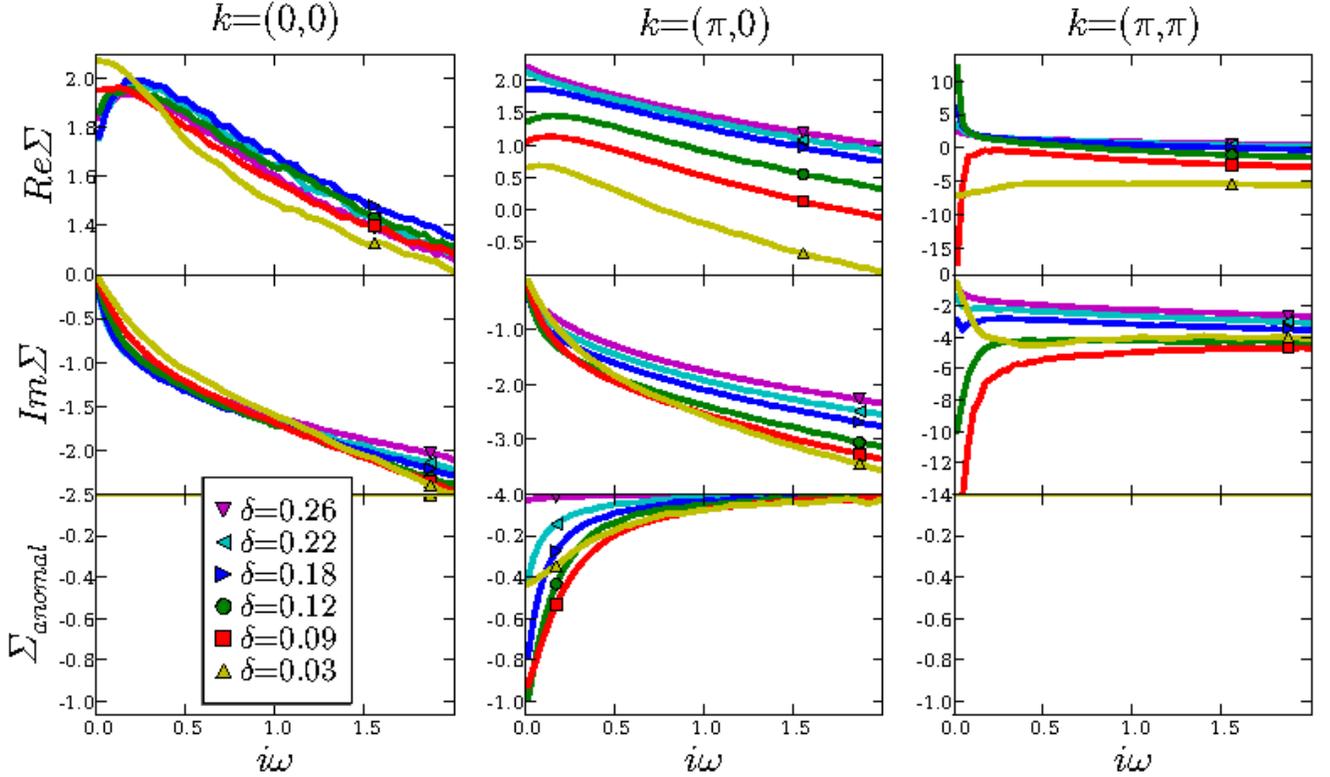}
\caption{ (Color online)
  Same as in Fig.~\ref{Self_tj_100} but at lower
temperature $T=0.005t$ in the superconducting state. The bottom
row shows the anomalouse self-energy. The $(0,0)$ orbital barely
changes in superconducting state. On the other hand, the large
scattering rate in $(\pi,0)$ orbital is severly reduced in the
superconducting state and the orbital becomes Fermi liquid like.
Large scattering rate in the normal state is now replaced by a
large anomalous component of self-energy (peaked around $\delta\sim
0.15$ - see Fig.~\ref{imSigctqmc}). Finally, the pole in
$(\pi,\pi)$ self-energy sharpens and the orbital remains Mott
insulating in most of the doping range considered. A pole of the
cluster self-energy is accompanied by a line of  zeros of the
Green's functions in certain parts of the momentum space
\cite{tdstan,tudor} and persist  in the superconducting state. }
\label{Self_tj_200} \end{figure}

\begin{figure}
\includegraphics[width=0.99\linewidth]{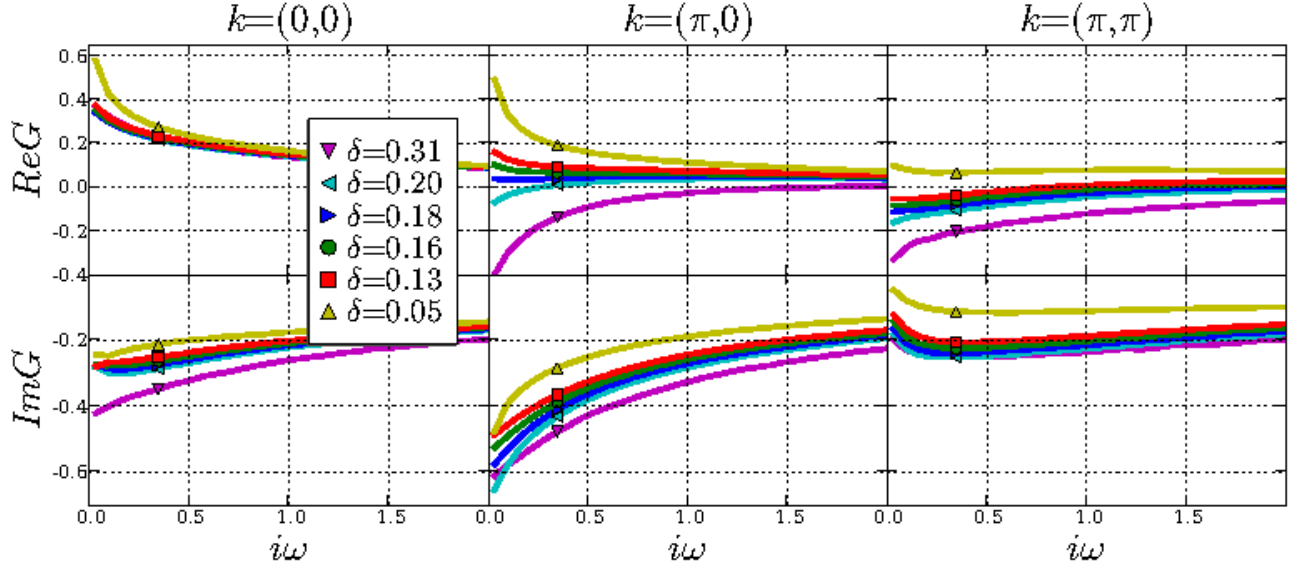}
\caption{ (Color online)
  C-DMFT cluster Green's functions at $T=0.01t$ in
the normal state of the t-J model obtained by CTQMC. The real part of
the Green's function vanishes for particle-hole symmetric
situation while its positive when spectral weight below $E_F$ has
"largest weight" and vice versa. The $(0,0)$ orbital does not
change much with doping and remains close to half-filling. The
$(\pi,0)$ orbital gives most of the weight at the Fermi level
(has largest imaginary part at zero frequency) and remarkably
becomes particle-hole symmetric at the doping level slightly
larger than the optimally doped level ($\delta=0.18$). The
$(\pi,\pi)$ orbital is gapped for all doping levels. }
\label{Gf_tj_100} \end{figure}

\begin{figure}
\includegraphics[width=1.\linewidth]{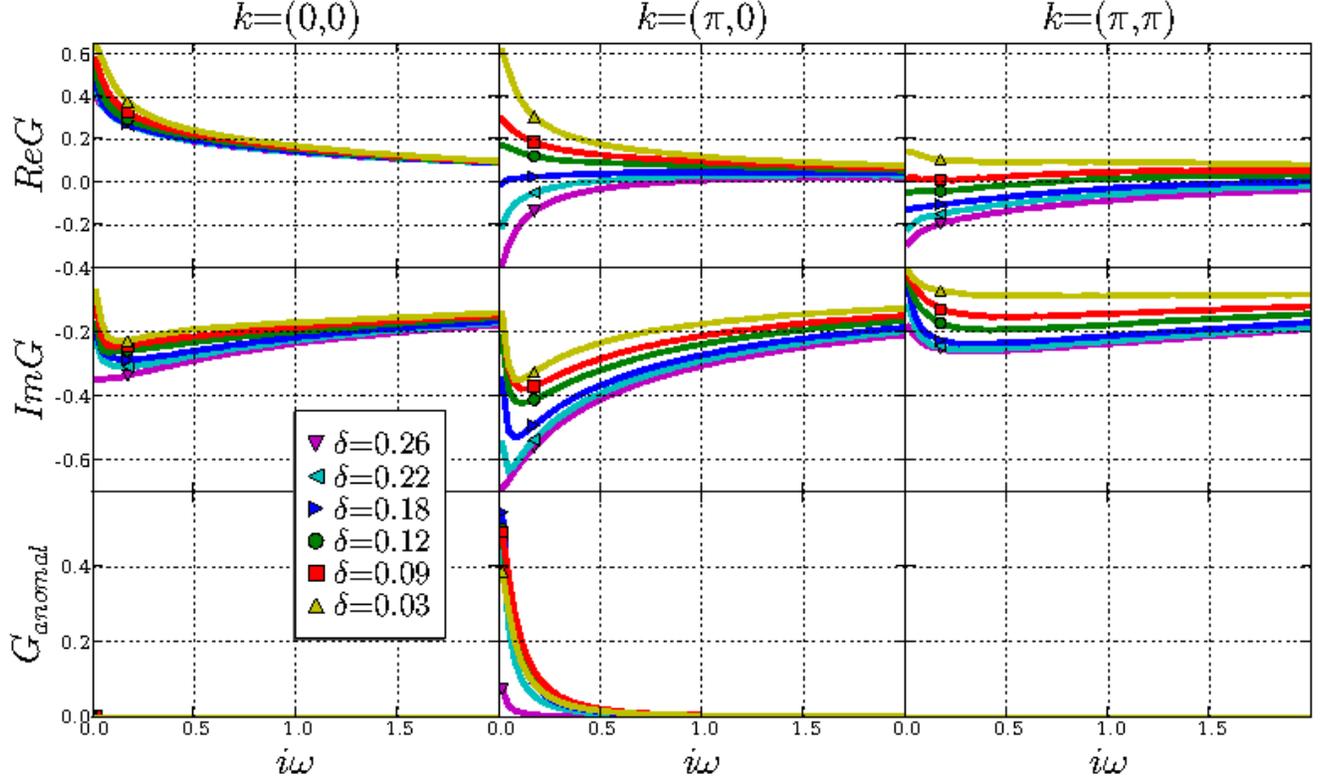}
\caption{
  (Color online)
C-DMFT cluster Green's functions of the t-J
model at $T=0.005t$ in the superconducting state obtained by CTQMC.
Superconducting gap opens in particular in $(\pi,0)$ orbital.
Particle-hole symmetry of this orbital is again evident from the
real part of the Green's function being close to zero around
optimal doping (blue curve with triangles pointing right). }
\label{Gf_tj_200} \end{figure}

\begin{figure}
\includegraphics[width=1.\linewidth]{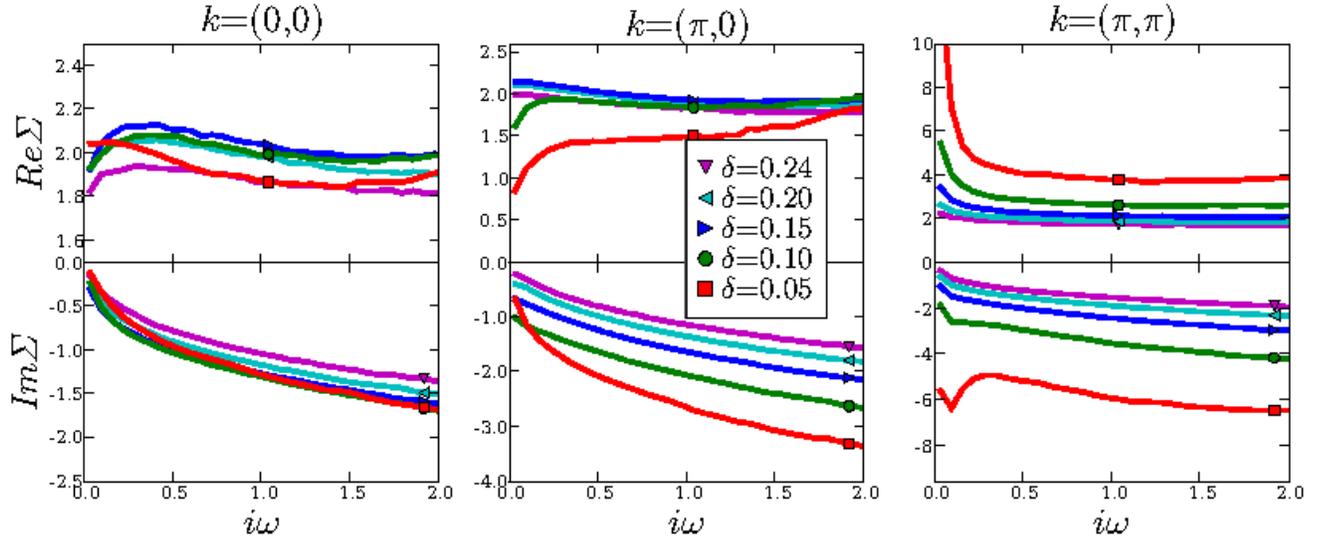}
\caption{
  (Color online)
  Hubbard model cluster self-energies in C-DMFT
obtained by CTQMC at $T=0.01t$ and $U=12t$ in the normal state.
Just like in the t-J model, the $(0,0)$ orbital is Fermi-liquid
like in the whole doping regime while $(\pi,0)$ orbital is
coherent only in the underdoped and overdoped regime. At optimal
doping (in the Hubbard model optimal doping is around $\delta\sim
0.1$) the scattering rate is largest. The important difference
appears in the $(\pi,\pi)$ orbital. The $(\pi,\pi)$ self-energy
is the largest self-energy of the system just like in the t-J
model. Contrary to the t-J model, the pole in the $(\pi,\pi)$
self-energy on the real axis, which appeares in the t-J model
around $\delta=0.1$, is now at zero doping. The self-energy of
the $(\pi,\pi)$ orbital thus monotonically grows when approaching
the Mott insulator.
}
\label{Sig_hub_100}
\end{figure}

\begin{figure}
\includegraphics[width=1.\linewidth]{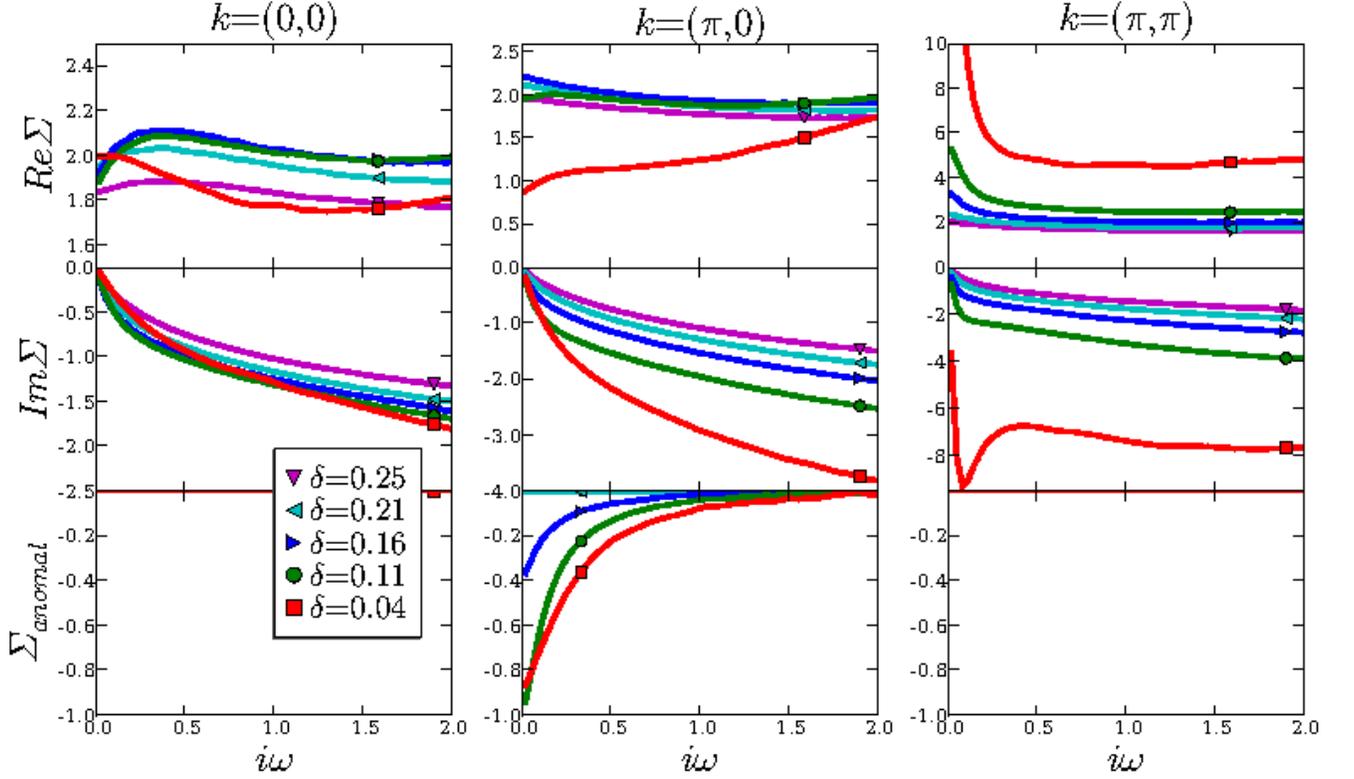}
\caption{
  (Color online)
  Similar than Fig.~\ref{Sig_hub_100} but at lower
temperature $T=0.005t$ in the superconducting state. Just like in
the t-J model, the $(\pi,0)$ orbital, which is representative of
the nodal part of the self-energy, becomes coherent in the
superconducting state and the anomalous self-energy is largest
around $\delta\sim 0.1$ where the scattering rate is largest in
the normal state. The $(\pi,\pi)$ self-energy sharpens with
decreasing temperature just like in the Hubbard model showing
that this orbital is in the Mott-insulating state in the
underdoped and optimally doped regime. } \label{Sig_hub_200}
\end{figure}

\begin{figure}
\includegraphics[width=1.\linewidth]{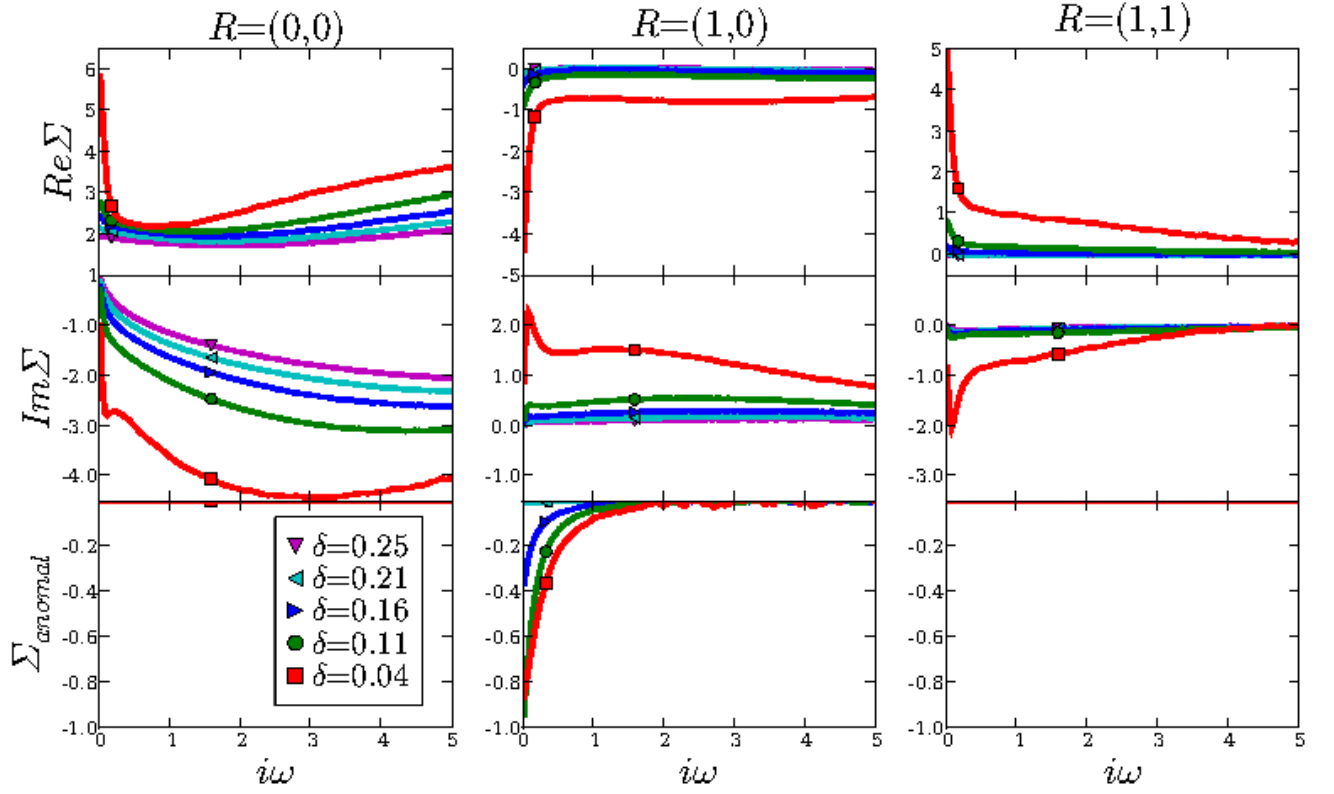}
\caption{ (Color online) The on-site and short range self-energy of the Hubbard
model in the superconducting state at $T=0.005t$. The on-site
self-energy is the largest and its imaginary part vanishes for
all finite dopings. The reason is that the pole in the
$(\pi,\pi)$ self-energy is now at zero doping. The non-local
components of the self-energy vanish rather rapidly with doping. }
\label{Sig_hub_re_200} \end{figure}

\begin{figure}
\includegraphics[width=0.99\linewidth]{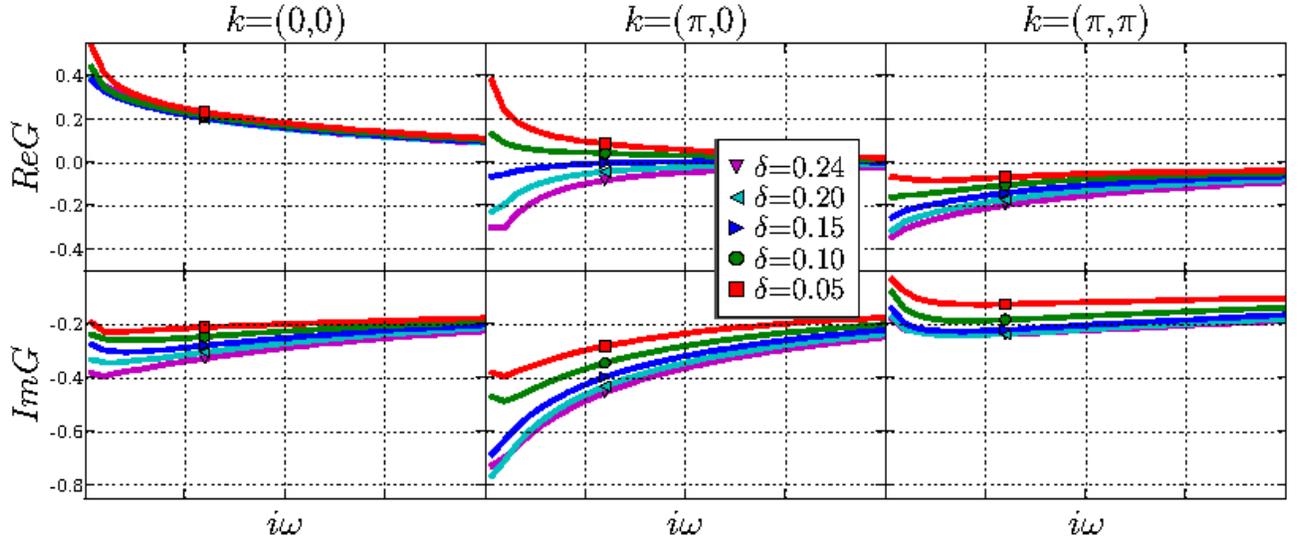}
\caption{ (Color online) Hubbard model Green's function in normal state at
$T=0.01t$. The physics is the same as in Fig.~\ref{Gf_tj_100} for
the t-J model. The only difference is that the particle-hole
symmetric point of the $(\pi,0)$ orbital appears around
$\delta\sim 0.12$ which is again slightly above the optimal doped
level.  } \label{Gf_hub_100} \end{figure}

\begin{figure}
\includegraphics[width=1.\linewidth]{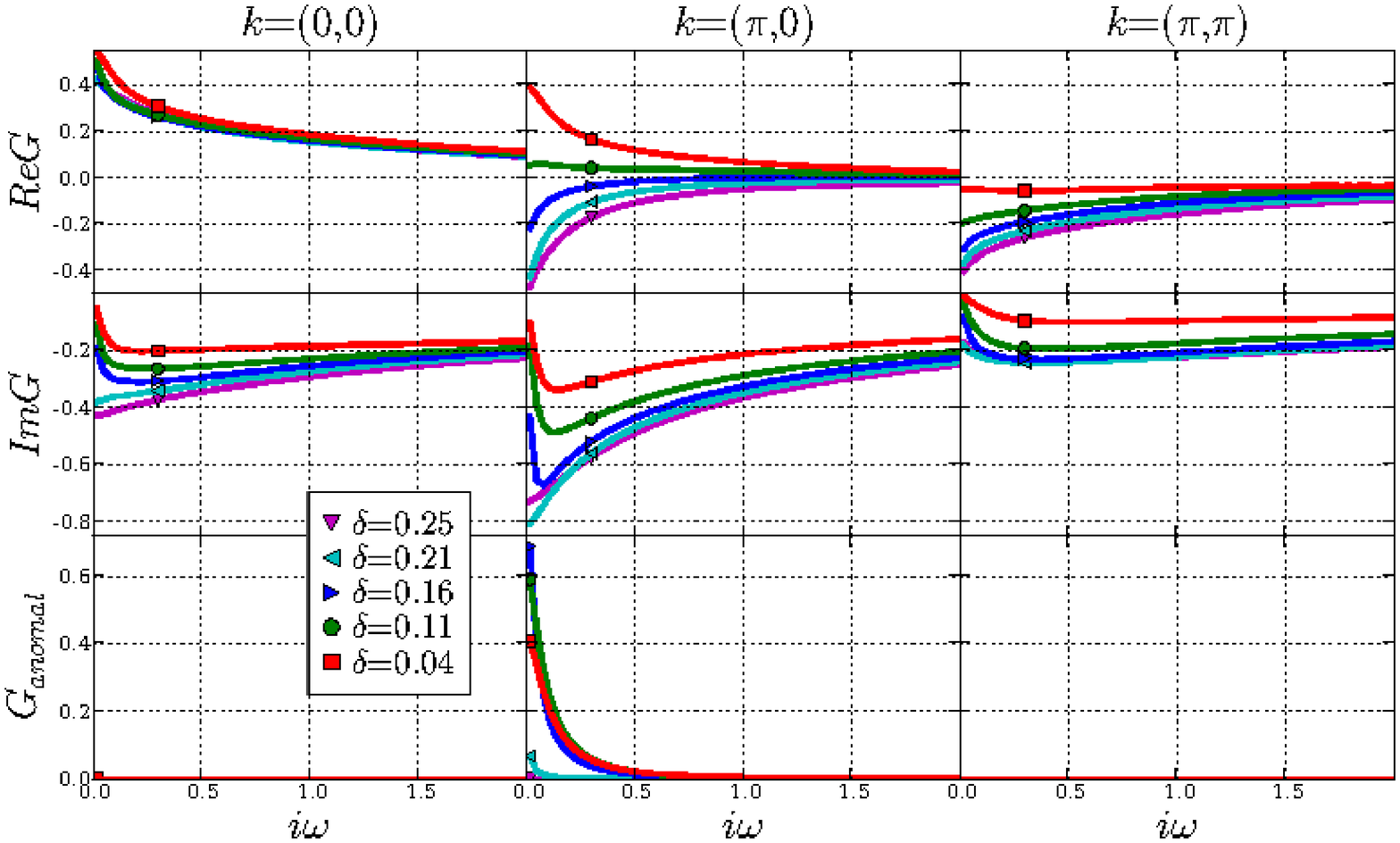}
\caption{
  (Color online)
  Similar than Fig.~\ref{Gf_hub_100} but for lower temperature
  $T=0.005t$ in the superconducting state.
} \label{Gf_hub_200} \end{figure}
\end{widetext}

In this section we discuss cluster quantities.  As discussed in
section \ref{formalism1}, in both C-DMFT and DCA formalism local
quantities, such as cluster self-energies and cluster Green's
function are diagonal in the cluster momentum basis.  Consequenctly,
the physical behavior of the system within the cluster DMFT approach
on a plaquette can be summarized in the four cluster quantities
$\Sigma_{0 0}$, $\Sigma_{\pi 0}$, $\Sigma_{0 \pi}$ and $\Sigma_{\pi
\pi}$, corresponding to the eigenvalues of the matrix containing
onsite, nearest-neighbor and next-nearest neighbor cluster self-energy
introduced in reference \onlinecite{Civelli}.  These cluster self energies
in the cluster momentum basis should not be interpreted as the lattice
self energies evaluated at 4 momentum points.

In the next few figures, we present low temperature self-energies for
the t-J model on the imaginary axis obtained using CTQMC impurity
solver. Figure~\ref{Self_tj_100} contains the data in the normal state
and Fig.~\ref{Self_tj_200} same quantities deep in the superconductings
state.

Starting from the low temperature $T=0.01t$ normal state solution
shown in Fig.~\ref{Self_tj_100} one notices large momentum
differentiation at small doping. The three orbitals evolve  very
differently with changes in   doping and  temperature as we will
show in the following. The $(0,0)$ component has Fermi-liquid
frequency dependence  with relatively small scattering rate at
zero frequency and small monotonic decrease of the real part of
the self-energy with increasing doping. The two degenerate
orbitals $(0,\pi)$ and $(\pi,0)$ are distinctly different from
the $(0,0)$ orbital. The scattering rate around optimal doping
$0.12 \lessapprox \delta \lessapprox 0.22$ remains large (of
order unity) even below the transition to superconducting state
$T\sim 0.01t$. We notice in passing that it becomes increasingly
difficult to converge the C-DMFT equations in the metastable
normal state around optimal doping. Critical slowing down is
observed which might be a signature of  local (cluster)
instability which might occur at zero temperature and might even
preclude the continuation of a translationally invariant normal
state solution down  to zero temperature.

The $(\pi\pi)$ self-energy in Fig.~\ref{Self_tj_100} is far the
largest among all four and, except at very small and very large
doping, it does not show any signature of coherence. At $\delta\sim
0.1$ it has a clear pole at zero frequency. From the above plot we can
see that a pole is on the real axis and it is located above the Fermi
level at small doping and crosses Fermi level around $\delta\sim 0.1$
and finally gets negative in the optimal and overdoped regime. A very
sharp pole on the real axis described above is indeed confirmed by the
NCA calculation. The consequence of the pole in self-energy is
appearance of zeros of the Green's function as discussed in
Ref.~\onlinecite{tdstan}. Physically it means that some states in
momentum space are damped and gapped even at very low temperature.  Figure
\ref{Self_tj_200} at lower temperatures show that this behavior
persist to temperatures much lower than $T_c$. Hence even in
superconducting state the large luttinger Fermi-surface is not
recovered.
The  antinodal fermions are  strongly damped and
gapped even in the superconducting state.   This is related to the occurrence of
Fermi arcs and  lines of zeros  of the Greens function \cite{Dzyalo} as noticed in
Ref.~\onlinecite{tdstan,tudor,georges1D}.
This phenomena was first noticed microscopic studies of coupled
ladders \cite{Lader1,Lader2} and related proposals also appeared in
recent phenomenological models of high Tc's \cite{Phenom1,Phenom2,Philips}.
However, in these studies, the location of the lines of zeros is tied
to the unklapp surface, while in the cluster DMFT the lines of zeros
is a dynamical object which evolves in a highly non trivial way with
doping.

The pole in $(\pi\pi)$ self-energy crosses the Fermi level at a
critical doping (or at least becomes very large at low energies) that
we denote by $\delta_1^c$. The existence of a pole in the self
energy appears also in Hubbard model, with an important difference. In
the Hubbard model, the pole is always below $E_F$ and therefore this
"critical" doping $\delta_1^c$ is zero. We also want to mention that
at small $U=6$ (below the Mott transition of the undoped system) in
the Hubbard model the above mentioned pole seems to be absent or at
least substantially reduced. This substantiates the idea that the
lines of zeros in the Green function appear only above a critical
coupling.

Fig.~\ref{Self_tj_200} shows cluster self-energies at a  lower
temperature, i.e., $T=0.005t$ in the superconducting state. The
$(0,0)$ orbital does not change very dramatically execpt that
becomes more coherent. On the other hand, the $(\pi,0)$ orbital
does show a dramatic effect. The huge scattering rate is now
replaced by the large anomalous component of the self-energy
while the scattering rate is severely reduced. The peak in
anomalous self-energy seems to track $T_c$ and coincides with the
point of maximal scattering rate in the normal state. We will
call this doping $\delta_2^c$ since it corresponds to the avoided
critical point identified  in Ref.~\onlinecite{qcp}. Finally, the
$(\pi,\pi)$ component of the self-energy sharpens with reducing
the temperature and the pole at $\delta_1^c\sim 0.1$ is even more
apparent. This result is quite surprising  because the
superconducting state would have been  expected to be more
coherent. As we show above, coherence  is only restored in three
of  the four orbitals,  while the $(\pi\pi)$ orbital remains
gapped.  Hence the Fermi surface  underlying the normal state
does not contain the Luttinger volume at small doping.

In Fig.~\ref{Gf_tj_100} and \ref{Gf_tj_200} we show the cluster
Green's functions of the t-J model in the normal state and the
superconducting state at lower temperatures.
The cluster Green's functions describe a coarse grained average of the
lattice Green's function over some parts of the Brillouin zone.  It is
evident from Fig.~\ref{Gf_tj_100} that the $(\pi,0)$ orbital contains
most of the spectral weigh (largest imaginary part of $G_{\pi0}$) over
the whole doping regime considered here.  The $(\pi,\pi)$ orbital is
clearly gapped since the real part of self-energy is too big to pick
up any states inside the band as was previously observed in the
extended DMFT study of the same model \cite{tJEDMFT,tJEDMFT1}. The
important message is contained in the real part of $(\pi,0)$ cluster
Green's function. The real part measures the particle hole asymmetry
of the orbital. It would vanishe if the orbital is perfectly
particle-hole symmetric. As one can see in Figs.~\ref{Gf_tj_100} and
~\ref{Gf_tj_200} the $(\pi,0)$ orbital has "more weight" below $E_F$
in underdoped regime and more weigh above $E_F$ in overdoped
regime. Remarkably it becomes almost particle-hole symmetric in the
region of optimal doping. The exact point of particle-hole symmetry is
close to $\sim 0.18$ which is just slightly above the point of maximal
$T_c$ and maximal anomalous self-energy. Fig.~\ref{Gf_tj_200}
demonstrates that this remarkable symmetry persist even in the
superconducting state where the gap appears in all the orbitals.

We now compare  the previous findings with the corresponding
quantities in the Hubbard model displayed in figures
\ref{Sig_hub_100}, \ref{Sig_hub_200}, \ref{Gf_hub_100} and
\ref{Gf_hub_200}. The Hubbard model at $U=12t$ has roughly the
same  superexchange as the one used in the previous study of the
t-J model  $J\sim 0.3t$ and therefore we expect similar physical
behaviour. We will demonstrate below  that indeed this is the
case, and highlight some quantitative differences between the two
models, such as the  numerical values of the critical dopings for
cluster quantities $\delta_{1}^c$ and $\delta_{2}^c$ .

Fig.~\ref{Sig_hub_100} shows the four cluster Green's functions at
$T=0.01t$ in the normal state. When the off-diagonal long range
order is allowed, the system starts to develop anomalous
components in the optimal doped regime at this temperature just
like in the t-J model at the same temperature. In the metastable
normal state, the $(0,0)$ orbital is again most coherent orbital
and not very sensitive to doping. On the other hand, the
$(\pi,0)$ orbital is clearly coherent for small and large doping
and the scattering rate around $\delta\sim 0.1$ is of the order
unity.  The point of maximum scattering rate and maximum
anomalous self-energy in the Hubbard model is however slightly
shifted towards lower doping, (relative to the t-J model) i.e.,
$\delta_2^c\sim 0.1$.

The $(\pi,\pi)$ orbital is again the one with far the largest
self-energy and scatering rate.  In the Hubbard model  the pole on
the real axis crosses zero exactly at zero doping, hence
$\delta_1^c=0$. However even at optimal doping $\sim 0.1$ the
real part of the self-energy is so large that the orbital is
almost completely  gapped.

Fig.~\ref{Sig_hub_200} demonstrates that the pole in $\Sigma_{\pi\pi}$
does not disapear in the superconducting state.  This was also the
case in the t-J model, and it is therefore a robust feature of the
approach to the Mott insulator within CDMFT.  The physical
interpretation is that part of the underlying Fermi-surface remains
gapped even in the superconducting state.
The $(\pi,0)$ orbital becomes coherent when entering the
superconducting state. Its imaginary part, at low frequencies is
maximal around $\delta_2^c$.

The cluster self energies in the cluster site representation contain
useful information about the range. For example, it has been argued
that near the Mott insulator, they become long ranged, while the
cluster cumulant remains short ranged \cite{tdstan,tudor}.  In
Fig.~\ref{Sig_hub_re_200} we show the onsite, nearest neighbor and
next-nearest neighbor self-energy, the actual output of the C-DMFT
scheme. These are related to the eigenvalues shown above through the
following linear relation
\begin{eqnarray}
  \Sigma_{R=(0,0)} = \frac{1}{4}\left( \Sigma_{00} +\Sigma_{\pi 0}+\Sigma_{0\pi}+\Sigma_{\pi\pi}\right)\\
  \Sigma_{R=(1,0)} = \frac{1}{4}\left( \Sigma_{00} -\Sigma_{\pi 0}+\Sigma_{0\pi}-\Sigma_{\pi\pi}\right)\\
  \Sigma_{R=(0,1)} = \frac{1}{4}\left( \Sigma_{00} +\Sigma_{\pi 0}-\Sigma_{0\pi}-\Sigma_{\pi\pi}\right)\\
  \Sigma_{R=(1,1)} = \frac{1}{4}\left( \Sigma_{00} -\Sigma_{\pi 0}-\Sigma_{0\pi}+\Sigma_{\pi\pi}\right)
\end{eqnarray}

On the heavily overdoped side of the Hubbard model $\delta>0.16$
presented in Fig.~\ref{Sig_hub_re_200} it is clear that the only
relevant quantity is the on-site self-energy which justifies use
of the single site DMFT in the overdoped site of the system. In
the underdoped regime, however, the nearest neighbor as well as
next-nearest neighbor self-energies are large and give rise to
qualitatively different results than those of  a single site
DMFT. They renormalize  the nearest neighbor and next-nearest
neighbor hopping and induce a substantial next nearest neighbor
hopping even for the model with vanishing bare $t'$. Furthermore
they distort the Fermi surface and cause variation of coherence
across the Fermi surface as we will show below.

Finally, the cluster Green's functions for the Hubbard model are
shown for two temperatures $T=0.01$ and $T=0.005$ in
Figs.~\ref{Gf_hub_100} and ~\ref{Gf_hub_200} at $U=12t$. Again we
notice qualitatively similar behaviour than those found in the
t-J model. The $(\pi,\pi)$ orbital is gapped in both normal and
superconducting state. The $(\pi,0)$ orbital contains most of the
spectral weight and becomes particle-hole symmetric slightly
above optimal doping around, i.e., around $\delta\approx 0.12$.
This particle hole-symmetry persists in the superconducting state.

We now turn to the real frequency information. In most of what
follows, we show results for the t-J mode, except when explicitly
stated otherwise.

\begin{figure}
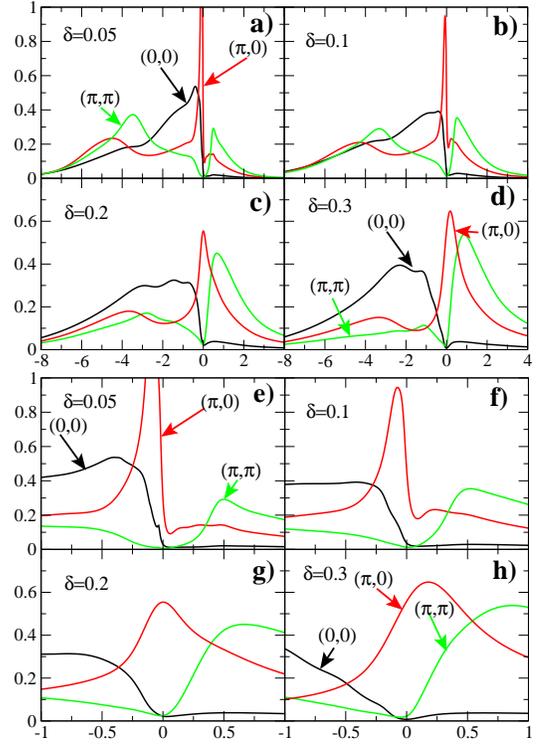

\includegraphics[width=0.8\linewidth]{Gloc.eps}\\
\includegraphics[width=0.8\linewidth]{Gloc_zoom.eps}
\caption{
  (Color online)
  Evolution of the EC-DMFT cluster spectral functions of the
  t-J model with doping in normal state at $T=0.025t\sim T_c$ using NCA as
an impurity solver. The upper panel shows the spectra in the interval
between $[-8t,4t]$ where Hubbard band is clearly observed. The lower
panel shows the region near the Fermi level.  } \label{Gloc}
\end{figure}

Figure \ref{Gloc} shows the evolution of the CDMFT cluster spectral
functions as a function of frequency for few doping levels. Notice
that due to symmetry $(\pi,0)$ and $(0,\pi)$ spectral functions
coincide. At zero doping (not shown) all four orbitals are half-filled
and the system is in Mott insulating state.

Upon  doping the system, the $(\pi,\pi)$ orbital gets emptied
first but in a very unusual way. Although its occupancy gets much
smaller than unity and therefore one would naively expect large
number of hole carriers in this band, it remains basically gapped
for arbitrary doping as we have established above on the basis of
the CTQMC results. This is very unusual since one naively expects
the orbital to be gapped only at an integer filling. Only at very
large doping $\delta > 0.3$ the self-energy of this orbital
approaches the other three self-energies so that the self-energy
becomes momentum independent and therefore local. At this large
doping, the $(\pi,\pi)$ orbital is essentially empty and we can
think of this orbital as an Anderson impurity model in the empty
orbital regime.

The $(0,0)$ orbital is also very inert in the whole doping range.
Its density of states at the Fermi level is small while its
occupancy only slightly decreases with increasing doping. The
orbital remains close to half-filling with very small number of
charge carriers induced in this band.

Finally, the the $(0,\pi)$ (and $(\pi,0)$) components have sharp
spectral features with very strong doping dependence. In going
from $\delta=0.3 $ to $\delta =0.1$ we observe the narrowing of
the quasiparticle width reminiscent of the single site DMFT,
however a qualitative feature of CDMFT is that at smaller dopings
this narrowing of the width is arrested, as a result of the
presence of exchange effects as seen in slave boson studies
\cite{Grilli} and in the large $N$ limit of the t-J model
\cite{Gabi_large_N}.

At low doping, the spectral function develops a pseudogap on the
scale of $J$ with most of the coherent spectral weight below the
Fermi level and a small fraction of it above the Fermi level.
This is a general feature of  the approach to the Mott transition
in  cluster DMFT and has been seen in earlier studies
\cite{Jarrel-pseudogap,tJEDMFT,tJEDMFT1,Stanescu-Philips,Kyung}.

The important message contained in figure ~\ref{Gloc} is that the
momentum differentiation at small doping is very large. The $(\pi\pi)$
orbital remains gapped at all dopings. It is in the Mott insulating
state at low doping and becomes empty in the overdoped regime, hence
it undergoes a band insulator to Mott insulator transition with
decreasing doping. Most of the dynamical information of the active
degrees of freedom representing the electrons close to the Fermi
surface of the lattice model is however contained in the $(0,\pi)$ and
$(\pi,0)$ components.

\begin{figure}
  \includegraphics[width=0.8\linewidth]{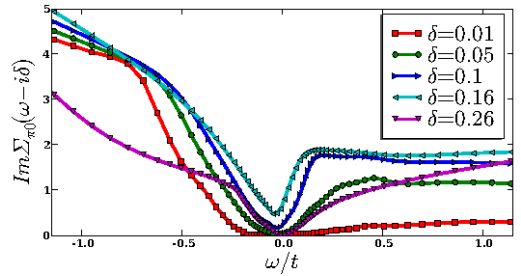}
  \caption{
    (Color online)
    EC-DMFT cluster self energies on real axis in normal state of the
    t-J model computed by NCA
    at $T=0.025t\sim T_c$. }
  \label{Self_real}
\end{figure}

The frequency dependence of the cluster $(\pi,0)$  self-energy and
its evolution with doping on real axis is shown in
Fig.~\ref{Self_real}. At small doping, the hole-like scattering
rate ($\omega<0$) is large while the electron-like ($\omega>0$)
is small.  Around optimal doping, the self-energy is roughly
linear in frequency, however, with large zero-frequency value. In
this regime, there is still a large particle-hole asymmetry in
the scattering rate. While the hole-like part is linear with
relatively small slope down to $\omega=-0.5t$, the electron part
is increasing only in the small region up to $\omega=0.15t$ with
larger slope. Only in the strongly overdoped system, the
self-energy becomes roughly particle hole symmetric at low
frequency.  This particle-hole asymmetry in scattering rate can
be contrasted with the approximate particle-hole symmetry in the
one-particle green's function at optimal doping. The combination
of the real-part of the self-energy and the band-structure, leads
to approximate recovery of this symmetry in the local
one-particle spectra at optimal doping although it is absent in
the scattering rate.

\subsection{ Zero frequency Quantities }
\begin{figure}
\includegraphics[width=0.85\linewidth]{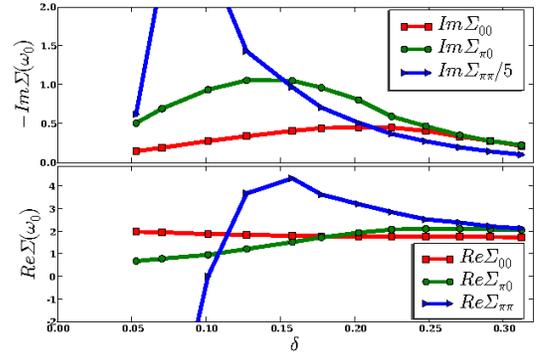}
\caption{(Color online)
  C-DMFT cluster self energies at the lowest Matsubara
point $\Sigma(i\omega_0)\sim \Sigma(\omega=0)$ versus doping in
CTQMC at $T=0.01t$ in normal state of the t-J model. }
\label{Self}
\end{figure}

Further insight into the nature of the CDMFT solution can be obtained
by examining the cluster self-energies at zero frequency. In
Fig.~\ref{Self} we display the CTQMC self-energy for the t-J model at
the lowest Matsubara frequency as a function of doping. In the
overdoped side, the real parts of all four self-energies merge
therefore the self-energy becomes local. The single site DMFT is
adequate.  The coherent quasiparticle peak at the Fermi level is
formed and arrises mainly from the $(\pi,0)$ orbital. The reason is that the
non-interacting density of states for $(0,0)$, $(\pi,\pi)$ and
$(0,\pi)$ orbital extends roughly between $[-4t,0]$, $[0,4t]$ and
$[-2t,2t]$, respectively (see Fig.~\ref{tbDOS}). For the momentum
independent self-energy, the Friedel sumrule dictates that the
effective chemical potential $\mu-\Sigma(0)$ is at the corresponding
non-interacting chemical potential $\mu_0$ which is slightly below
zero frequency. The $(\pi,\pi)$ orbital is therefore empty being a
band insulator-like.  At smaller doping, this orbital acquires
enormous real part of $\Sigma'(0)$ which pushes effective chemical
potential $\mu-\Sigma'(0)$ far below the band edge of the
tight-binding Hamiltonian. This orbital is therefore in the Mott
insulating state for smaller dopings. The insulating state in this
orbital does changes the nature from band-like to Mott like insulator.

For the $(0,0)$ orbital, the effective chemical potential is
close to its upper band edge. The non-interacting density of
states at the band edge for this orbital is small (see Fig.~\ref{tbDOS}) and only a very
small number of charge carriers are doped into the orbital.
Therefore it remains close to Mott insulating state with small
scattering rate at the Fermi level.

The $(\pi,0)$ orbital is slightly less than half-filled in the
doping range considered here, and the real part of self-energy
smoothly increases with doping (see Fig.~\ref{Self}) such that
the effective chemical potential $\mu-\Sigma'(0)$ is positive in
the underdoped side (carriers are hole like) and negative in the
overdoped side (carriers are electron like). Close to the optimal
doping, effective chemical potential is close to zero which makes
orbital approximately particle-hole symmetric at low frequency
(see Fig~\ref{Gloc}).

The $(\pi,\pi)$ self-energy acquires a pole on real axis around
$\delta_1^c \sim 10\%$ doping which can be identified in
Fig.\ref{Self} as a divergent point of $\Sigma_{\pi\pi}^{''}(0)$
and zero of $\Sigma_{\pi\pi}'(0)$.

\begin{figure}
\includegraphics[width=0.8\linewidth]{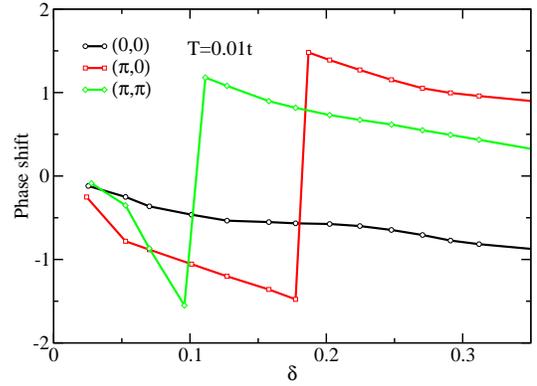}
\caption{(Color online)
  Phase shift (see definition in the text) versus doping for the four orbitals of the
  cluster obtained by CTQMC within C-DMFT method in the t-J model.
} \label{phase_shift1} \end{figure}

Figure \ref{phase_shift1} describes the low energy phase shift in
each orbital. It is defined by $\delta_{\vK}= \arg(G_{\vK}(i
0+))$.  Phase shifts are defined modulo $\pi$.  Notice two
important features: at very small doping the phase shifts in all
channels are close to zero. They confirm the picture suggested in
Ref.~\onlinecite{qcp} in which the cluster degrees of freedom are
weakly affected by the surroundings.
It is reminiscent of the RKKY phase of the two impurity Kondo
model. The system reaches the unitarity limit, as the phase shift
crosses $\pi/2$ in the $(0,\pi)$ channel near $\delta=0.18$. This
is because the real part of the cluster Green's function at zero
frequency vanishes around optimal doping as shown in
Figs.~\ref{Gf_tj_100} and \ref{Gf_tj_200}.

\begin{figure}
\includegraphics[width=0.8\linewidth]{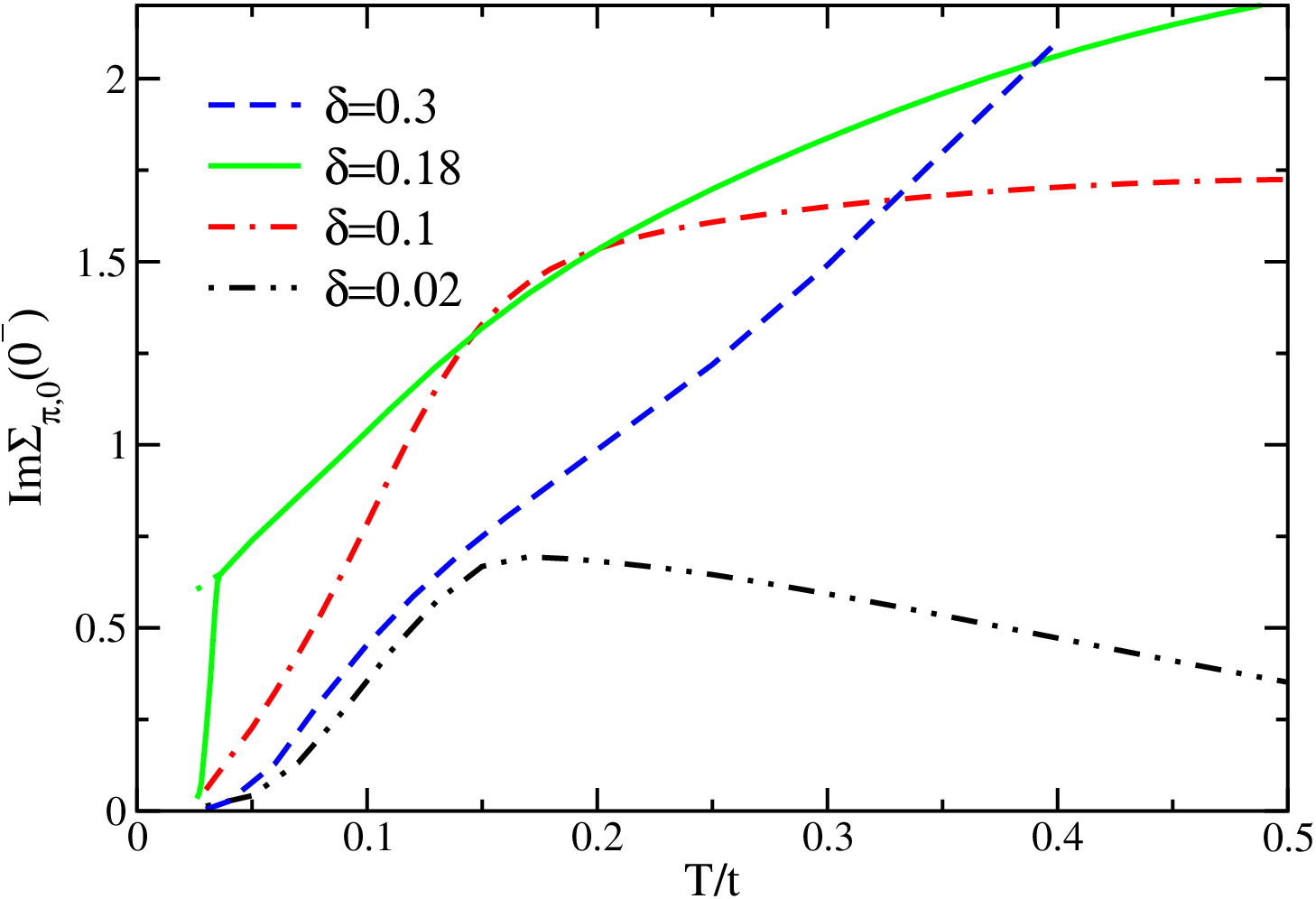}\\
\includegraphics[width=0.77\linewidth]{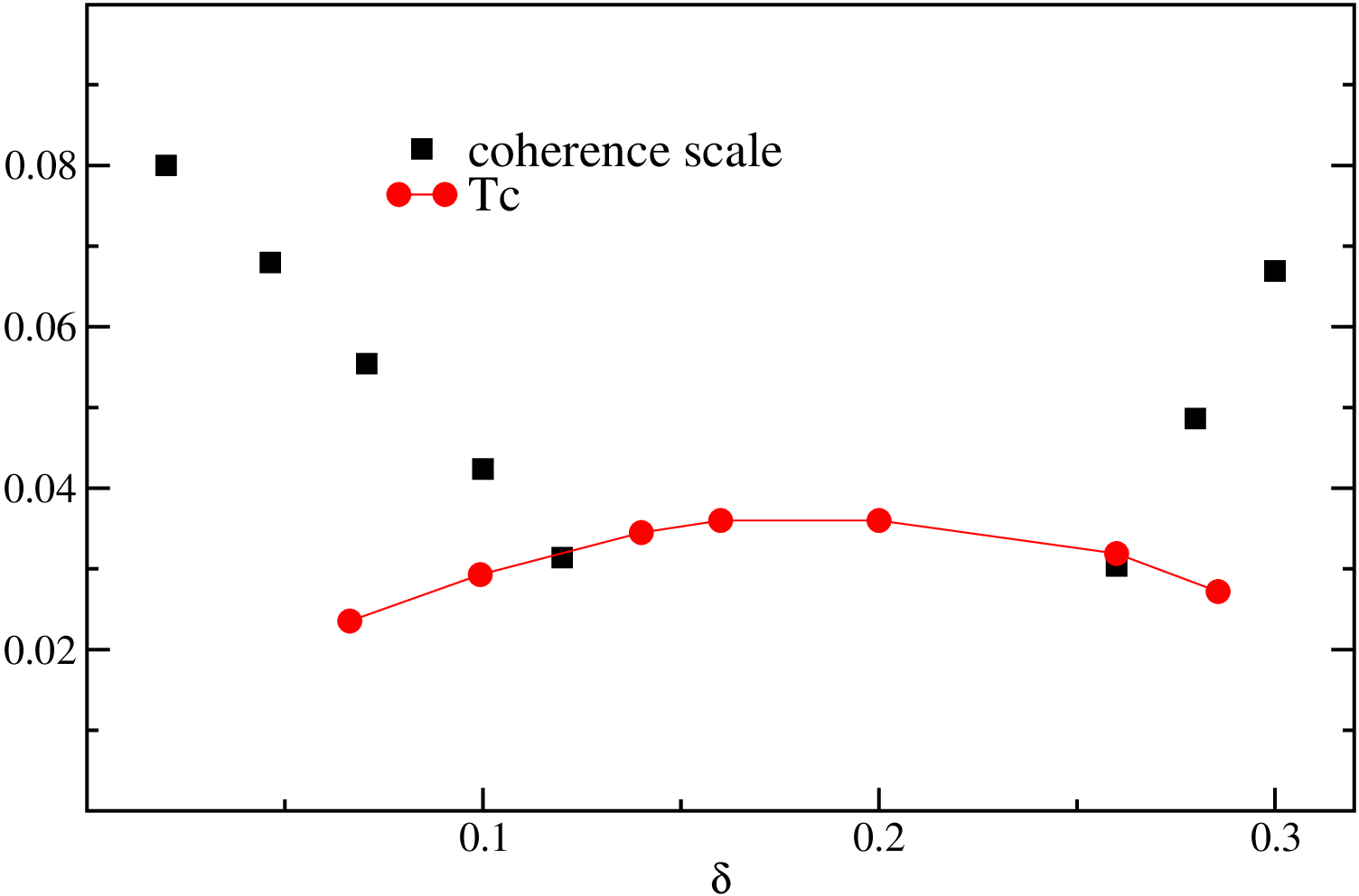}
\caption{
  (Color online)
upper panel:  The imaginary part of the cluster $(\pi,0)$
self-energy at zero frequency as a function of temperature for
few doping
levels. Computed with EDCA using NCA as the impurity solver.\\
lower panel: An estimation of the coherent scale in the normal
state of the t-J model (black dots) and transition temperature to
superconducting state (red dots) within EDCA using NCA as
the impurity solver. }
\label{phased}
\end{figure}

The  indication for the existence of an anomaly  around
${\delta_2}^c$  is seen most clearly  in  the imaginary part of
the real frequency electron self-energy at $(0,\pi)$ evaluated at
zero frequency. We display  EDCA-NCA calculations of it in the
upper panel of Fig.~\ref{phased} . At large and small doping the
scattering rate is small as expected for a Fermi liquid.
Remarkably it becomes very large in the region near optimal
doping when the critical temperature is maximal. This doping
point corresponds to the above defined critical doping
$\delta_2^c$ which is in NCA around 0.18. The transition to the
superconducting state severely reduces the scattering rate
eliminating the traces of the underlying critical behavior.  A
coherence scale, estimated from the scattering rate, is plotted
in the bottom panel of Fig.~\ref{phased} and shows it tends to
vanish close to the point of maximal superconducting transition
temperature.

\begin{figure}
\includegraphics[width=0.8\linewidth]{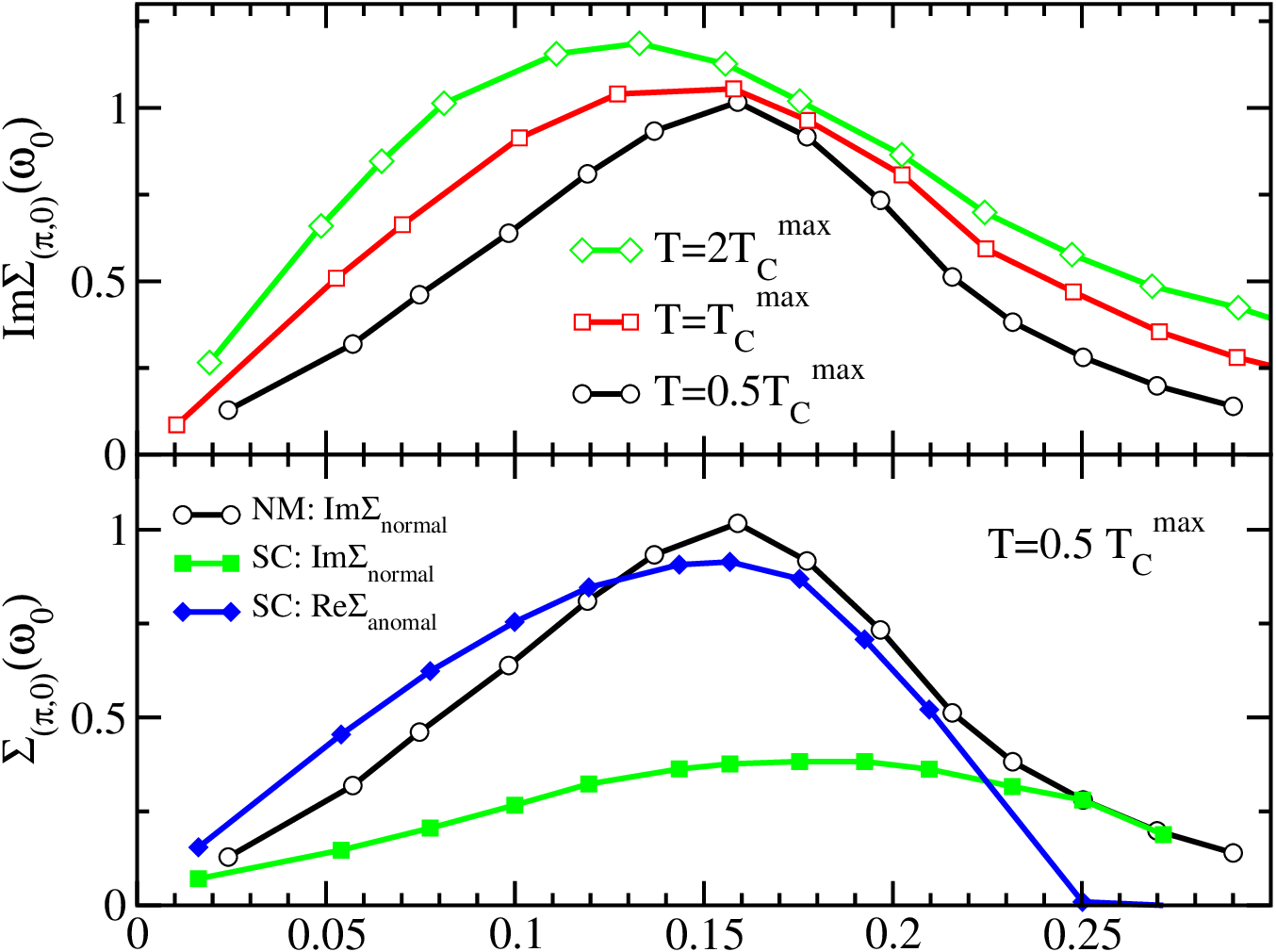}
\caption{
  (Color online)
upper panel: Imaginary part of $(\pi,0)$ cluster self-energy at the lowest
Matsubara frequency $i\pi T$ versus doping for three different
temperatures obtained by CTQMC for the t-J model. The scattering rate
is peaked at optimal doping.\\
lower panel: The large imaginary part of
self-energy of the normal state (black curve with circles) is
severely reduced in superconducting state (green curve with
squares). The scattering rate is peaked at the point of maximal
anomalous self-energy (blue curve with diamonds) tracking the
point of the highest $T_C$. } \label{imSigctqmc}
\end{figure}
Figure~\ref{imSigctqmc} show the CTQMC results for the scattering
rate within CDMFT  and confirms the incoherence of the optimally
doped system. The imaginary part of the self-energy at the first
Matsubara point is small for both, the underdoped and overdoped
system, while it is peaked at optimal doping. The peak is
slightly shifted with temperature and, if the normal state is
continued below the superconducting transition temperature, the
peak of scattering rate coincides with the maximum of the
anomalous self-energy which traces maximum of the transition
temperature (see Fig.~\ref{Self_anomal}). The scattering rate is
severely reduced in the superconducting state when  off-diagonal
long-range order is allowed in the calculation.

\section{Superconductivity, Tunneling Density of States, Fermi Arcs and Nodal Quasiparticles}
\label{supra}

The superconducting state is characterized by an order parameter $\langle
c_{\vk\uparrow} c_{-\vk\downarrow} \rangle ={\cal F}_{\vk}(\tau=0)$
and by the presence of a frequency dependent
anomalous component to the self energy. In Nambu notation, the
self-energy in $(\pi,0)$ and $(0,\pi)$ orbital takes the
following form \begin{equation} \underline{\Sigma}_\vK(i\omega)=
  \left(\begin{array}{cc}
    {\Sigma}_{\vK\up}(i\omega) & \Sigma^{an}_\vK(i\omega)\\
    \Sigma^{an}_\vK(i\omega) & -{\Sigma}_{-\vK\down}(-i\omega)
  \end{array}\right).
\end{equation}
and the corresponding Green's functions is
\begin{equation}
\underline{G}_\vK(i\omega)=
 \left(\begin{array}{cc}
    {\cal G}_{\vK\up}(i\omega) & {\cal F}_{\vK}(i\omega)\\
    {\cal F}_{\vK}^\dagger(i\omega) & -{\cal G}_{-\vK\down}(-i\omega)
  \end{array}\right).
\end{equation}

The sign of the anomalous components chosen by the system is
$\Sigma^{an}_{\pi 0}=-\Sigma^{an}_{0\pi}$. Within C-DMFT,
this is precisely the nearest neighbor self-energy and its
lattice analog (using  the original C-DMFT periodization
\cite{CDMFT}) takes the form $\Sigma_\vk = \frac{1}{2}(\cos{k_x} -
\cos{k_y})\Sigma^{an}_{0 \pi}$.

\begin{figure}
\includegraphics[width=0.99\linewidth]{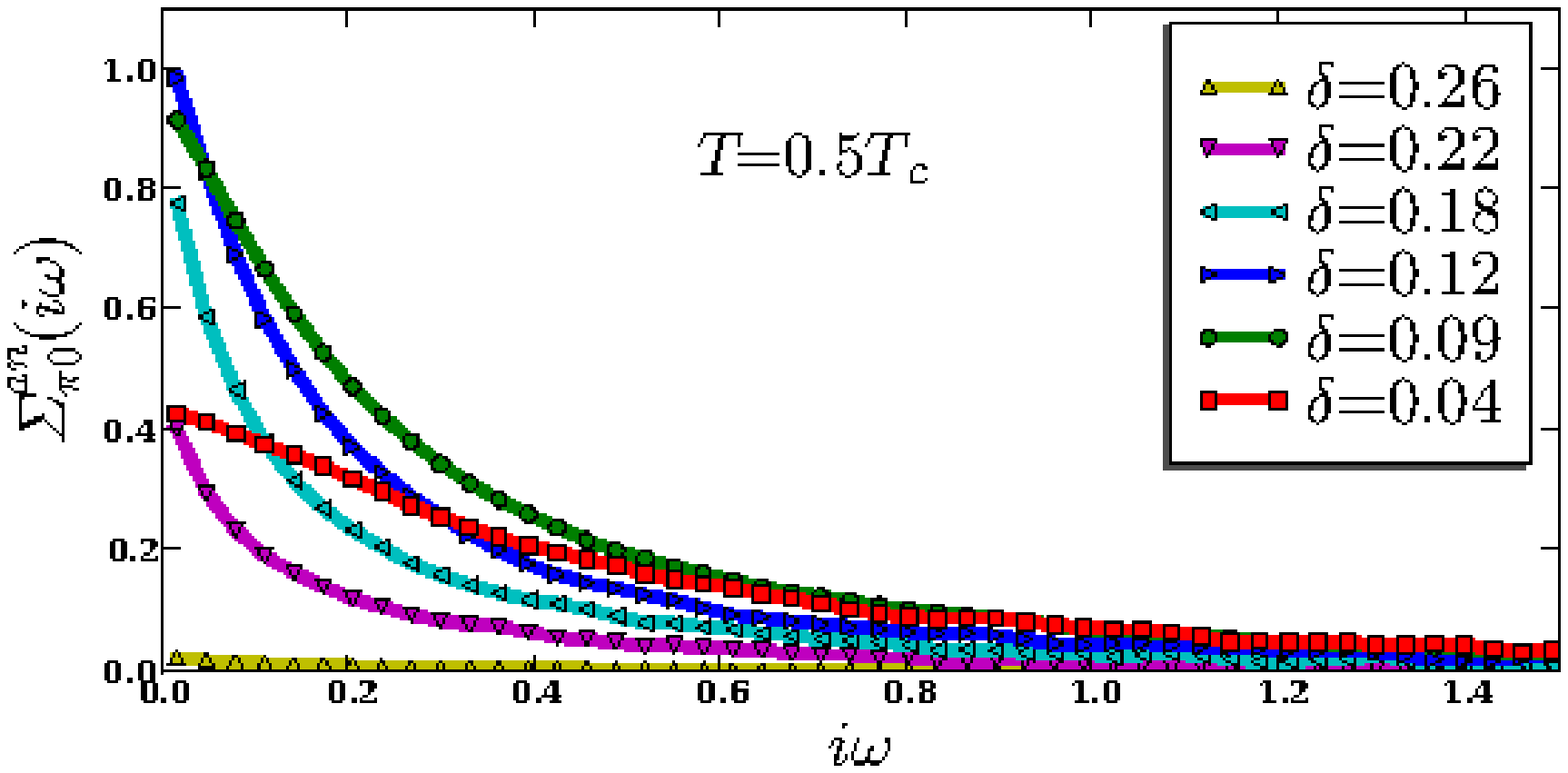}\\
\includegraphics[width=0.99\linewidth]{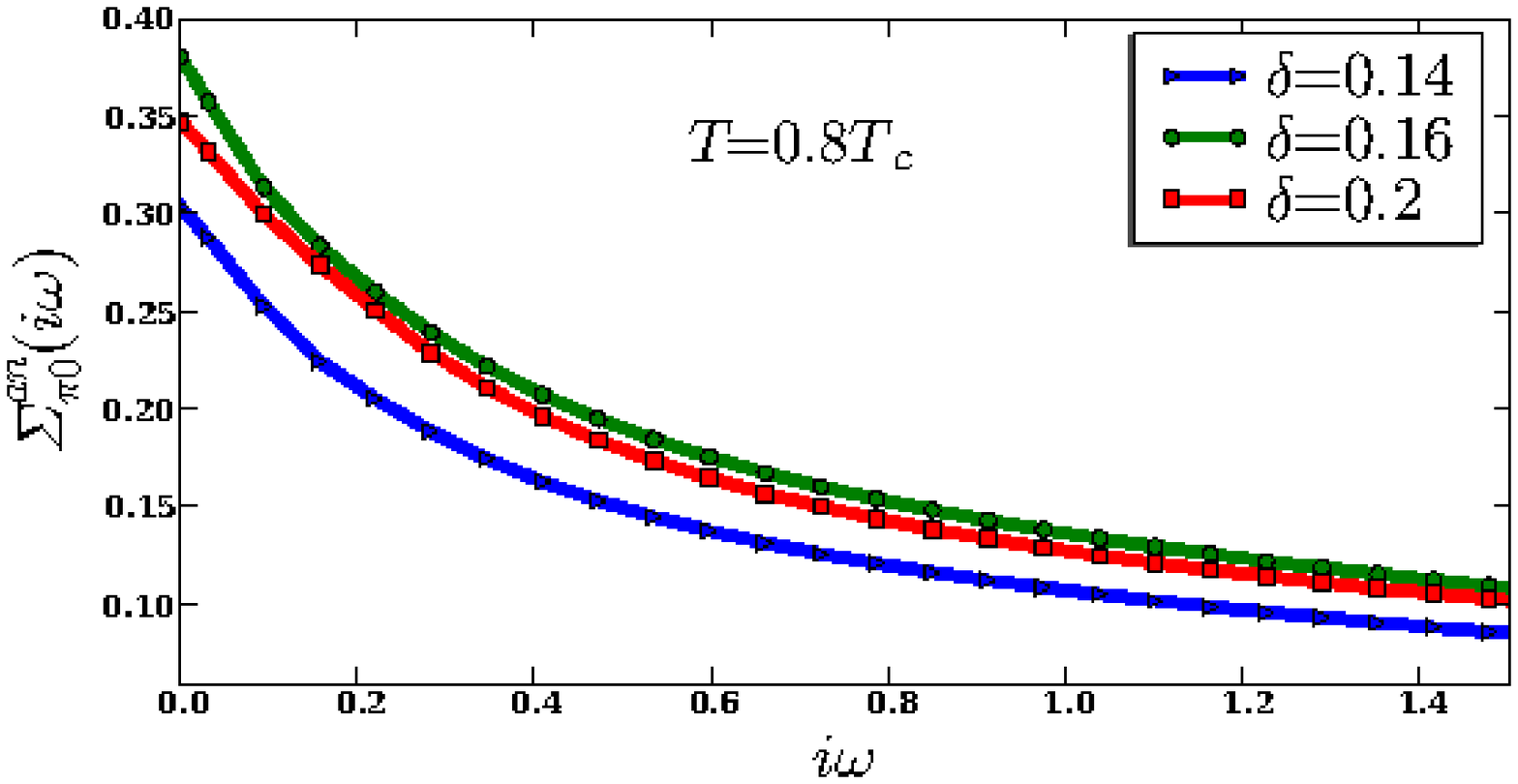}
\caption{(Color online)
  upper panel: C-DMFT  t-J model anomalous self-energy
computed by CTQMC at
$T=0.005t\sim 0.5 T_c^{max}$.\\
lower panel: t-J model anomalous self-energy within EDCA
computed by NCA at $T = 0.033 t \sim 0.9 T_c^{max}$ }
\label{Self_anomal} \end{figure}

The anomalous self-energy $\Sigma^{an}_{\pi 0}$ is plotted in
Fig.~\ref{Self_anomal}. The upper part of the figure shows the
CTQMC results within C-DMFT while the lower part shows the
NCA-results within EDCA. In both cases, the function is
monotonically decreasing with imaginary frequency and is largest at
optimal doping. Furthermore, at the low values of the Matsubara
frequency the anomalous self energy exhibits a fast upturn and
sublinear frequency behavior that gets less pronounced as the
doping is reduced. This trend is likely due to the reduction of
density of states in the pseudogap region.

The anomalous self energy obeys a spectral representation
\begin{equation} \Sigma^{an}_\vk(i{\omega_n}) =
\Sigma^{an}_\vk(\infty) - \int \frac{d\omega}{\pi}
\frac{\Im{{\Sigma^{an}_\vk}}(\omega)}{i \omega_n - \omega}
\nonumber \end{equation}

The infinite frequency value of the self energy vanishes in the
Hubbard model but is non zero in the $t-J$ model and is related to the
order parameter of the system $\cF_\vq(\tau=0) $ through the following
exact relation:
\begin{eqnarray} \Sigma^{an}_{\vk}(\infty) =
-\frac{3}{(1+\delta)^2}\sum_\vq J_{\vk-\vq} \cF_\vq(\tau=0).
\end{eqnarray} where $\cF_\vq(\tau=0)=\langle c_{\vq\uparrow}
c_{-\vq\downarrow}\rangle$.

Notice that simpler mean field theories of the t-J model such as the
slave boson mean field theory \cite{Liu} assume only the static,
frequency independent anomalous self energy. Other approaches based on
the equation of motion for the Hubbard operators \cite{Viktor} capture
a frequency dependent order parameter but neglect the static infinite
frequency component. A similar analysis of the pairing interaction,
has recenty been carried out for the ladders in
Ref.~\onlinecite{Poilblanc}.

The existence of a finite value of the anomalous self energy of the
t-J model at infinite frequency, should be interpreted as the
existence of a non zero value for anomalous self energy in the Hubbard
model at a scale of the order $U$.

The value of the anomalous self-energy at zero-frequency and low
temperature and the gap (defined as the distance between the positive
and negative energy peaks in the tunnelling density of states divided
by two) is similar in all versions of the cluster DMFT. For the
parameters used in our study, ($J/t=0.3$, near optimal doping) the
anomalous self energy is of the order of unity at low temperature (see
the upper panel of Fig.~\ref{Self_anomal}).

\begin{figure}
\includegraphics[width=1.0\linewidth]{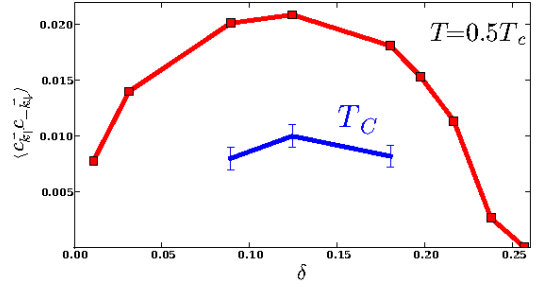}
\caption{(Color online)
  Order parameter in C-DMFT computed with CTQMC at
  $T=0.5 {T_c}_{max}$. The critical temperature (in units of $t$)  for few doping values
  is also displayed.
}
\label{bcs}
\end{figure}

On the other hand, $T_C$ , the superconducting order parameter, and
the value of anomalous $\Sigma^{an}(\infty)$ are more sensitive
quantities, and differ between the various cluster schemes.  The
schemes with higher $T_C$ (extended versions of CDMFT) show slower
decrease of the anomalous self-energy, larger infinite frequency
component of the anomalous self energy and larger value of the
superconducting order parameter. In C-DMFT the maximum value of the
order parameter is around $0.02$ (see Fig.~\ref{bcs}) which is
approximately 8 times smaller than maximum achieved in EDCA .
Consequently, the static pairing in C-DMFT is very small while it
reaches almost $1/3$ in extended versions of the cluster DMFT (both in
EDCA and and EC-DMFT), i.e., the magnitude of the anomalous
self-energy at infinity as compared to the value at zero shown in
Fig.~\ref{Self_anomal}.


From the anomalous Greens function we can extract the
order parameter, i.e., the anomalous Green's function at equal time
$\cF_{\pi0}(\tau=0)$.
The order parameter versus doping as obtained by the CTQMC and
C-DMFT is shown if Fig.~\ref{bcs}. It has a dome-like
shape and tracks the value of the critical temperature, just like
in BCS theory. In the same figure, we also display critical
temperature $T_C$ at optimal doping.   Due to critical slowing down in the region of
transition, many DMFT iterations are needed to  determine
the critical temperature.

\begin{figure}
\includegraphics[width=1.0\linewidth]{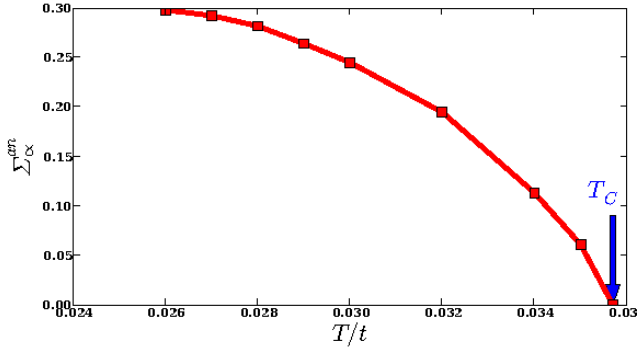}
\caption{
  (Color online)
  Anomalous self energy $\Sigma^{an}(\omega=\infty)$ as a
  function of temperature within EDCA using NCA as the
  impurity solver at optimal doping $\delta = 0.18$. }
\label{bcs2}
\end{figure}

The temperature dependence of the related quantity, the anomalous
self-energy at infinity, computed with NCA is shown in
Fig.~\ref{bcs2}. It has a clear BCS-like temperature dependence
saturating to a value of the order of $\sim 0.3$ which is around
$1/3$ of the zero frequency value.

\begin{figure}
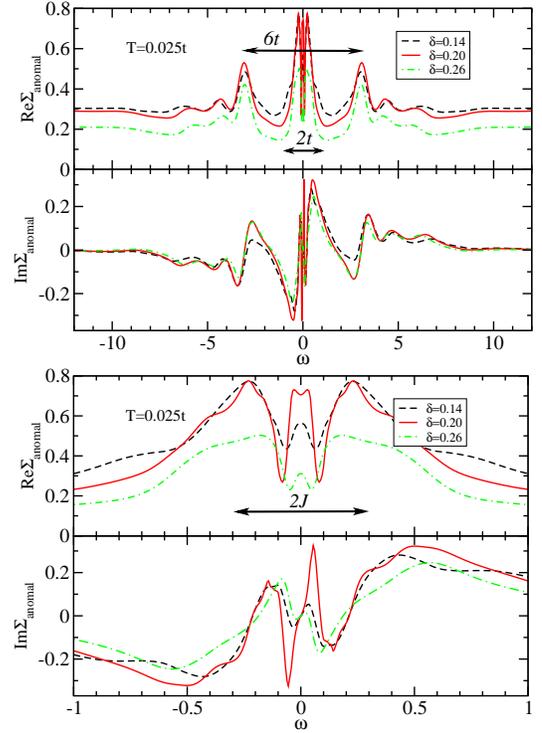

\includegraphics[width=0.8\linewidth]{sig_anomal.eps}\\
\includegraphics[width=0.8\linewidth]{sig_anomal_zoom.eps}
\caption{(Color online)
  Anomalous self energy on real axis within EDCA using
  NCA as the impurity solver. The top panel shows the extended region
  in frequency space while the lowest panel zooms in the low frequency
  part.}
\label{eliashberg}
\end{figure}

\begin{figure}
\includegraphics[width=1.\linewidth]{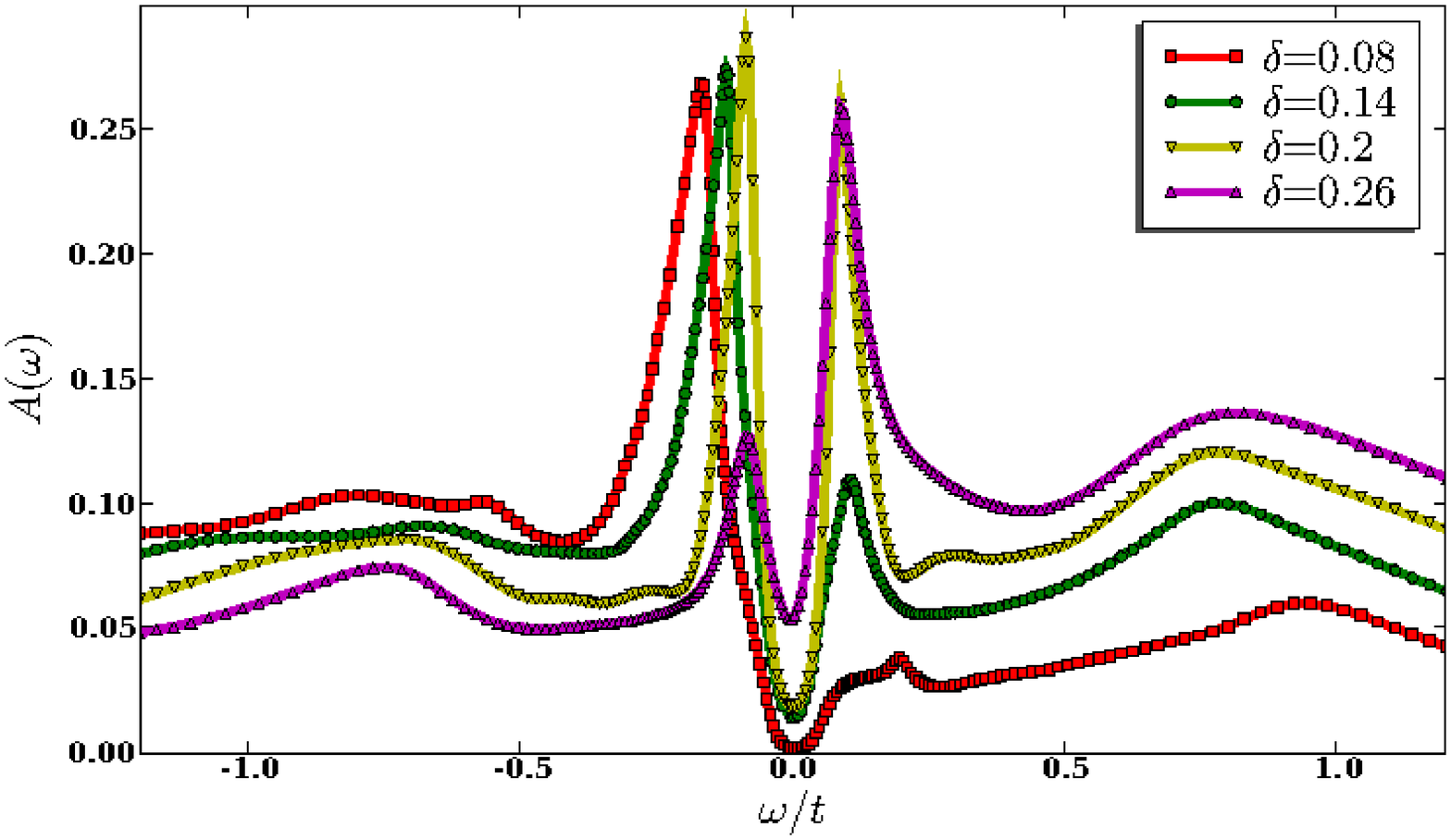}\\
\includegraphics[width=1.\linewidth]{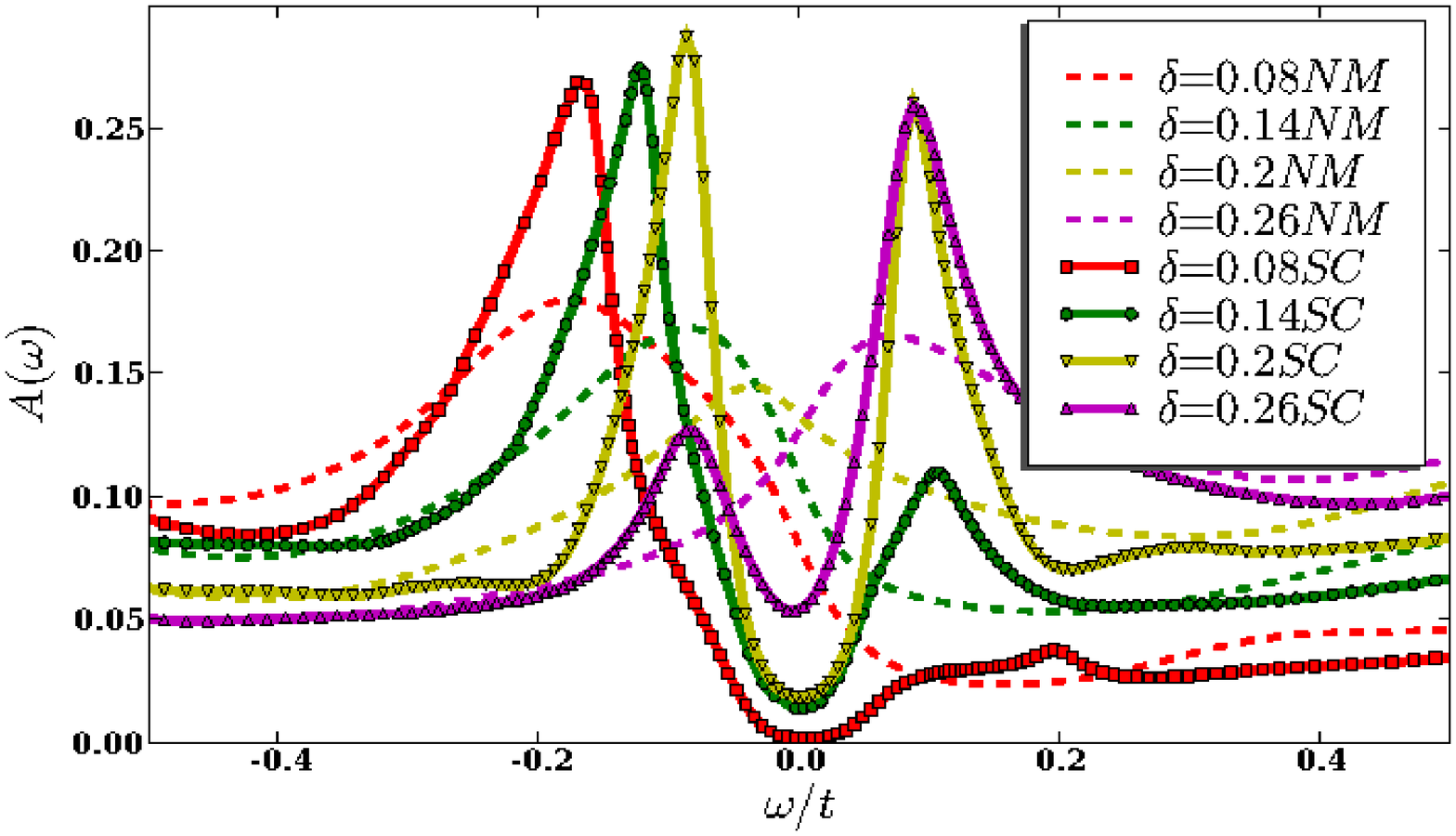}
\caption{(Color online)
  Tunneling density of states (local spectral function)
in EDCA  obtained with NCA at constant $T=0.025t$.
The lower pannel is a  blow up  of the low energy regions. The dotted lines
show the normal state spectral function for the same doping and
temperature.
}
\label{dos}
\end{figure}

Using NCA we can we can examine directly $\Sigma^{''}(\omega) $ and
$\Sigma^{'} (\omega)$ on real axis. There are several features in the
frequency dependence of the anomalous self energies depicted in figure
\ref{eliashberg} which exhibit noticeable departures from the standard
Migdal Eliashberg theory of superconductivity.  First, the real part
of the self energy does not change sign.  This indicates that the
interaction is attractive over the whole frequency range. There is no
characteristic energy corresponding to $\omega_{Debye} $ where the
interaction turns from attractive to repulsive.  Furthermore, the
spectral function displays significant spectral weight not only at a
scale of order J, but also at the scale of order t, extending all the
way to very high frequencies.  Several scales can be clearly
identified in the anomalous self-energy: the size of the SC gap in
one-particle spectra $\sim 0.1t $ (see Fig.~\ref{dos} and the
discussion of the figure later in this section ), the spin exchange
$J$, the hopping $t$ and a scale of the order of half the bandwidth
$\sim 3t$.

\begin{figure}
\includegraphics[width=0.8\linewidth]{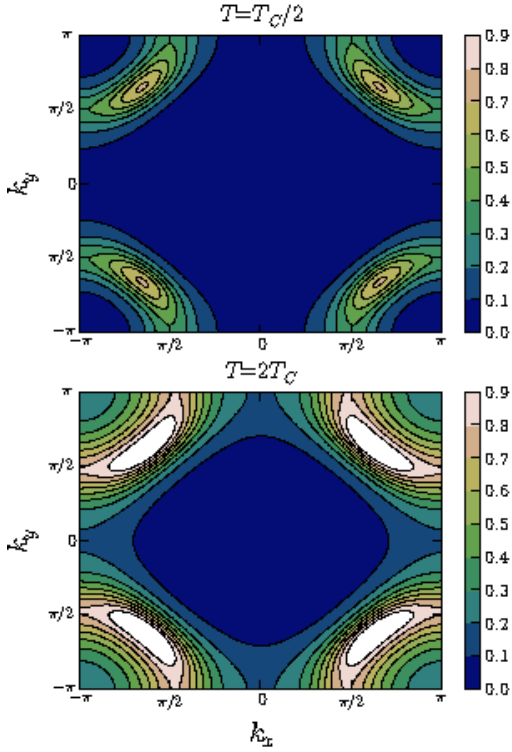}
\caption{(Color online)
  Momentum resolved spectral function in underdoped regime
  ($\delta=0.09$) at zero frequency above and below $T_C$ in
  (metastable) normal state. We use CTQMC and C-DMFT. }
\label{Ak}
\end{figure}
It is useful to momentum resolve the one particle spectra at low
energies to understand the origin of the low energy quasiparticle
excitations in the system. This requires the choice of a periodization
scheme.
For simplicity we use the cumulant periodization scheme
introduced in reference \onlinecite{tdstan,tudor,new-review}.  A
more detailed discussion of the periodization problem will be
given elsewhere \cite{Marcello-new}. Here we focus on
the temperature dependence, which require the finite temperature
techniques described  in  this paper.

The results are shown in Fig.~\ref{Ak}. As shown in earlier work
\cite{Parcollet,Civelli}, C-DMFT is able to produce Fermi arcs in the nodal
region. The advantage of the CTQMC technique relative to other
solvers, is that it allows to investigate, for the first time within CDMFT, the
temperature dependence of the arcs. As shown in Fig.~\ref{Ak}
the Fermi arcs shrink with decreasing temperature, reminiscent of
recent experiments \cite{ArcsSchrink} on cuprates.
The physical mechanism for the formation of the arcs, and their
shrinking with decreasing temperature is the shift in the real
part of the momentum dependent self energy, which is enhanced in
the cumulant periodization. While the validity of this
periodization down to zero temperature, with the consequent
formation of lines of zeros and Fermi pockets, is at this point a
conjecture that deserves further study, there is no question that
the formation of the arc and their temperature dependence, at
finite temperatures is a robust property of the cellular DMFT
treatment and is visible in other periodizations.
Therefore, the results of this paper together with the earlier zero
temperature results of Ref.~\onlinecite{tdstan} are consistent, at the
qualitative level, with both the recent De Haas Van Alven measurements
\cite{taillefer} and photoemission measurments \cite{kanigel}. With
decreasing temperature the Fermi arcs, evolve into a small pocket at a
finite distance from a line of zeros which darkens one side of the
pocket.
%
\begin{figure}
\includegraphics[width=0.85\linewidth]{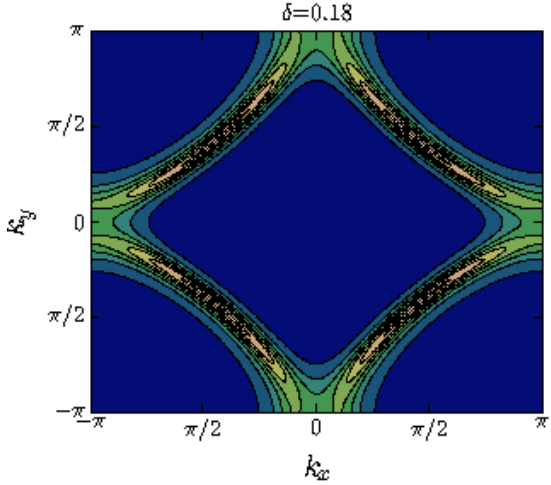}
\caption{(Color online)
  Momentum resolved spectral function in optimally doped regime
  in superconducting state ($T=T_c/2$) at zero frequency. We use
  CTQMC and C-DMFT. }
\label{Ak2}
\end{figure}

The arcs are increased with doping and they develop into a banana
shape structure. The Fermi surface at optimal doping in
superconducting state is displayed in Fig.~\ref{Ak2}. Notice the sharp
quasiparticles in nodal region and gap in the antinodal region.

We now turn to another observable, the superconducting tunneling
density of states and its doping dependence displayed in
Fig.\ref{dos}.  This quantitity has been extensively investigated
experimentally \cite{Fisher}.  On a broad energy scale, there is
considerable particle hole asymmetry in those curves and the
positive frequency part decreases as we underdoped.  This is
expected on very general grounds for a doped Mott insulator
\cite{assym1,assym2,assym3,assym4,assym5}.

Remarkably, around the optimal doping $\delta\sim 0.18$ the curves are
more particle hole symmetric at low frequencies.  The asymmetry in the
superconducting state evolves from the asymmetry of the underlying
normal state. To confirm this we plot the density of states of the
underlying normal state with dashed lines in the lower panel of figure
\ref{dos}. It is clear from figure \ref{dos} that the same magnitude
of the assymetry is present in the normal state local density of
states as in the superconducting tunneling density of states.

The low energy slope of the tunnelling density of states is only weakly
doping dependent, as was shown in the CDMFT  exact diagonalization
study of the Hubbard model \cite{Marcello-submitted}.

Besides the considerable particle hole asymmetry at low doping , there
are several features in figure \ref{dos} which are in qualitative
agreement with experiments \cite{Arpes_review}. For example the
dip-hump feature in the tunneling density of states, in the unoccupied
part of the spectra.

Another surprising aspect of the tunneling is that the increase in the
gap with decreasing doping is correlated with a {\it decrease} in the
intensity of the coherence peaks. This is the oppositve of what is
expected for a BCS superconductor where the growth in coherence peaks
correlates with an increase in the superconducting gap. This
observation can also be understood in terms of the two gap picture.
The gap in the tunnelling density of states, (maximum between the
coherence peaks) is controlled by the gap originating from the normal
component of the self energy. This gap increases with decreasing
doping. On the other hand the degree of coherence is controlled by the
anomalous self energy which decreases with decreasing doping as shown
in fig~\ref{dos}.

The two gap picture of the cuprates,   has recently
emerged from the analysis of numerous experiments
\cite{ram-twogap,Deutcher, Tacon,Tanaka}. It is also part
of various  phenomenological pictures of cuprate superconductors
\cite{Phenom1,Phenom2,Tramble_phenom}. This picuture has been recently put on a
microscopic basis by  Variational Cluster Approach studies
\cite{Hanke} and C-DMFT studies of the Hubbard \cite{Marcello-submitted}.

In the optimally doped
regime the gap value is of the order of $\Delta \sim 0.09t$.
This value was obtained from fig. \ref{dos} but similar values result
from analytic continuation
of CTQMC data to real axis.
As discussed above, the critical
temperature $T_C$ strongly depends on cluster scheme employed. Using
the maximum $T_C$ of each scheme, we can determin the ratio
$2\Delta/T_C$ at optimal doping. In EDCA $T_C^{EDCA} \sim 0.036t$ and
$2\Delta/T_C\sim 5$, in EC-DMFT $T_C^{EC-DMFT} \sim 0.026t$ and
$2\Delta/T_C\sim 7$ and $T_C^{C-DMFT} \sim 0.01t$ therefore
$2\Delta/T_C\sim 18$.  In conventional superconductors described by
BCS theory, this ratio is universaly equal to $2\Delta/T_C=3.5$ but
increases in the strong coupling Eliashberg theory. The cluster DMFT
superconductivity is thus in the very strong coupling limit when
compared to conventional superconductors. Recent experiments on
Bi2212 \cite{DoTc} seems to suggest that the ratio $2\Delta/T_c$ is close to
$8.0$ being somewhere between the two limits of extended and
non-extended version of the CDMFT schemes.

\begin{figure}
\includegraphics[width=0.99\linewidth]{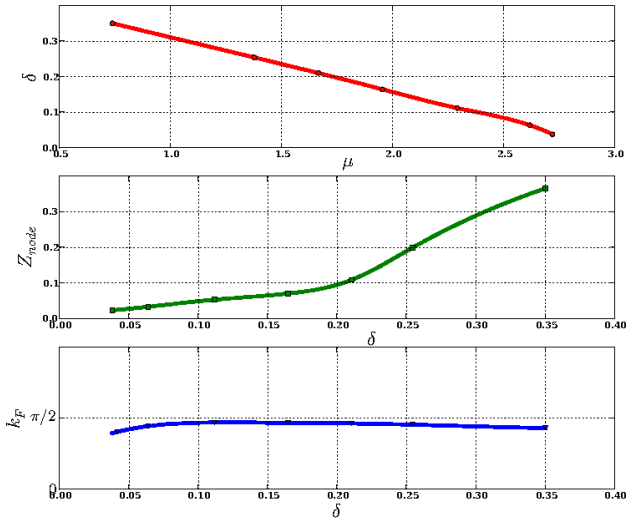}
\caption{(Color online)
  upper panel: doping versus the chemical potential for the
  Hubbard model at $T=0.005t$. It shows linear dependence and downturn
  at small doping. This could point to divergence of the
  compressibility zero doping.\\
  middle panel: the nodal quasiparticle
  residue $Z$ versus doping. It is slowly increasing in the underdoped
  and optimally doped system and increases rather rapidly in the overdoped
  system.\\
  lower parnnel: fermi momentum along the nodal direction
  versus doping.  }
\label{n_mu}
\end{figure}

\begin{figure}
\includegraphics[width=0.99\linewidth]{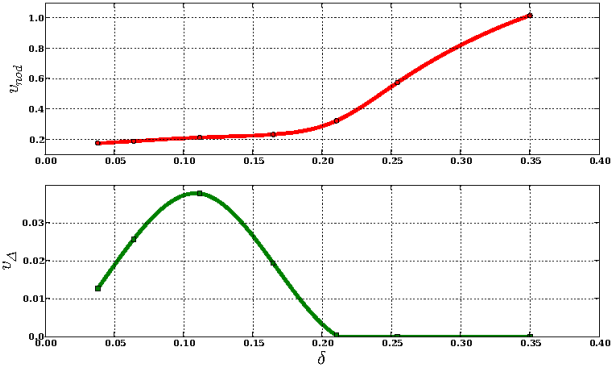}
\caption{(Color online)
  upper panel: the nodal velocity perpendicular to the Fermi
  surface versus doping. It is almost flat in the underdoped and
  optimally doped system and increases rapidly in the overdoped
  system.\\
  lower pannel: the nodal anomalous velocity parallel to the
  Fermi surface versus doping has a dome-like shape with the peak
  around optimal doping.  }
\label{veloc}
\end{figure}

In Figs.~\ref{veloc} and \ref{n_mu} we present some insights into the
nodal quasiparticles of the Hubbard model as obtained from the CTQMC
results shown in Figs.~\ref{Sig_hub_200} and ~\ref{Gf_hub_200}.  The
self-energy in the nodal region is obtained  from the self-energy
periodization  \cite{CDMFT}, i.e.,
\begin{eqnarray}
\Sigma(\vk) &=&
\frac{1}{4}[
\Sigma_{R=(0,0)} + \Sigma_{R=(1,0)}\cos k_x
\nonumber\\
&+& \Sigma_{R=(0,1)}\cos k_y +\cos(kx)\cos(ky)\Sigma_{R=(1,1)}]
\end{eqnarray}
This allows us to determine  the position of the
Fermi momentum $\mu-\epsilon_{\vk_F}-\Sigma_{\vk_F}(\omega=0)=0$
and quasiparticle renormalization amplitude
$Z=1/(1-d\Sigma(\vk_F)/d\omega)$.  In Fig.~\ref{n_mu} we plot
$Z_{nodal}$ and $\vk_F$ along the nodal direction in the
superconducting state where the coherence is established and
quasiparticles are well formed. Fermi surface is close to
$(\pi/2,\pi/2)$. The renormalization amplitude $Z$ is very slowly
increasing with doping in the underdoped and optimally doped
regime but has a fast upturn once the normal state gets more
Fermi-liquid like.

The evolution of the nodal velocities at very low temperatures
and its consequences for the
superconducting gap in the nodal as well as antinodal region was
recently studies in Ref.~\onlinecite{Marcello-submitted} using
exact diagonalization as the impurity solver. Here we confirm all
the qualitative trends in the doping dependence of these
quantities,  using the CTQMC solver.

In Fig.~\ref{veloc} we plot quasiparticle Fermi velocity perpendicular
to the Fermi surface and anomalous velocity parallel to the Fermi
surface in the nodal region. The velocities are defined by
\begin{eqnarray}
  v_{nodal} &=& Z_{nodal}\left(\frac{d\epsilon_\vk}{dk_{\bot}}+
  \frac{d\Sigma_\vk}{dk_{\bot}}
  \right)\\
  v_{\Delta} &=& Z_{nodal}\frac{\Sigma^{anomal}_\vk}{dk_{\|}}
\end{eqnarray}
It is clear from the Fig.~\ref{veloc} that the nodal velocity is
almost constant in the underdoped, optimally doped and lightly
overdoped regime, compatible with observation in Ref.~\onlinecite{Zhou}. The
anomalous velocity, however, is of dome-like shape and tracks the
critical temperature. The anomalous velocity measures the slope of the
superconducting gap at the node and its downturn in the underdoped
regime suggest that the superconducting gap at the node decreases with
decreasing doping. This surprising result is in accordance with
recent Raman experiments \cite{Tacon} and angle resolved
photoemission measurements \cite{Tanaka} showing that the
superconducting gap at the node in the deeply underdoped regime indeed decreases.

\section{Optical Conductivity}
\label{optics}

\begin{figure}
\includegraphics[width=0.8\linewidth]{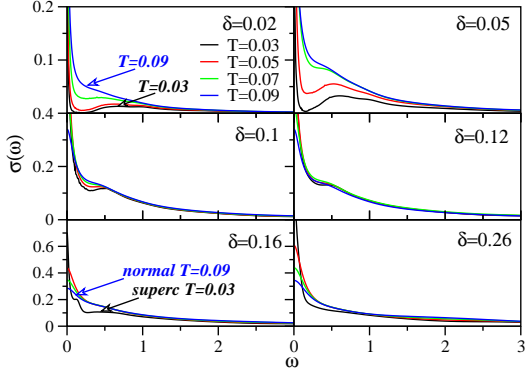}
\caption{(Color online)
  Optical Conductivity at different dopings and temperatures
within EDCA using NCA for the t-J model. The optical conductivity
evolves from a sum of two distinct features at low dopings to a broad
Drude feature at high dopings.  }
\label{DCA-optics}
\end{figure}

We now turn to the optical conductivity, which we display in figure
\ref{DCA-optics} for the t-J model.  This quantity has been investigated both
theoretically and experimentally over the last twenty years. For
reviews see Refs.~\onlinecite{review7,review5,review2}. The integrated
spectral weight is a measure of the number of carriers, and its
evolution with doping has attracted considerable attention
\cite{Bontemps_redistribution,Tinkman,Bontemps_kinetic,Molegraf,Timusk,Boris,controversy,our_with_vanderMarel,optics-Haule}.

The starting point of a theory of the optical conductivity is the Kubo formulae,
\begin{widetext}
\begin{equation}
  \sigma(i\om_n)=\frac{1}{\om_n}
  \left[
    \om_p^2 \delta_{\a\b}-
    e^2 \sum_{\vk\sigma\vk'\sigma'}
    v_{\vk}^\a v_{\vk'}^\b
    \int_0^\beta e^{i\om_n\tau}
\langle T_\tau
  c^{\dagger}_{\vk,\sigma}(\tau)c_{\vk,\sigma}(\tau)
  c^{\dagger}_{\vk',\sigma'}c_{\vk',\sigma'}
\rangle
    \right]
\end{equation}
where the plasma frequency is evaluated from the expectation value of
the projected kinetic energy and the operators $c $ and $ c^\dagger$
are projected fermions of the t-J model.

In principle the evaluation of the optical conductivity within CDMFT
requires the evaluation of the vertex function, since current vertex
corrections are non vanishing in plaquette C-DMFT.  However for DCA in
a plaquette, we have shown that these corrections vanish
\cite{optics-Haule}. This suggest, that as a first step in
investigating optical conductivity we can neglect current vertex
corrections, and evaluate the conductivity from a convolution of the
Greens functions
\begin{eqnarray}
  \sigma(\om) = \frac{i \omega_p^2}{\om}-\frac{i e^2}{\om} \sum_{\vk \sigma}
  v_\vk^2\; \int\frac{dx}{\pi}
  f(x)\left[\cG_{\vk}(x+\om+i\delta)\cG_{\vk}^{''}(x)+\cG_{\vk}^{''}(x)
  \cG_{\vk}(x-\om-i\delta)\right.\nonumber\\
  \left. +
  \cF_{\vk}^\dagger(x+\om+i\delta)\cF_{\vk}^{''}(x)+\cF_{\vk}^{\dagger ''}(x)
  \cF_{\vk}(x-\om-i\delta)
  \right]
\label{not_vertex}
\end{eqnarray}
\end{widetext}

Within C-DMFT, in the regime where the NCA solvers can be used,
the f-sum rule
\begin{equation}
 \int_{0}^{\infty}\sigma'(x)dx = \frac{\pi e^2}{4}
 \sum_{\vk,\sigma,\alpha=(x,y)}
\left[{d^2\varepsilon_\vk}/{dk_\a^2}\right] n_{\vk\sigma}
\label{sumri}
\end{equation}
is obeyed within a few percent, suggesting, that even for
C-DMFT, where the vertex corrections are non vanishing, the
corrections introduced by this effect are small.  Notice that the
right-hand side of the Eq.~(\ref{sumri}) is proportional to the
kinetic energy of the low energy model if this model contains nearest
neighbor hopping only.

Formula (\ref{not_vertex}) depends on the momentum dependent Greens
function and therefore on the periodization scheme used and the
cluster method employed.  The qualitative features discussed in this
paper and the behavior of the integrated quantities are common to all
methods.

The optical conductivity has been modeled as either a one component
or a two component system via an extended Drude analysis
\cite{Basov_two_comp,Basov_Eff_mass}. The two component
parametrization consists of a Drude peak and a mid infrared feature.

The cluster DMFT results for the optical conductivity of the t-J model
are shown in figure
\ref{DCA-optics}. We show the evolution of the optical conductivity
with doping at various temperatures.  In the very underdoped regime,
there are clearly two components to the optical conductivity with an
optical pseudogap, opening as a function of temperature. On the
other hand beyond $\delta=0.1$ one can describe the optics in terms of
one broad feature which narrows as the temperature is reduced.

\begin{figure}
\includegraphics[width=1.\linewidth]{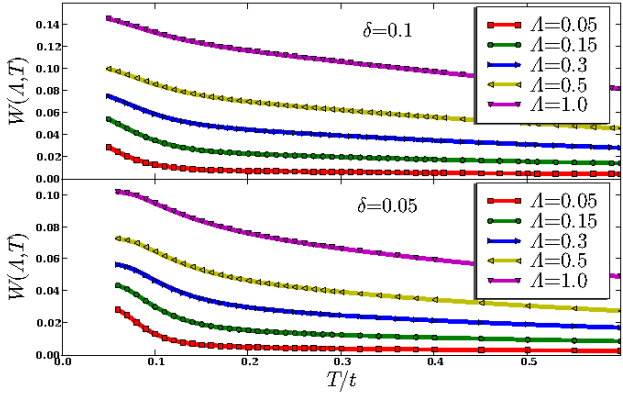}
\caption{(Color online)
  Integrated spectral weight in normal state as a function of
  temperature in underdoped regime for few different cutoff
  frequencies. EDCA and NCA was used.}
\label{Neff}
\end{figure}

It is customary to parameterize the optical conductivity in terms of a
generalized Drude model
\begin{equation}
\sigma(\omega) = \frac{\omega_p^2}{4\pi}\frac{1}{\frac{1}{\tau(\omega)}+i\omega\frac{m^*(\omega)}{m}}
\end{equation}
This parameterization relates the real and imaginary part of the  optical conductivity  {\it in
a given energy range} to  two functions,
$\frac{1}{\tau(\omega) \omega_p^2}$ and
$ \frac{m^*(\omega)}{m \omega_p^2} $ via
\begin{eqnarray}
&&\frac{1}{\tau(\omega)\omega_p^2} = \frac{1}{4\pi}\frac{\sigma'}{\sigma'^2+\sigma''^2}\\
&&\frac{m^*(\omega)}{m \omega_p^2} = \frac{1}{4\pi}\frac{1}{\omega}\frac{\sigma''}{\sigma'^2+\sigma''^2}\\
\end{eqnarray}
The quantity $ \omega_p^2 $ is determined from a requirement involving
the energy range in which the parameterization is used, namely
\begin{equation}
   \frac{\omega_p^2}{8}=\int_0^\Lambda \sigma'(\omega)d\omega \label{plasma}
\end{equation}
where $\Lambda$ is the high energy cutoff.

\begin{figure}
\includegraphics[width=0.8\linewidth,angle=-90]{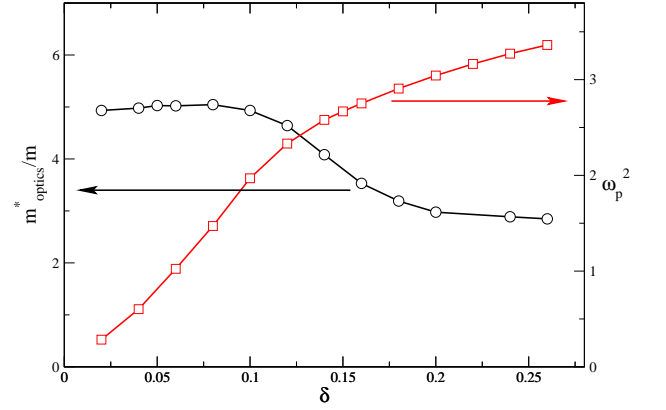}
\caption{(Color online)
  effective mass and plasma frequency as a function of doping.
Obtained from optical conductivity of Fig.~\ref{DCA-optics}.}
\label{plasma-mass}
\end{figure}
Figure \ref{plasma-mass} describes the evolution of the plasma
frequency and effective mass versus doping in the t-J model.
The plasma frequency vanishes at half-filling and linearly increases
at low doping.
The optical mass is weakly doping dependent, and changes from
approximately $3$ in overdoped regime to $5$ in underdoped regime with
largest slope at optimal doping. Weak doping dependence of the
effective mass of the same magnitude was pointed out in
Ref.~\onlinecite{Basov_Eff_mass}.

Given a parametrization of the optical conductivity as a sum of a few
poles, the optical mass measures the ratio of the total spectral
weight compared to the weight in the zero energy pole, representing
the Drude peak. If the transitions between the upper and lower Hubbard
band of the Hubbard model are included in $\omega_p$, i.e.,
$\Lambda>U$, than $\omega_p$ is finite approaching the Mott transition
and consequently the optical mass diverges. On the other hand,
excluding transitions between the Hubbard bands, results in $\omega_p$
vanishing as the Mott transition is approached. In the t-J model, the
upper Hubbard band is projected out, therefore the optical mass is always
finite.
As long as the transitions into the upper Hubbard band are excluded,
the plasma frequency $\omega_p$ of cluster DMFT and single site DMFT
are not too different. Notice, however, that $m^*/m$ is enhanced in
cluster DMFT relative to single site DMFT (not shown) because
superexchange transfers optical weight from the low energy to the
intermeditate energy range $\sim J$.

The optical spectral weight $\omega_p^2$ is in general function of
temperature and cutoff $\Lambda$,
i.e.,
\begin{equation}
  \frac{\omega_p^2}{8} =  W(\Lambda,T)
\end{equation}
In experiment, the cutoff is usually chosen such that the interband
transitions are absent ($\Lambda\sim$ 1eV).  The interband transitions
or transitions into the upper Hubbard band are absent in the t-J model
therefore this requirement is taken into account automatically.

The optical pseudogap which separates the two components of
spectra and is seen as a dip at the scale of $J$ in
Fig.~\ref{DCA-optics} is quite large in the underdoped system
$\delta\sim 0.05$. One could expect that the integral spectral
weight $W(\Lambda)$ for small enough $\Lambda\sim J$ might start
to decrease below a certain characteristic temperature of a
pseudogap. However, as shown in Fig.~\ref{Neff} there is no sign
of such a decrease for any cutoff frequency $\Lambda$ or any
temperature. Although the pseudogap  gap  clearly increases  with
temperature, the Drude peak more than compensates for this
spectral weigh loss and W incresases as T decresases.

\begin{figure}
\includegraphics[width=0.9\linewidth]{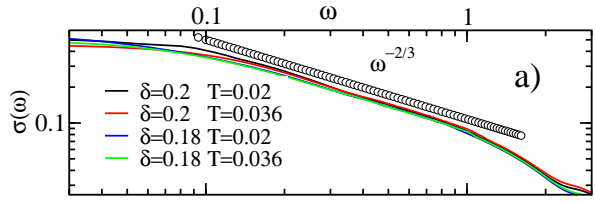}
\caption{(Color online)
  The optical conductivity $\sigma(\omega)$ of the t-J model
is proportional to $\omega^{-2/3}$ in the intermediate frequency
region for optimally doped system.  }
\label{powerlaws}
\end{figure}
Near optimal doping, the optical conductivity displays remarkable
power laws in an intermediate asymptotic regime.  These power laws
were first pointed out by N. Bontemps group in
Ref.~\onlinecite{Bontemps_powerlaw}.  The power laws, and the
possibility to a connection to an underlying quantum criticality, has
been a subject of several recent experimental papers
\cite{VanDerMarelScience}.  CDMFT provides a natural explanation for
these anomalies~\cite{optics-Haule}. These powerlaws were seen in
exact diagonalization of much larger systems \cite{Zemljic},
indicating again the power of the cluster DMFT when it can be compared
with available exact results. The power of the optical conductivity is
very close to $2/3$ as seen in figure \ref{powerlaws}, but an analytic
derivation of this result is not available.

\begin{figure}
\includegraphics[width=1.\linewidth]{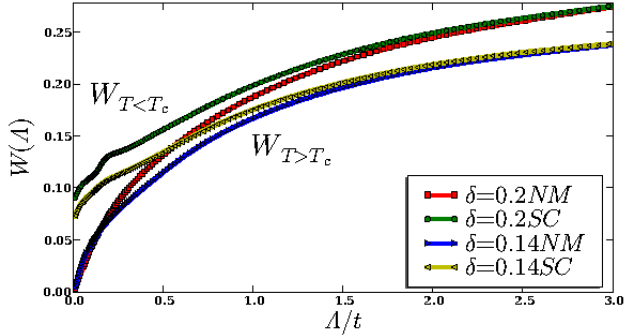}
\caption{(Color online)
  Superconducting and normal integrated spectral weight
  ($N_{eff}$) as a function of cutoff frequency. Optical spectral
  weight which collapses to delta function in superconducting state
  comes from a very extended energy interval ($\sim 3t$). We employed
  EDCA and NCA.
}
\label{sumrule}
\end{figure}

A surprising aspect of the physics of strongly correlated materials,
is that low energy phenomena affects the spectra of the material over a
very large energy scale.  This general phenomena is illustrated in
Fig.~\ref{sumrule}, which shows the integral of optical spectral
weight $W(\Lambda)$ in the normal and the superconducting state.  Low
energy phenomena like the onset of superconductivity which involves a
scale of a fraction of $J$, involves redistribution of optical weight of
the order of $4t \approx 1eV$ which is 40 times more than the gap
value.  A theoretical insight from our calculation is that the high
frequency redistribution of weight comes from the anomalous Greens
function $\cF*\cF$ in Eq.~(\ref{not_vertex}) and hence can not
be observed in the density of states or ARPES measurements.
The large range of redistribution of spectral weight has been also measured
on cuprates and pointed out in Ref.~\onlinecite{Bontemps_redistribution,Molegraf}.

\begin{figure}
\includegraphics[width=0.8\linewidth]{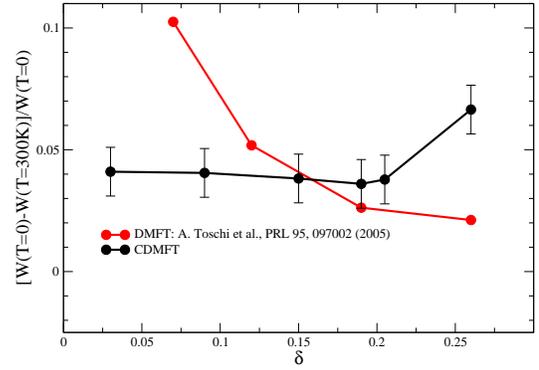}
\caption{(Color online)
  Difference between the low temperature and $300K$ optical
  spectral weight integrated up to $\Lambda=6t$. The cluster data are
  computed within EDCA for the t-J model.
  The single site DMFT data
  are reproduced from Ref.~\onlinecite{Toschi} and correspond to
  the Hubbard model with the same
  superexchange interaction.
  The error bars are due to extrapolation of spectral weight to zero
  temperature from finite temperature results ($T_{min}\sim 0.5 T_c$).
}
\label{single_vs_cluster}
\end{figure}
It is useful to compare the results for the temperature dependence of
the integrated spectral weight of cluster DMFT with those of single
site DMFT as reported by Toschi et.al.~\cite{Toschi} . These are
displayed in figure \ref{single_vs_cluster}.  There are two important
observations, first the sign and the order of magnitude of the effect
is similar in single site and in cluster DMFT. This indicates that
local quantities can be reliably computed in this framework, and do
not change dramatically as the sizes of the cluster is
increased. Second, the doping dependence of this low energy kinetic
energy difference $W(0)-W(300K)$ has opposite slope in cluster DMFT
than in single site DMFT. An interesting question is whether the
existent experimental data agrees better with the single site or
cluster DMFT. It seems that the results in
Ref.~\onlinecite{Ortolani,Syro,Molegraaf} are in better agreement with
the plaquette DMFT, but a more detailed comparison between theory and
experiment, is left for future studies using  the more realistic band
structure of each compound and a  more precise periodization scheme.

\begin{figure}
\includegraphics[width=0.8\linewidth]{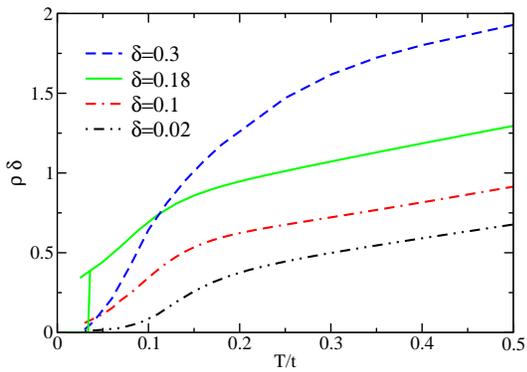}
\caption{(Color online)
  Resistivity versus temperatures in EDCA
  using NCA as the impurity solver.
} \label{resistivity} \end{figure}

We can also compute the  temperature dependent resistivity,
using EDCA in which the vertex correction vanish
\begin{eqnarray}
  \frac{1}{\rho(T)} = e^2 \sum_{\vk \sigma}
  v_\vk^2\; \int\frac{dx}{\pi}
  \left(-\frac{df(x)}{dx}\right)
  \left[{\cG^{''}_{\vk\sigma}(x)}^2+{\cF_{\vk\sigma}^{''}(x)}^2\right].
\end{eqnarray}

Notice that while the scattering rates at zero frequency tend to
saturate at high temperature (see Fig.~\ref{phased}), the
resistivities do not as seen in Fig~\ref{resistivity}.

Notice that the scaling of the resistivity with the number of
holes is approximately obeyed, and that the maximum amount of
linearity is obtained near optimal doping.  More detailed
comparison with experiments will require  a more realistic
modelling of the band structure and a detailed investigation of
the dependence of this quantity on  periodization sheme used.

\begin{figure}
\includegraphics[width=0.9\linewidth]{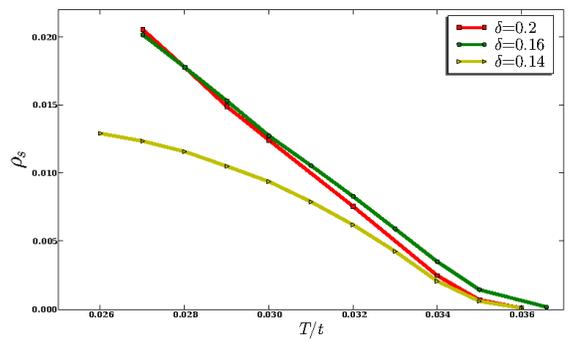}
\caption{(Color online)
  Superconducting stiffness versus temperature, as obtained
  from optical conductivity using NCA and EDCA.
}
\label{stiffness}
\end{figure}

Finally, since we have access to both the real and imaginary part of
the optical conductivity we can compute the superconducting stiffness,
defined as the strength of the delta function peak in the
superconducting state.
Its temperature and doping dependence close to $T_c$ is displayed
in figure \ref{stiffness}. In optimal and overdoped regime, stiffness
is linear function of temperature close to the transition while it is
substantially reduced in the underdoped regime due to opening of the
pseudogap. Similar trend was found in cuprates as pointed out in
Ref.~\onlinecite{stife}.

\begin{figure}
\includegraphics[width=0.9\linewidth]{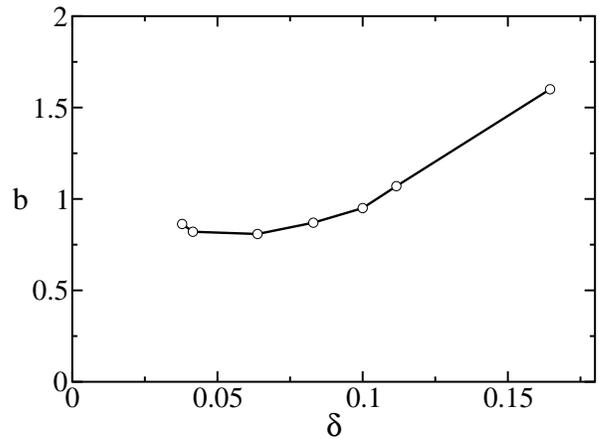}
\caption{(Color online)
  Coefficient of the linear term of the superconducting stiffness
  $\pi(\rho_s(0)-\rho_s(T))/T$ versus doping for the Hubbard model using CTMQC.
}
\label{cstiffness}
\end{figure}

With NCA, we are not able to reach sufficiently low temperatures to
address the crucial issue of the doping dependence of the linear term
of the superfluid stiffness. Instead we use the techniques of section
\ref{supra} to evaluate the low temperature behavior of this quantiy
\begin{eqnarray}
\rho_s(0)-\rho_s(T) = \frac{b}{\pi} T = e^2 \frac{2 \log 2 }{\pi^2}
(k_B T)\frac{Z_n^2 v_0^2}{v_F v_\Delta}
\end{eqnarray}
where $Z_n$, $v_\Delta$, and $v_F$ were evaluated in section \ref{supra}
and $v_0$ is the band fermi velocity.  The coefficient $b$ can be
evaluated directly from imaginary axis data of CTQMC and is plotted in
Fig.~\ref{cstiffness}.

CDMF captures the weak dependence of $b$ on doping, which was a
subject of intensive experimental investigations \cite{Broun,
Boyce}. More detailed studies of this quantity in C-DMFT,
including vertex corrections, and more investigations of the
periodization dependence of this quantity,
as well as the related B1g slope of the Raman scattering \cite{Tacon}
is certainly warranted.

\section{Magnetism Superconductivity and Spin Resonance}
\label{magnetism}

In this section we turn to the magnetic properties, starting from the
cluster magnetic quantities \cite{Bourgers,Tranquada,review5}.
As in the rest of the paper, we confine ourselves to the study of
 minimal   models,  in this section the t-J model   with $t'=0$.
  Notice however, that it is
known from numerous studies that the presence of a next nearest
neighbor hopping $t'$ affects significantly the region of
stability, of the magnetism, and can suppress it altogether
\cite{Civelli}.

\begin{figure}
\includegraphics[width=0.8\linewidth,angle=-90]{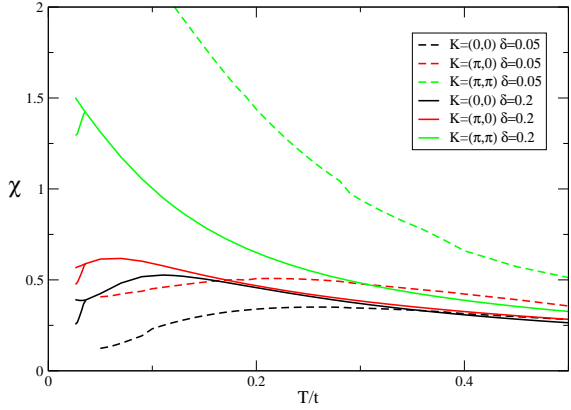}
\caption{(Color online)
  The cluster spin susceptibilities of the t-J model versus temperature at two
  different dopings obtained by EDCA and NCA.}
\label{chi1}
\end{figure}
The static cluster susceptibilities are displayed in figure
\ref{chi1}. These are coarse grained versions of the momentum
dependent magnetic susceptibility, over the different regions of the
Brillouin zone of the size of one quarter of the first Brillouin zone.
While the cluster susceptibilities are relatively smooth functions of
doping, they clearly demonstrate that the spin fluctuations in
different regions of the Brillouin zone have dramatically different
behavior with doping and temperature.  The $(\pi,\pi)$ component,
dominated by the $\chi_{\pi \pi}$ susceptibility strongly increases as
we approach half filling.  In contrast the uniform component, $\chi_{0 0}$
decrease as doping is reduced, a signal of the
opening of the pseudogap.  The same is true of $\chi_{\pi 0}$.  Hence,
an interesting property of the pseudogap state is the increase of
staggered magnetic fluctuations with the opening of the pseudogap. A
similar contrast between the staggered and uniform response, is seen
in their temperature dependence.  We see that while the uniform response
decreases with temperature in the underdoped regime, the staggered
response increases.

\begin{figure}
\includegraphics[width=1.0\linewidth]{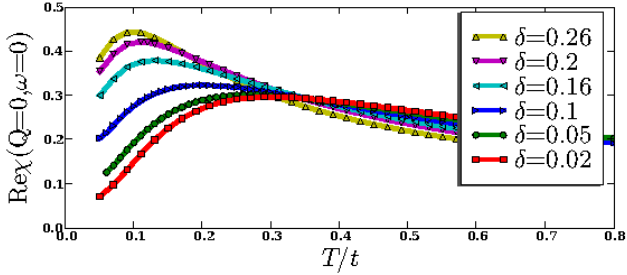}
\caption{(Color online)
  The $\vQ=0$ cluster spin susceptibility versus temperature
  at different dopings for the t-j model within EDCA.
}
\label{chi}
\end{figure}
We now proceed to uniform spin susceptibility shown in
Fig.~\ref{chi}. The $\vq=0$ susceptibility at zero doping displays the
characteristic behavior of the Heisenberg model, with a Curie like
behavior at high temperatures and a broad maximum at a scale of the
order of $J$, as the spins begin to form singlets. The main effect of
doping is to reduce the effective exchange. Experimentally the shift
of the minima in the susceptibility is seen\cite{Nakao,YKubo}, but it
occurs faster than in figure \ref{chi}.  The effective spin-exchange
will be reduced by the addition of a negative next-nearest
neighbor hopping $t'$ to the model.

\begin{figure}
\includegraphics[width=0.8\linewidth]{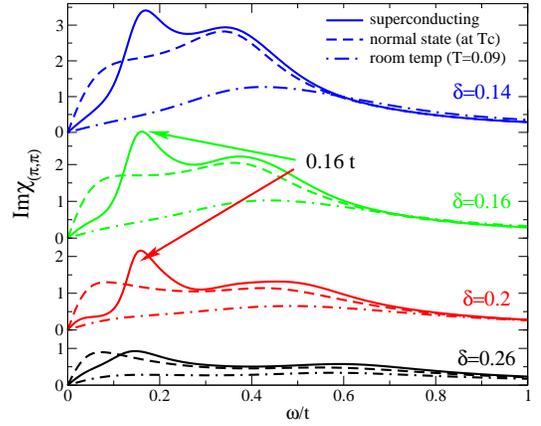}
\caption{(Color online)
  The dynamical spin susceptibility at $\vq=(\pi,\pi)$ for few
  different doping levels and three different temperatures:
  superconducting state, normal state at the transition temperature
  and at room temperature. The pronounced peak is formed in SC
  state at $0.16t\approx 48\,\mathrm{meV}$ and a broad peak in normal state is
  around $100-140\,\mathrm{meV}$.
  Susceptibility at normal temperature is
  much smaller and the peak moves to higher frequencies.
  The resonance is strongest at the  optimally
  doped system.  It disappears quickly in the overdoped side and
  somewhat more slowly  in the underdoped side. Results are obtained
  with EDCA and NCA.
}
\label{susc40}
\end{figure}
We now turn to the frequency dependence of the imaginary part of the
$(\pi,\pi)$ susceptibility probed in neutron scattering experiments.
As shown in figure \ref{susc40}, a pronounced peak in the $(\pi,\pi)$
spin response at frequency $0.16t$ in the optimally doped regime can
be observed when entering the superconducting state. The position of
the peak is temperature independent, but depends weakly on doping
tracking the critical temperature.  Our results are in qualitative
agreement with experiment, for example the resonance energy scales
with doping like $5 T_c$ and its position does not depend on
temperature \cite{Bourgers}.
In addition we see a broader peak around $0.35-0.45t$ extending to
very high frequencies of order of $t\approx 300\,\mathrm{meV}$
which also gains some weight  when entering the superconducting state.

Cluster  methods coarse grain  the momentum dependence. In the
plaquette case, the coarse graining is done
 over $1/4$ of the Brillouin zone centered at
$(\pi,\pi)$ therefore it is reasonable to compare our results
with the $\vq$ integrated susceptibility from
Ref.~\onlinecite{Dai} where the two features, present in the mean
field theory, $35\,\mathrm{meV}$ resonant peak as well as broader
peak around $75\,\mathrm{meV}$ extending up to
$220\,\mathrm{meV}$ were observed.

\begin{figure}
\includegraphics[width=0.9\linewidth]{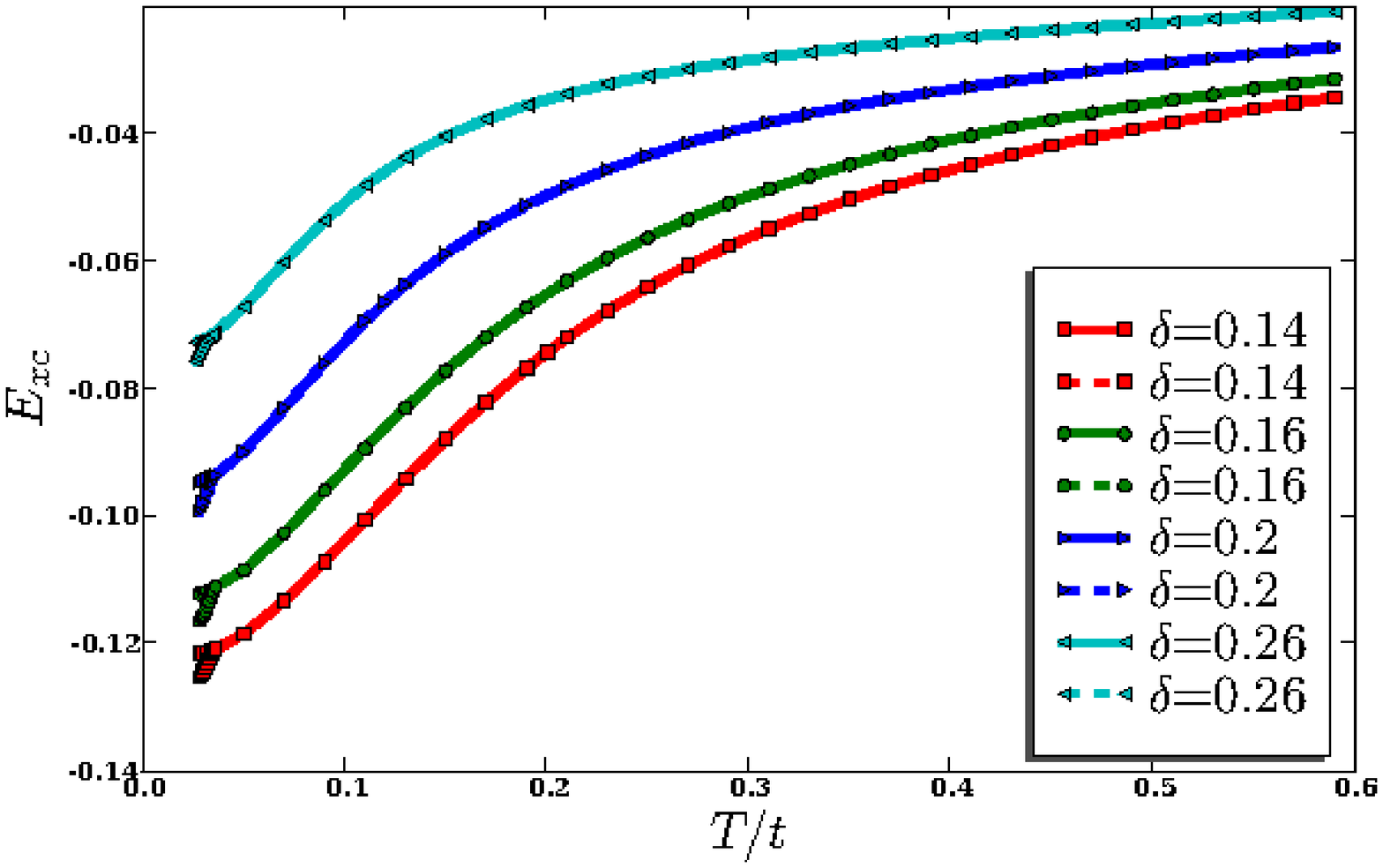}\\
\includegraphics[width=0.9\linewidth]{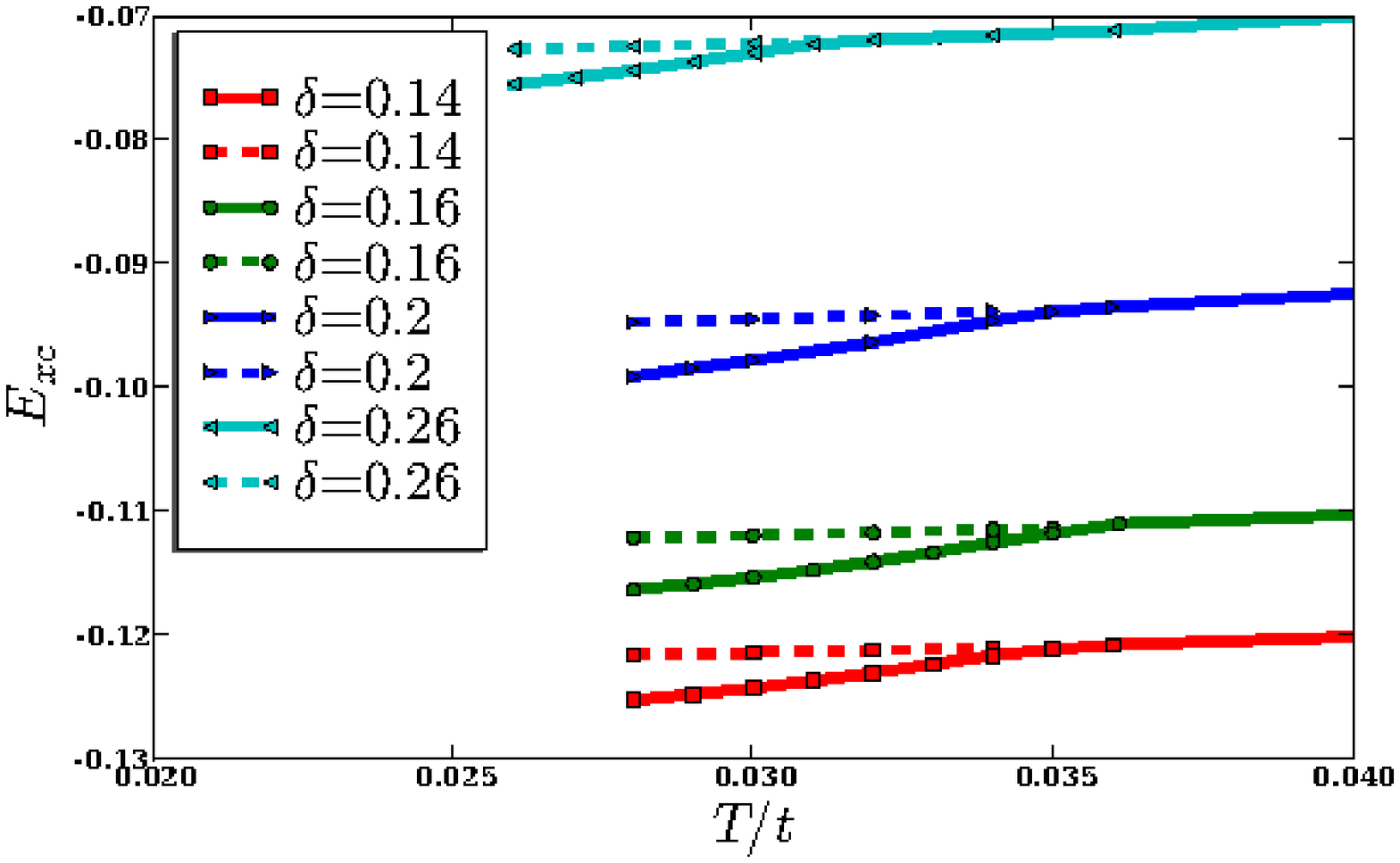}
\caption{(Color online)
The exchange energy versus temperature for few doping levels obtained
by the NCA in EDCA. The lower panel is a blowup of the low
temperature regime. The dotted lines correspond to metastable normal
state below $T_c$ while the full lines continue into the
superconducting state.
}
\label{ExcEne}
\end{figure}
The exchange energy of the t-J model can be expressed as an integral
of the spin susceptibility\cite{Scalapino}
\begin{equation}
E_{xc} = \frac{3 J}{\pi} \int {d^2}\vq\, d\omega\, b(\omega) \Im[{\chi}(\vq, \omega)] ( \cos q_x + \cos q_y )
\end{equation}
Using this equation, one can elucidate the origin of the
superconducting condensation energy, and the relative contribution of
the different features of the spectral function.

Clearly, an important contribution to superconducting condensation
energy arises from the incoherent features of the spin spectral function
(around the frequency $0.4-0.5t$) rather than from the spin resonance.

The exchange energy as a function of temperature is shown in
Fig.~\ref{ExcEne}. At high temperature, spins are disordered and the
exchange energy is negligible. At temperature below $J$ the singlets
are formed and the exchange energy noticeably decreases, especially in
the underdoped regime. At $T_c$ the exchange energy decreases further
and gives far the largest contribution to the condensation energy of
the t-J model, as shown in Ref.~\onlinecite{optics-Haule}. The exchange
energy mechanism, observed in cluster DMFT study, is thus in agreement
with the strong coupling magnetic mechanism for the superconductivity.

The spin resonance has been viewed from two different perspectives
(see Ref.~\onlinecite{Bourgers} and references therein):
i)
starting from electronic quasiparticles and their residual
interactions in a d-wave superconductor, residual interactions form a
particle hole bound state with spin one, which is identified as the
spin resonance.  ii) Alternatively starting from a disordered quantum
spin system, one can identify the spin resonance as a massive spin one
excitation, which becomes massless as one approaches the magnetically
ordered phase.

The cluster EDMFT equation \ref{SCC2} reconciles both points of view
in a unified approach, since the equations for the spin susceptibility
contain both the exchange interaction characteristic of the insulator
$J(q)$, as well as the quasiparticle contribution described by the
spin cumulant $M$ Eq.~\ref{SCC2}.

The appearence  of the spin resonance requires  the dramatic decrease
of the anomalously large scattering rate in the normal state which is
strongly reduced when the electrons condense to form d-wave pairs,
avoiding criticality at low temperatures.
The resonance, however,
appears only in the superconducting state and is not present in
the normal state.

\section{ Pseudoparticle interpretation, Connection with other Mean Field Theories}
\label{pseudoparticles}

\begin{figure}
\includegraphics[width=0.9\linewidth]{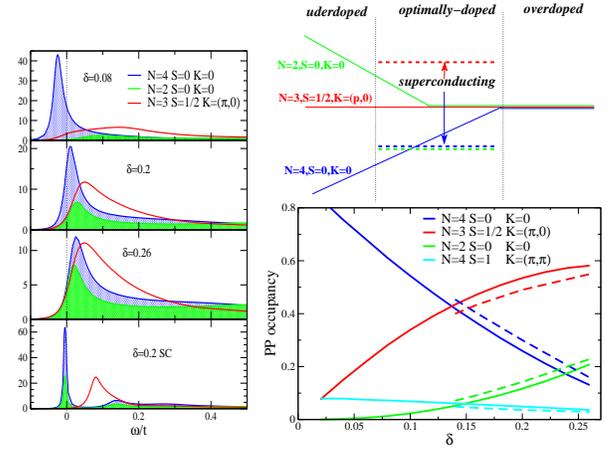}
\caption{(Color online)
  left: Pseudoparticle spectral functions for the
three most important pseudoparticles: ground states for N=4, N=3 and
N=2 sectors. Right-top: Sketch of pseudoparticle threshold energies
which can be interpreted as the effective many-body levels in normal
and superconducting state. Right-bottom: Pseudoparticle occupancies
versus doping for most important pseudoparticles. The full lines
correspond to the normal state while the dashed lines correspond to
the superconducting state.  }
\label{pseudop}
\end{figure}

In this section we give interpretation of physical observables in
terms of pseudoparticle (eigenstates of the cluster) spectral
functions. This is an alternative insight into a rich physics
contained in the solution of cluster DMFT equations on a plaquette.

Pseudoparticle creation and annihilation operators were introduced as
mathematical entities representing the atomic eigenstates of the
plaquette immersed in the Cluster DMFT medium.  We have found that out
of the large number ($3^4$) of pseudoparticles that we introduced,
very few of them are important for reproducing the low energy part of
physical observables. For example, more than 95\% of the one-particle
spectral function at the Fermi level comes from a few convolutions of
pseudoparticles in Eq.~(\ref{NCA_G}), within NCA approach.  This
constraint is however not present for high energy part of the spectra
such as Hubbard bands where
contribution of most of pseudoparticles can be identified.  The ground
state and the low laying excitations are much more restricted and are
a superposition of only a few atomic states. In the plaquette these
important states are:
\begin{eqnarray}
\Gamma_4    &=&  | N=4, S=0, K=0\rangle\\
\Gamma_{4'} &=& | N=4, S=1, K=(\pi\pi)\rangle\\
\Gamma_{3\sigma(\pi 0)} &=& | N=3, S=1/2, Sz=\sigma, K=(\pi,0)\rangle\\
\Gamma_{3\sigma(0 \pi)} &=& | N=3, S=1/2, Sz=\sigma',K=(0\pi)\rangle\\
\Gamma_2 &=& | N=2, S=0, K=0\rangle
\end{eqnarray}
where $N$ is the number of electrons in the cluster eigenstate, $S$
and $S_z$ are the total spin and its $z$ component of the cluster
eigenstate, and $K$ is momentum of the cluster eigenstate.

Notice that although only few cluster eigenstates contribute to the
ground state, the wave function is still highly nontrivial since it is
a product state of an infinite number of states in the bath with the
few atomic eigenstates of the impurity.
This surprising result of restriction to a few cluster eigenstates could
be beneficial to devise useful approximations while extending C-DMFT to
larger clusters in the future. In this paper we exploit this fact to
give a simple interpretation of the different doping regimes of the
t-J model.

Fig.~\ref{pseudop}-left shows the evolution of the three most important
pseudoparticle spectral functions from the underdoped to the overdoped
regime.

At small doping, the cluster is mostly occupied by the singlet state
with one particle per site and zero momentum $\Gamma_4=| N=4, S=0, K=0\rangle$,
(half filled singlet). This pseudoparticle has the largest occupancy as
shown in Fig.~\ref{pseudop}.  It describes a system locked in a short
range singlet state as a consequence of the strong superexchange
interaction.

The cluster electron spectral function describes the process of
addition and removal of an electron from the cluster at frequency
$\omega$. Within NCA, it is constructed from the convolution of two
pseudoparticles with different cluster occupation $N$ and $N+1$, or
$N-1$ , with the frequency restricted between zero and $\omega$ as
described by Eq. (\ref{NCA_G}).  The necessary
condition for a peak of the one-particle spectral function at the
Fermi level is that at least two pseudoparticle spectral functions
share a common threshold and are strongly peaked at the same threshold.

In the \textit{underdoped regime}, the thresholds of all other pseudoparticles
except $\Gamma_4$ are significantly shifted with
reference to the half filled singlet, a pseudogap results in the one
particle spectra in the underdoped regime. This gap in threshold
energies severely limits the possible decay processes of the electron
resulting in a low electronic scattering rate. This is the
plaquette-impurity model of a few holes propagating in a sea of
singlets.

At large doping, i.e. in the {\it overdoped regime}, where the
Kondo scale is dominant we obtain the standard DMFT description
of a strongly correlated Fermi liquid. As is well known
from the study of the Fermi liquid regime of the single impurity
Anderson model, all pseudoparticles develop thresholds (X-ray
singularity) at the same frequency which is related to the Kondo
temperature of the problem. In our plaquette DMFT, all three
important pseudoparticles (half-filled singlet, doublet with one
hole per plaquette $\Gamma_{3\sigma\vK}$ and singlet with two
holes per plaquette $\Gamma_2$) have a power law divergence at
the same threshold frequency at zero temperature
(Fig.~\ref{pseudop}) which is a standard
signature of the Kondo effect. Hence, the one particle spectral
function begins to develop the Kondo-Suhl resonance at the Fermi
level since the convolution between the doublet
$\Gamma_{3\sigma\vK}$ and half-filled singlet $\Gamma_4$ (or
$\Gamma_2$ singlet) state is large at low frequency. The
one-particle spectral function is peaked slightly above the Fermi
level.  Notice that while we cannot follow the formation of the
Kondo resonance to very low temperatures due to the well known NCA
pathologies, we can clearly see the onset the Fermi liquid
behavior in Fig.~\ref{Gloc}.  The overdoped regime characterized
by the common threshold of pseudoparticles is distinctively
different from the underdoped regime, where only important state
is the half-filled singlet $\Gamma_4$ and the
$\Gamma_{3\sigma\vK}$ doublet. The latter has a very little
spectral weight in the region of the singlet peak.

{\it The Transition Region, Normal State:} In the optimal doped
regime, the Kondo scale and the superexchange compete giving rise to a
regime with very large scattering rate and consequently a small
coherence scale.

Surprisingly the evolution of the spectral function with doping is
such that the optimally doped regime is approximately particle hole
symmetric.  As shown in Fig.~\ref{pseudop}-right-top the threshold of the N=2
cluster ground state and N=3 cluster ground state (doublet) merge
first resulting in a Kondo-like contribution to the electron spectral
function. This contribution is peaked above the Fermi level in a
one band model  below   half-filling, in a Fermi liquid regime.
The half-filled singlet however remains the lowest state in
energy and still gives a significant contribution to the electron
spectral function. The later contribution is peaked below the Fermi
level and keeps a pseudogap-like shape.  Adding  the two
contributions to the electron spectral function  restores
the particle-hole symmetry in the density of states both in normal and
superconducting state at optimal doping (see Fig.~\ref{Gloc}g and
\ref{dos}).  The approximate restoration is important, because it is
known that cluster of impurities such as the two impurity Kondo model
\cite{VarmaJones}, have a critical point only in the particle hole
symmetric case \cite{Jones_Kotliar_Millis}.

Notice that the point of maximum scattering rate in
Fig.~\ref{phased} coincides with the merging of the thresholds of
the pseudoparticles (see Fig \ref{pseudop}). Around the same doping level,
an  approximate particle hole symmetry is restored in one-particle
Green's function (see Fig.~\ref{Gf_tj_100} and \ref{Gloc}).
Hence the term avoided cluster quantum multicriticality describes
better the phenomena observed in this study since to reach the quantum
critical point both the particle hole symmetry and the ratio of Kondo
to RKKY coupling need to be varied.

{\it Transition into the superconducting state:} The degeneracy
responsible for the strongly incoherent metal with large scattering
rate at the Fermi level is lifted by the superconductivity avoiding
the critical point.  This dramatic reduction of scattering rate in
going from the normal to the superconducting phase, depicted in figures
\ref{phased} and \ref{imSigctqmc}, highlights how anomalously incoherent the normal
state at optimal doping is, and how those anomalies are removed by
superconductivity.

This fact has also a natural interpretation in terms of
pseudoparticles.  Fig.~\ref{pseudop}-left shows that both important
singlet pseudoparticles (for $\Gamma_4$ and $\Gamma_2$) develop a very
sharp peak at the same threshold frequency and, at the same time, their
occupancy increases (see Fig.~\ref{pseudop}-bottom) upon condensation,
indicating that electrons are locked into singlets with zero
momentum. A gap opens between the singlets and doublets which gives
the gap in the one-particle density of states. Because of this gap in
the pseudoparticle thresholds, the large imaginary part of the
electron self-energy does not persist in the superconducting state
(see also Fig.~\ref{phased}). Notice however that in superconducting
state the pseudoparticles are strongly mixed and the off-diagonal
spectral function $A_{\Gamma_4 \Gamma_2}$ also develops a pole at the
same threshold as $A_{\Gamma_4}$ and $A_{\Gamma_2}$. The off-diagonal
spectral function $A_{\Gamma_4 \Gamma_2}$ describes the creation of a
Cooper pair on the cluster $G_{\Gamma_4 \Gamma_2} = \langle
0|a_{\Gamma_4}^\dagger(\tau) a_{\Gamma_2}|0\rangle$ and therefore
diverges at low temperature at the same threshold frequency.

Since the density of states is composed of two almost equally
important contributions, i.e., the convolution of the doublet with
both singlets ($\Gamma_4$ and $\Gamma_2$), the superconducting gap is almost
particle hole symmetric in the optimally doped regime with half-width
of the order of $0.1t$. When the doping value is changed from its
critical value, the asymmetry in the superconducting density of states
appears.  The magnitude of the asymmetry is the same as the asymmetry
of the corresponding normal state spectra and comes from the fact that
the occupancy and therefore importance of the $\Gamma_4$ singlet
exceeds the importance of the $\Gamma_2$ singlet (see Fig.~\ref{pseudop}-bottom).

Finally we comment on the role of the triplet pseudoparticle.  The
spin susceptibility comes almost entirely from the convolution of the
half-filled singlet with the half-filled triplet ($\Gamma_4$ with
$\Gamma_{4'}$). The later develops a peak at an energy $0.16t$ upon
condensation which results in the resonance in the spin
susceptibility shown in Fig.~\ref{susc40}.

It is interesting to derive the form of a low energy Hamiltonian
involving the pseudoparticles in question. The conservation of charge,
spin and cluster momentum considerably restricts the form of this
Hamiltonian.  If we assume it is of the Kondo form, it takes the
following form
\begin{widetext}
\begin{eqnarray}
H = \sum_{\Gamma} \epsilon_{\Gamma}  a^{\dagger}_\Gamma a_\Gamma
+ \sum_{k \vQ \sigma} \epsilon_{k \vQ} c^{\dagger}_{k \vQ \sigma} c_{k \vQ \sigma}
+  J_1 a^{\dagger}_{\Gamma_4} a_{\Gamma_2} \sum_{k k'  \vQ \sigma'
  \sigma} \epsilon_{\sigma, \sigma'} c_{k \vQ \sigma}  c_{k \vQ
  \sigma'} +h.c
\label{Hpp}
\\
+ J_2 \sum_{k\sigma k'\sigma',  \vK,\vK'\in[(0,\pi),(\pi,0)]} a^{\dagger}_{\Gamma_{3\sigma\vK}} a_{\Gamma_{3\sigma'\vK'}}
{c^{\dagger}}_{k' \vK'\sigma'} c_{k \vK\sigma} + h.c.
+\lambda\sum_\Gamma(a^{\dagger}_{\Gamma} a_{\Gamma}-1)
\nonumber
\end{eqnarray}
\end{widetext}
where $\Gamma$ runs over the relevant low energy pseudoparticles.
$\epsilon_{\sigma, \sigma'}$ is an antisymmetric tensor and the $\vQ$
runs over cluster momenta. Here $c^\dagger_{k\vQ\sigma}$ operators
create an electrons in the bath with cluster
momenta $\vQ$ and spin $\sigma$. The operators $a^\dagger_\Gamma$
create a pseudoparticle on the cluster (see Eq.~\ref{pseudo_def}).

This Hamiltonian contains the competition of the particle hole and
particle particle channels for pairing with the baths of conduction
electrons, and the approach to criticality is controlled by the
variation of the on site energy $\epsilon_{\Gamma}$ which should be
identified with the pseudoparticle thresholds.  It would be very
intersting to investigate this impurity model with the tools and the
perspective of Ref.~\onlinecite{Lorenzo}.
It is clear that superconductivity will add magnetic field like
terms
proportional to $\langle a^{\dagger}_{\Gamma_4} a_{\Gamma_2} \rangle$
$ \sum_{k k' \vQ \sigma' \sigma} \epsilon_{\sigma, \sigma'} c_{k \vQ
\sigma} c_{k \vQ \sigma'}$.  These terms should be strongly relevant,
and moves the system away from criticality.

\begin{figure}
\includegraphics[width=0.9\linewidth]{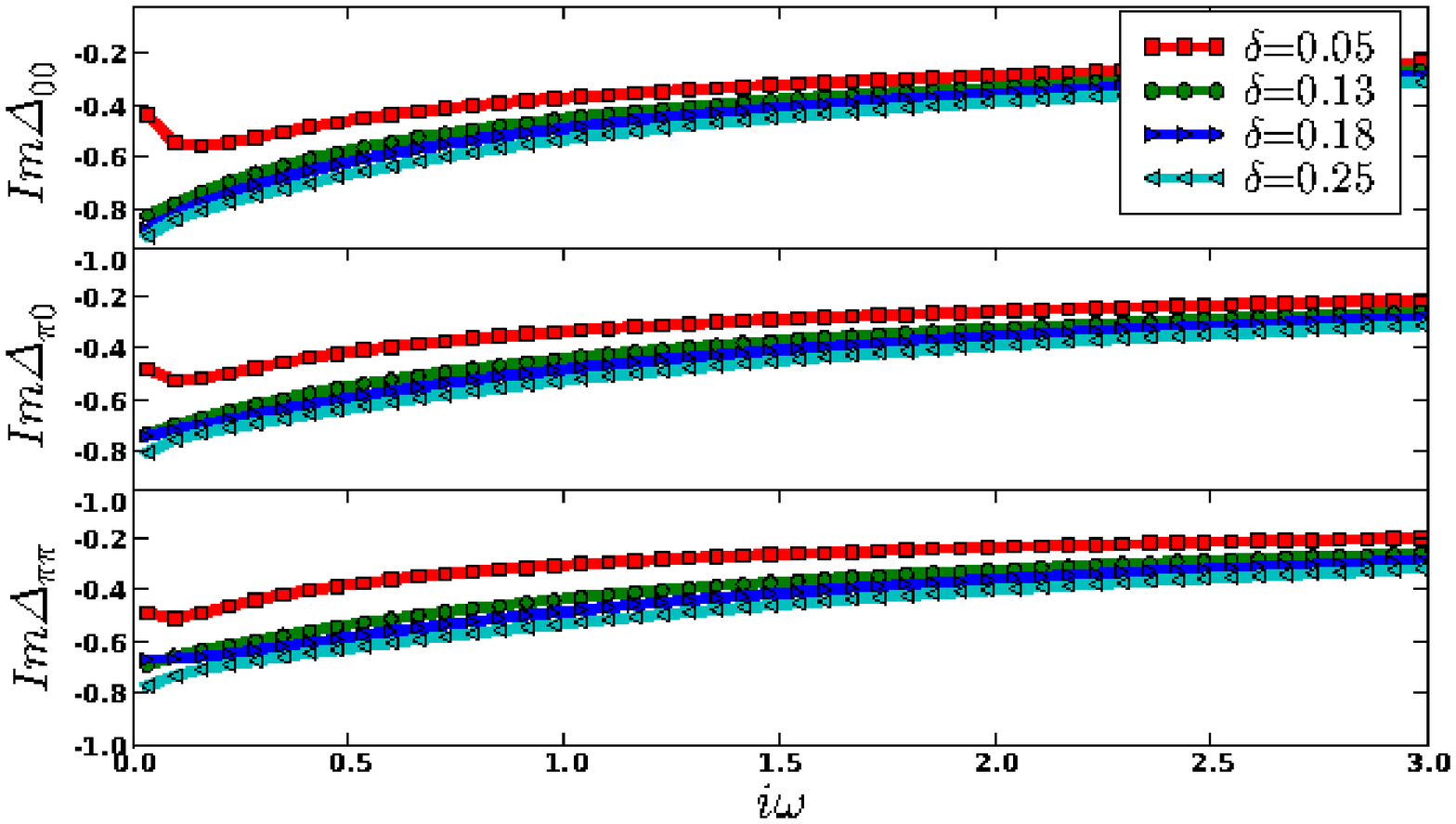}\\
\includegraphics[width=0.9\linewidth]{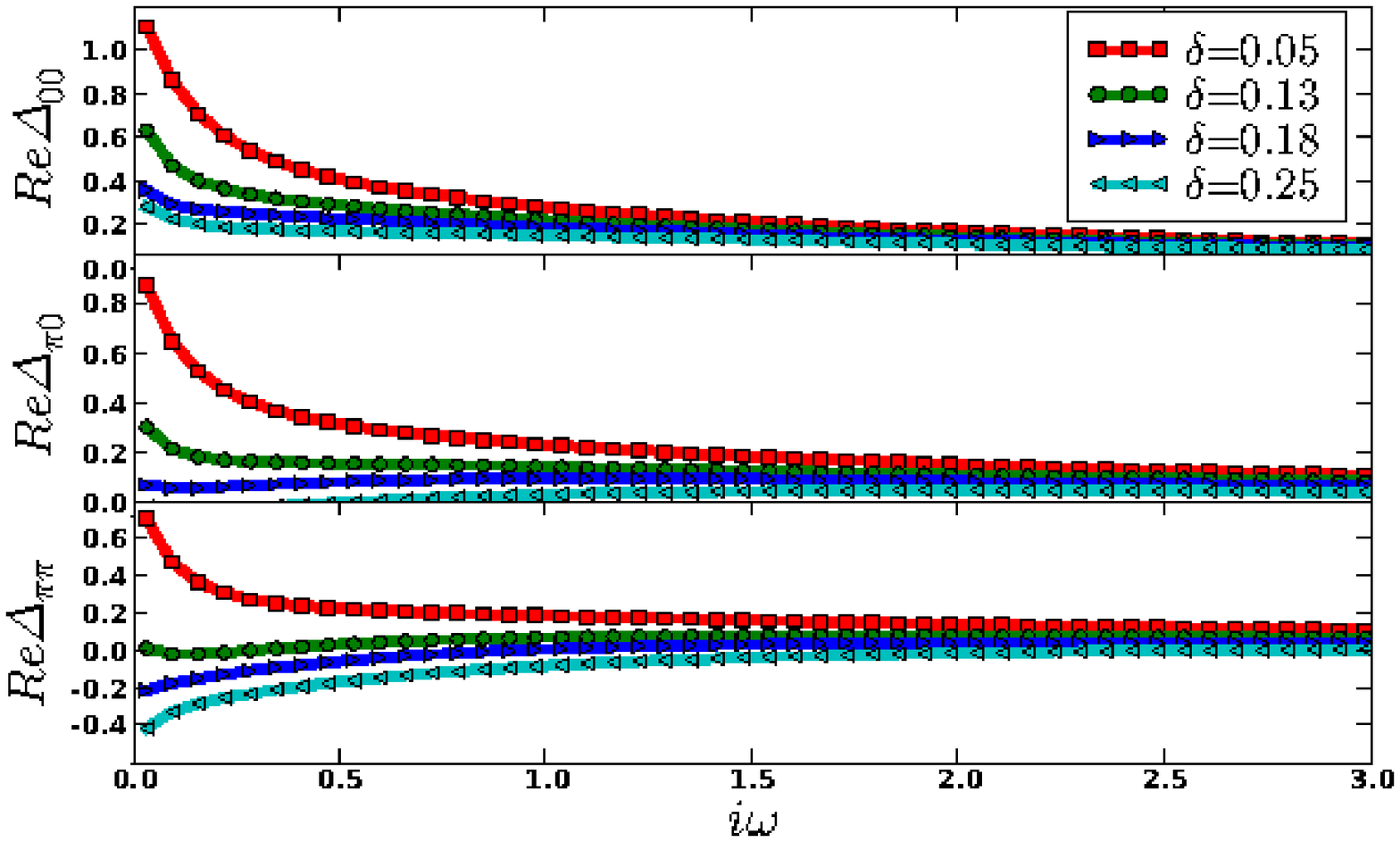}\\
\includegraphics[width=0.9\linewidth]{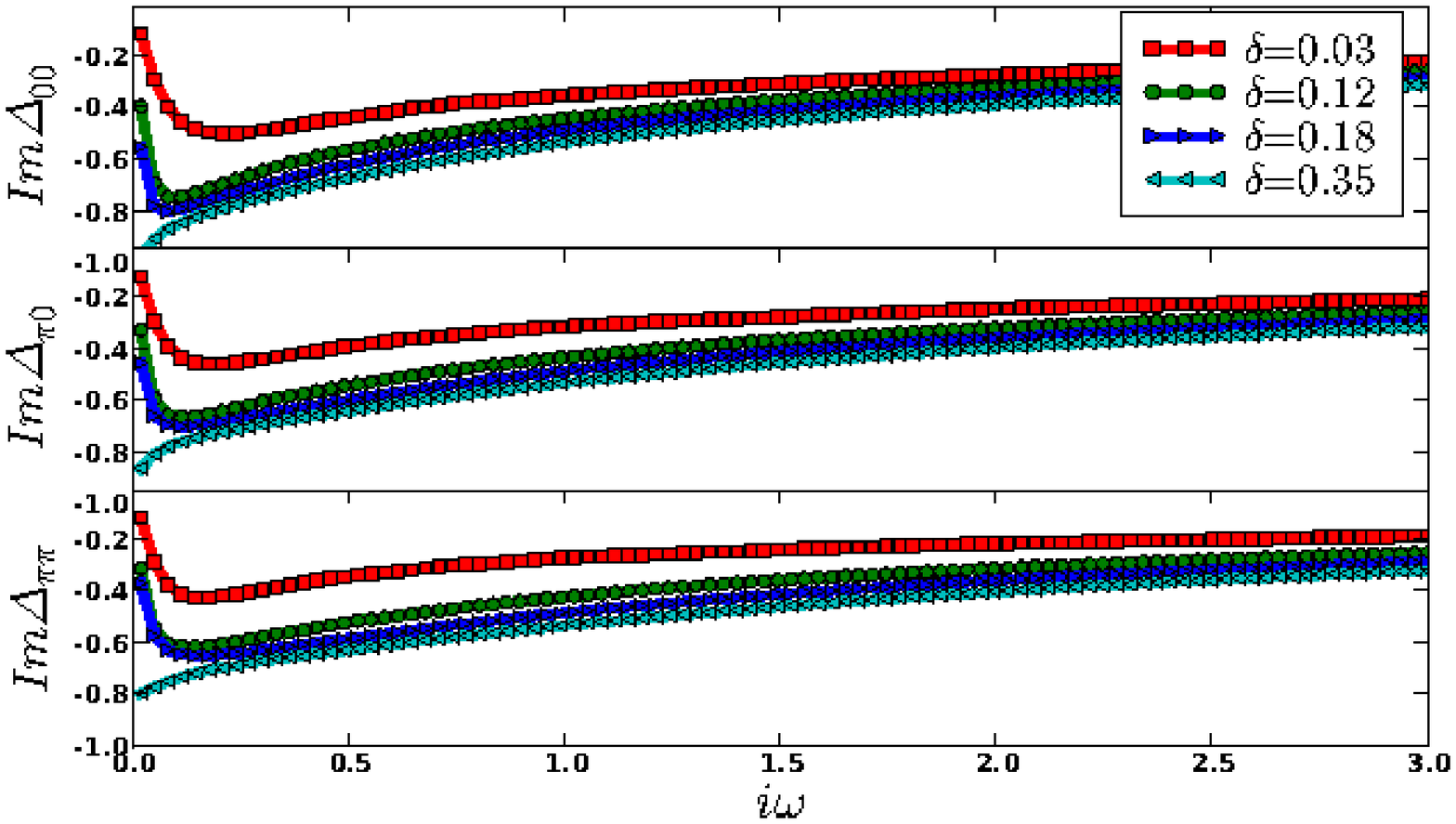}\\
\includegraphics[width=0.9\linewidth]{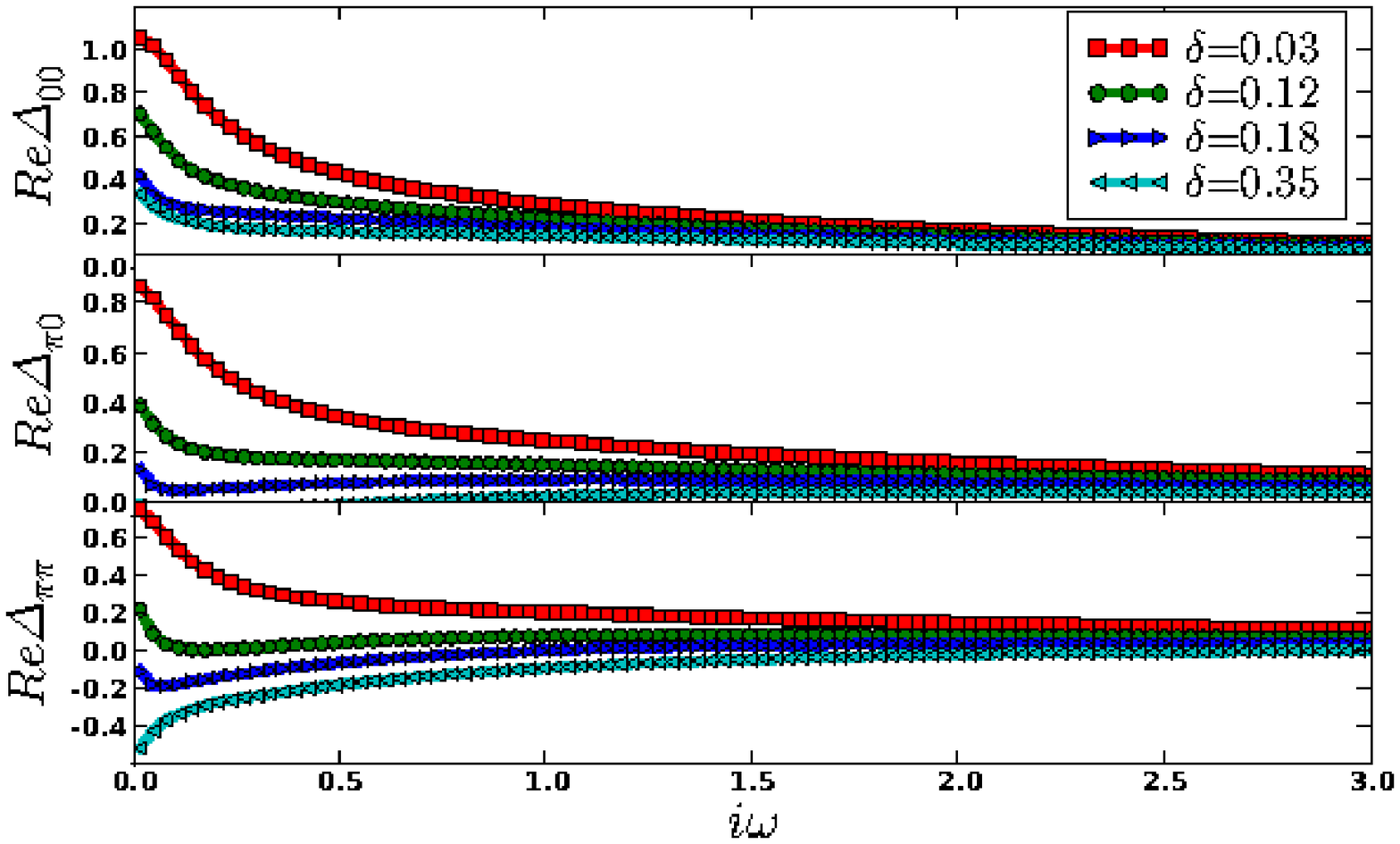}
\caption{(Color online)
  First panel: Imaginary parts of the cluster
hybridization functions for various dopings in normal state at
$T=0.01t$ using C-DMFT and CTQMC. Second panel: Real parts of the same hybridization
functions  in normal state at $T=0.01t$. Third panel: Imaginary
parts of the same hybridization functions in superconducting
state at $T=0.005t$. Fourth panel: Real part of the same
hybridization functions in superconducting state at $T=0.005t$.
The self-energies of the cluster show strong momentum dependence
while hybridizations are only weakly momentum dependent.
Furthermore, there seems to be no indication of any criticality
in the hybridization functions such as the formation of a gap. }
\label{hybrid} \end{figure}

Within CDMFT, the cluster of few sites ($2\times 2$ in our case)
hybridizes with the Weiss field $\Delta$, defined in Eq.~(\ref{SCC1}).
In single site DMFT, this effective medium drives the Mott
transition. On the Bethe lattice within single site DMFT, it is
proportional to the local Green's function $\Delta = t^2 G$ and
therefore becomes gapped in the Mott insulating state while it remains
gapless in the metallic phase. Hence due to the DMFT self-consistency
condition this quantity shows a very strong doping dependence.

Within cluster DMFT the effective medium is only weakly doping
dependent and the evolution with doping is smooth (see
Fig.~\ref{hybrid}) in the doping range considered here.  Moreover,
this quantity shows very mild momentum dependence as opposed to strong
momentum dependence of self-energy shown in
Fig.~\ref{Self_tj_200}. For examples, the $(0,0)$ and $(\pi,\pi)$
Green's functions show almost no spectra at low frequency (are almost
gapped) while the hybridization functions of these two orbitals are
very similar to $(0,\pi)$ hybridization function which contains most
of the low frequency spectral weight. The mild and smooth doping
dependence of hybridization functions leads us to believe that the
proximity to quantum cluster criticality, which manifests itself in
large scattering rate and vanishing coherence scale, is driven by the
impurity model itself rather than the self consistency condition.

The picture here, is based entirely on a finite temperature
analysis, and is in the spirit of the DMFT approach, where we
approach the strong correlation problem starting from high
temperatures.

It is important to continue the normal solutions of the plaquette DMFT
equations to very low temperature to clarify the mathematical source
of the criticality that we observe at higher temperatures.  The
critical point could occur exactly at $T=0$, as proposed by Capone
et. al.\cite{CaponeM1} in the context of the two band Hubbard model with
inverted Hund rule exchange, and by us in ref
\cite{qcp}.   The quantum critical point  may
exist in a impurity model with a fixed bath or might require the DMFT
self consistency condition.  Alternatively, there may be a finite
second order endpoint of a first order line, as  found in
DMFT lattice models related to the two impurity model \cite{Moeller}.
Notice also that power laws in an intermediate asymptotic regime,
without an obvious underlying quantum critical impurity model have
also been found in impurity model related to frustrated magnets
\cite{georges}.

Still, the precise nature of the low temperature normal state phase
below $T_c$ is not essential for the validity of the CDMFT
description. What matters is that at very high temperature $ T > J$
single site DMFT is a good description of the system, and as we lower
the temperature we find a broad region of temperatures where the
plaquette reference frame correctly captures the physics of the
problem with its apparent criticality, even though at much lower
temperatures a more non local description will be needed. It is even
possible that the zero temperature solution of the DMFT equations do
not exist, in which case, unlike the standard BCS theory where the
superconductivity is viewed as an instability of a normal phase, we
would have a superconducting state that exist without an underlying
normal state.

There  is an important distinction between our views and those of
local quantum critical scenario based on single site  EDMFT
scenario \cite{Si}. In the latter case, the locality of the
quantum critical theory of the lattice  is asserted to be
reliable at $T=0$, while the results of the EDMFT equations in
two dimensions are known to be  less reliable as temperature is
raised
\cite{Pankov} . On the other hand, the results of
plaquette DMFT are expected to become more accurate as the
temperature is raised.

\section{Conclusion and Discussion}

In this paper we developed and applied a plaquette Dynamical Mean
Field Theory to understand the nature of the superconductivity near
the Mott transition. In relation to earlier works, we focused of low
but finite temperatures to allow a comparison with the underlying
normal phase. For this purpose advanced impurity solvers
were brought to bear on the solution of the CDMFT equations.

The idea of using a plaquette in a self consistent medium as a
reference frame to reconstruct physical correlations functions on a
lattice, while appealing, has several different implementations
through different cluster schemes. Here, we stressed the numerous
qualitative features which are common to all methods while pointing
out the few significant quantitative discrepancies that we found among
the different cluster methods in the course of our investigations.

The low temperature landscape of strongly correlated electron systems
can have many competing phases, for example conmensurate and
incomensurate condensates of charge spin and current. A first step
towards understanding this landscape is to follow the evolution of well
defined phases as a function of control parameters. In this paper we
focused on the superconducting and normal phase.  Other phases and the
competition with superconductivity can be studied with CDMFT
techniques, as was done for example in Ref.~\onlinecite{Capone_Nell} for the
conmensurate antiferromagnetism.

We find that the normal state  in the mean field theory has two
distinct regimes, which are naturally characterized in terms of
the  regimes of the impurity model.  At low doping, in the
immediate proximity of the Mott insulating state,  we have a
realization of the RVB picture of holes propagating in a sea of
spins with strong singlet correlations. In the impurity model
language that corresponds to the RKKY phase of the two impurity
model and its generalization to a plaquette. At high doping, we
have a regime with well formed quasiparticles with a Fermi
surface containing $1-\delta$ electrons. In the impurity model
language this corresponds to the Kondo regime of the  one
impurity model, and single site DMFT provides an adequate
description of its properties.

Plaquette CDMFT has three independent cluster self energies.  For
very large doping, only the local cluster self energy is non zero,
indicating the validity of single site DMFT. As doping is reduced,
$\Sigma_{\pi\pi}(i\omega)$ acquires a large real and imaginary
part. This is controlled by the existence of a pole which
approaches zero frequency at certain doping $\delta_{1}^c$
($\delta_{1}^c\sim 0.1$ in the t-J and $\delta_1^c = 0$ in the
Hubbard model). When combined with the cumulant periodization,
this anomalous growth, gives rise to a topological transition
associated to the formation of line of zeros in the Greens
function (line of poles in the self energy) at zero temperature
discussed in Ref.~\onlinecite{tdstan}. We call the doping at
which the topological transition of the Fermi surface happens
$\delta_{3}^c$. Notice however, that from a CDFMT perspective
which focus on the finite temperature description, an infinite
self energy is not necessary, and all that is required to
generate the pseudogap regime with its concomitant formation of
Fermi arcs, is a self energy which exceeds the bandwidth.

We identified another critical doping, $\delta_{2}^c$ associated with
a maximum in the scattering rate of the third cluster degree of
freedom $\Sigma_{0\pi}$.
This is an example of cluster quantum multicriticality.  Namely, a
mapping of a lattice model onto a quantum impurity model with a
critical point.  This critical point satisfies the following
conditions: a) it requires a cluster of impurities for its
existence, hence it has no analogy in single site DMFT, and b)
has at least two unstable (relevant) directions (for example the
ratio $J_{kondo}/J_{RKKY}$ and particle hole symmetry breaking in
the two impurity model Varma Jones critical point 
\cite{VarmaJones}).

At a critical doping $\delta_{2}^c$, there is an avoided critical
point in the normal phase, which is near the doping level with the
highest superconducting transition temperature.  Since $\delta_{2}^c >
\delta_{1}^c$, $\delta_2^c$ may lie very close to $\delta_{3}^c$ if
one adopts a periodization scheme along the lines of the cumulant
periodization, but this issue is left for future studies since it
strongly depends on the periodization scheme.  Looking at the
scattering rate and coherence temperature in the normal state solution
of CDMFT equations, we thus identified a critical doping
$\delta_{2}^c$, which could be related to the hidden quantum critical
point which has been hypothesized by many authors based on a large
body of experimental data \cite{Tallon,VarmaJones}.

We  have not analyzed the properties of the CDMFT quantum  impurity
model describing the normal state  at zero temperature.
It is even possible   that
the normal state solution simply does not exist at $ T=0$. These
points are largely academic  from the point of view of the finite
temperature  physics which we want to describe with CDMFT. The
manifestations of the possible quantum criticality are rapidly
removed by the onset of superconductivity. The electronic
lifetime  on the Fermi surface controlled by $ Im \Sigma(0,\pi)$
is dramatically reduced as the system goes superconducting.

One then arrives at a superconducting state, which inherits the normal
state gap, largely caused by $\Sigma_{\pi\pi}$, but with coherent
nodal quasiparticles characterized by a weakly doping dependent
velocity perpendicular to the Fermi surface. The velocity along
the Fermi surface $v_\Delta$ has a dome like shape and decreases in
the underdoped regime providing further support to  the two gap
picture  of the
superconducting state of the underdoped  cuprates
\cite{ram-twogap,Tacon,Tanaka,Hanke,Marcello-submitted}.

The superconducting phase is stabilized by the gain of superexchange
energy, namely improved spin-spin singlet correlations.
We resolved the frequency dependence of the anomalous Greens function
and the anomalous self energy, and found them to have a structure very
different from conventional phonon mediated superconductivity in the
Migdal Eliashberg theory. Since the superconducting state restores
coherence, long lived sharp excitations, Bogolubov quasiparticles and
a sharp spin mode which resembles the neutron "40 meV resonance",
emerge below $T_C$.

We extracted different observables such as  tunneling density of
states, optical conductivity, optical mass and plasma frequency,
integrated optical spectral weight superfluid stiffness and spin
susceptibility which  compare well  at a qualitative level with
experimental data on copper oxide materials.

We have shown that at $\delta_{2}^c$ (which occurs very near the
maximum in $T_c$) the coherence energy vanishes and the scattering
rate is maximal.  At this doping, an approximate particle-hole
symmetry in the one electron spectra is recovered, and approximate
power laws in physical quantities ($\sigma\propto\omega^{-2/3}$)
emerge in an intermediate frequency range.

Upon periodization, the large value of the non local self energies
turn Fermi surface into Fermi arcs \cite{tdstan}, and we studied the
evolution of the fermi arcs with temperature. We showed that within
C-DMFT fermi arcs shrink with decreasing temperature.

Our solution of the CDMFT on a plaquette, has many similarities
with the earlier studies of Anderson's Resonating Valence Bond
theory of high temperature superconductivity in the  slave boson
mean field theory formulation. This approach, correctly
predicted the  d-wave symmetry of the superconducting order
parameter and the presence of a pseudogap with the same symmetry
well above $T_c$\cite{Gabi}.

The similarity between the results of the CDMFT and slave  boson  approaches is not
accidental. Both methods  are mean field techniques  based on order
parameters that can be defined within a plaquette, and
capture the effects of the proximity to a Mott insulating state using
a small set of short range degrees of freedom.

Compared with slave boson mean field theory, CDMFT has additional
flexibility both in the frequency of the one electron spectral
function as well as in  its momentum dependence. One crucial
difference, is a much more pronounced momentum space
differentiation with very different electronic properties at the
nodes and at the antinodes.
This anisotropy, with the concomitant existence of two energy scales
in the superconductor, resolves the earlier problems of the RVB theory
related to the doping dependence of the linear term of the penetration
depth \cite{wen-lee}.
The need for the introduction of more
anisostropy in the microscopic theory had been anticipated by
experiments, and by the phenomenological analysis of Ioffe and
Millis \cite{Ioffe1,Ioffe2,Ioffe3}. Recent phenomenological models
\cite{Phenom1,Phenom2} have also generated a more pronounced
momentum dependence of the one particle spectra, and incorporated
in their approach a $v_{\Delta}$ that decreases with decreasing
doping. The main differences between these phenomenological
approaches and the more microscopic C-DMFT reside  in the
location of the lines of zeros of the Green function.  While in
Refs.~\onlinecite{Phenom1,Phenom2} the lines of zero lie on the
the umklapp surface,  in C-DMFT the lines of zeros are dynamical
entities with a location that evolve with doping.

CDMFT is an extension of single site DMFT an  approach that   has
been very successful in describing many aspects of the finite
temperature Mott transition.   Using a single impurity in a medium,
this method, has been able to describe several regimes near a Mott
transition. A Fermi liquid regime, at small $U$ and temperature, a
bad metal at temperature larger than a characteristic temperature
$T_{coh}(U)$, a Mott insulator at large $U$ and low temperature, and
a bad insulator when the temperature is high enough that the
Hubbard bands begin to merge.

By construction, single site DMFT assigns the same self energy to the
electronic states on the whole fermi surface. Hence  at a given
energy and temperature,  either  all the states at all  $\vk$ points
are coherent, or they are all incoherent.  This is not a good
description of the high temperature superconductors, which
therefore cannot be described with single site DMFT.

On the other hand, CDMFT allows the states in the nodal region to
be coherent quasiparticles while at the same time the states in
the antinodal region are highly incoherent and have a pseudogap,
i.e., $T_{coh}(nodal)\ll T_{coh}(antinodal)$. The self energy in
the nodal region could be compared to a single site DMFT in the
Fermi liquid regime with $U< U_{c2}$ and $T_c<T_{coh}(U)$ while
the antinodal self energy is more of a single site DMFT in the
bad insulator regime $U>U_{c2}$ and $T_c>T_{coh}(U)$.
Plaquette DMFT, offers a mean field picture of the lattice
problem, whereby the different cluster self energies and cumulants
describe different regions of momentum space with distinct
physical properties. A nodal region which is closest to a Fermi
liquid, an antinodal region which exhibits a  pseudogap, and an
intermediate region between the two, described by the $(0,\pi)$
self energy, which exhibits  the maximum scattering rate at
criticality.

This qualitative picture is only a crude caricature of the full CDMFT
solution, but it is a useful qualitative guide to understand how the
Fermi arcs originate from the proximity to the Mott insulator, and
above all, highlights why single site DMFT is inadequate in this
situation.

The objective of this work was to advance our understanding of the t-J
and Hubbard model as a "bare bones" model of the density driven Mott
transition. Important open problem is to incorporate and understand
how other effects, such as the effects of more realistic band
structure in the multi band model, the disorder and the electron
phonon interaction, which play an important role in cuprates, can
effect the solution of the model,

We presented a qualitative comparison with several experiments in
the copper oxide based materials, and given the limitations of
the model and of the methodology used, this comparison is very
encouraging and warrant future studies including more accurate
modeling and further methodological improvements.

Future studies should include a realistic band structure of the
copper-oxygen planes and additional Coulomb terms beyond the local
Hubbard $U$, that can be accomodated on the plaquette.  In addition to
$d_{x^2-y^2}$ copper band, it would be desirable to include
another copper band, namely, $d_{z^2}$ band which is coupled to apical
oxygen. Although the latter band is filled in the band structure
calculation, it comes close to the Fermi level.

Another important direction is to better momentum resolve one
particle and two particle quantities. The latter will require
advances in the analytic continuation techniques of QMC data, as
well as a better understanding of how to convert cluster
quantities into lattice observables in C-DMFT. Furthermore within
a cluster size, it is important to implement an optimal choice of
orbitals in CDMFT describing different momentum patches in the
Brillouin zone.
Functional approaches \cite{olivier,new-review}
as well as CDMFT inspired, modeling of experimental data along the
lines of Ref.~\onlinecite{Perali} can provide useful guidance in this
direction.

Mean field approaches clearly separates the short distance effects
contained in the theory, from long distance effects which will require
the introduction of fluctuations due to vortices and pair
fluctuations. The $T_c $ vs $\delta$ line in CDMFT should be
interpreted as being close to the Nerst line in the cuprate phase
diagram \cite{ong}. On the other hand, the true superconducting
critical temperature line, is strongly reduced relative to the CDMFT
on the underdoped side of the phase diagram to the effects of long
wavelength fluctuations of the order parameter, which require long
wavelength field theoretical techniques along the lines of
Ref.~\onlinecite{zlatko}

Finally other inhomogeneous phases, such as stripes, charge, bond, 
pair density waves, and other broken symmetries can appear as
secondary instabilities, and can be studied with our methods by
inserting a relatively local (restricted to a plaquette), but site
dependent self energies into the CDMFT functional.

\section{Acknowledgements}
We wish to thank M. Civelli for very enlightening discussion and
comparison of numerous data with comparative exact diagonalization
study.  Fruitful discussions with C. Marianetti, P. W\"olfle
A. Georges and O. Parcollet, C. Castellani and M. Capone are
gratefully acknowledged.  GK was supported by the NSF under grant DMR
0528969.

\end{document}